\documentclass[a4paper,11pt]{article}
\pdfoutput=1 
\usepackage{jcappub}
\usepackage{amsmath}
\usepackage{graphicx}
\usepackage{latexsym}
\usepackage{xspace}
\usepackage{color}
\usepackage{hyperref} 
\usepackage{bm}
\usepackage{relsize}
\usepackage{tabularx}
\usepackage{multirow}
\usepackage{amssymb}
\usepackage[table]{xcolor}
\usepackage{slashbox}
\usepackage{braket}
\usepackage{comment}
\usepackage[utf8]{inputenc}
\usepackage{slashed}
\usepackage{empheq}
\usepackage[T1]{fontenc} 

\allowdisplaybreaks

\makeatletter
\gdef\@fpheader{}
\g@addto@macro\bfseries{\boldmath}
\makeatother

\makeatletter
\renewcommand*\env@matrix[1][\arraystretch]{%
  \edef\arraystretch{#1}%
  \hskip -\arraycolsep
  \let\@ifnextchar\new@ifnextchar
  \array{*\c@MaxMatrixCols c}}
\makeatother

\newcommand{\ds}{\displaystyle}

\newcommand{\ie}{\textsl{i.e.~}}

\newcommand{\N}{\mathcal N}

\newcommand{\dd}{\mathrm{d}}

\newcommand{\sss}[1]{{\scriptscriptstyle{#1}}}

\newcommand{\uPl}{\mathrm{Pl}}

\newcommand{\usssPl}{\sss{\uPl}}

\newcommand{\Mp}{M_\usssPl}

\newcommand{\beq}{\begin{equation}}
\newcommand{\eeq}{\end{equation}}
\newcommand{\bea}{\begin{eqnarray}}
\newcommand{\eea}{\end{eqnarray}}

\newlength{\wsingfig}
\newlength{\wdblefig}
\newlength{\wquadfig}
\newlength{\wtriplefig}
\newcommand{\tsr}{_{\mu\nu}}                
\newcommand{\utsr}{^{\mu\nu}}               
\newcommand{\stsr}{_{ij}}                   
\newcommand{\ustsr}{^{ij}}                  
\newcommand{\Tsr}{_{IJ}}                    
\newcommand{\uTsr}{^{IJ}}                   
\newcommand{\cov}{_\text{cov}}

\subheader{}

\title{Explicit gauge-invariant variables in multifield inflation beyond linear order and Hamiltonian dynamics
}

\author[a]{Julien Grain,}
\emailAdd{julien.grain@universite-paris-saclay.fr}

\author[a]{Hugo Holland,}
\emailAdd{hugo.holland@universite-paris-saclay.fr}

\affiliation[a]{Universit\'e Paris-Saclay, CNRS, Institut d'Astrophysique Spatiale, 91405, Orsay, France}

\author[b]{Lucas Pinol}
\emailAdd{lucas.pinol@phys.ens.fr}

\affiliation[b]{Laboratoire de Physique de l'Ecole Normale Sup\'erieure, ENS, CNRS, Universit\'e PSL,
Sorbonne Universit\'e, Universit\'e Paris Cit\'e, F-75005, Paris, France}

\date{today}

\begin{document}
\sloppy

\abstract{
General relativity coupled to multiple scalar fields is a diffeomorphism-invariant constrained system.
Consequently, a naive counting of the perturbative degrees of freedom unavoidably overestimates the true number of physical modes propagating in the theory, as gauge redundancies and constraint equations remove non-dynamical ones.
While this problem has been solved for linear fluctuations, this work presents the first explicit calculation of all large-scale gauge-invariant phase-space variables in multifield inflation and at second order in perturbation theory, in a Hamiltonian language.
Building upon the well-known Sasaki-Mukhanov variables, we show how to construct a finite-dimensional basis of quadratic corrections which are invariant under gauge transformations.
Although our procedure is generic to any number of fields and at any scale, we restrict to super-Hubble scales for their explicit solution, which we deliver.
Henceforth, we prove that it is possible to recover the usual flat-gauge and comoving-gauge fluctuations as large-scale gauge-invariant combinations, making for a robust consistency check of the gauge-fixed procedure to connect theoretical predictions above the horizon to observations.
We derive the quadratic and cubic Hamiltonian of multifield inflation in a gauge independent manner, then we gauge fix our theory by going into the flat gauge,  and we show perfect agreement with the literature on this topic, usually based on a Lagrangian approach.
After these concrete steps, we propose a more formal proof of the existence of gauge-invariant variables at quadratic order, and we provide a sketch of the procedure that should allow to go to higher orders in perturbation theory.
}



\maketitle

\flushbottom

\section{Introduction}

Despite the undeniable success of cosmic inflation in describing the earliest moments of our Universe, its microphysical origin remains so far unknown. 
Although a sustained period of accelerated expansion of spacetime indeed solves the issues in the standard Hot Big Bang scenario,  no field from the Standard Model of particle physics can be responsible for it.
Physicists must then invoke additional fields with tailored properties; chief amongst which is the famous slow-roll model of inflation: a single scalar field canonically coupled to gravity and with canonical kinetic terms slowly descends its almost flat potential.
Since this occurs in the very early universe, at energy scales far beyond those probed in terrestrial experiments, these procedures do not contradict the current absence of detection of physics beyond the Standard Model on Earth. 
However, once the door to new physics has been opened, one is confronted with a tantalising amount of possible theories.
Surprisingly, many of them—single--field or multi-field, minimally coupled or not to gravity, with canonical kinetic terms or not—can provide enough accelerated expansion, leading to a strong model degeneracy.
In order to break those degeneracies, one must look into the detailed properties of inflation, such as the physics of fluctuations beyond the homogeneous and isotropic background evolution.


The description of cosmological fluctuations during inflation is a delicate exercise.
It requires the use of both quantum field theory and general relativity.
In practice, early universe cosmologists rely on a semi-classical treatment.
The background spacetime is treated as a classical (non-quantum) relativistic universe whose dynamics is dictated by the usual Einstein equations coupling it to the homogeneous and isotropic component of the matter content.
On top of this background, fluctuations develop both in the spacetime metric and in the matter content, resulting in two main difficulties.
On the one hand, general relativity being a constrained gauge theory, relativistic effects must be taken into account: gauge redundancies appear; matter perturbations back-react on spacetime perturbations; and constraint equations must be solved.
On the other hand, once the physically propagating degrees of freedom have been identified, those are promoted to quantum operators, leading to the aforementioned semi-classical treatment: a classical background and quantised fluctuations.
In this work, we will mostly address the first of these aspects.

The question of gauge issues has interested several generations of early universe cosmologists.
After pioneering work to tackle general gauge issues in linear cosmological perturbation theory in ~\cite{BardeenGauge,KodamaSasakiGauge}, Mukhanov and Sasaki independently proposed their eponymous gauge-invariant variables, tailored to single-field inflation, in~\cite{Mukhanov:1985rz,Sasaki:1986hm}.
The first review dedicated to this topic appeared shortly after~\cite{MUKHANOV1992203}, before extensions to multiple fields in~\cite{Taruya:1997iv, Gordon:2000hv, Hwang:2000jh, GrootNibbelink:2001qt}.
For more modern reviews, see also~\cite{Malik:2008im,Gong:2016qmq}.
The common ground of these references is to describe gauge transformations at the level of the fluctuations themselves, to explicitly construct gauge-invariant combinations, and to use those to make phenomenological predictions.
Alternatively, Ref.~\cite{Langlois:1994ec} proposed instead to identify the phase-space volume that is left invariant by gauge transformations in a Hamiltonian formulation.
Other works formulating gauge invariance at the level of the Hamiltonian include~\cite{Gong:2016qpq} with an interesting complementary approach based on the path integral formalism and~\cite{Grain_2026} where the justification for the separate universe approach was carefully investigated, both encompassing multiple scalar fields with non-canonical kinetic terms.

Leveraging these robust theoretical bases, it has become increasingly common to make predictions for inflationary linear fluctuations that can be readily compared to actual astronomical observations.
This theory-observations cycle culminated in the precise measurements of the Gaussian statistics of the primordial adiabatic scalar fluctuation---and constraints on the maximum amount of primordial tensor modes---by the Planck collaboration~\cite{Planck:2018jri}.
Since the end of the Planck mission, more recent Cosmic Microwave Background observations have even tightened the bounds on tensor modes which so far remain undetected, see, e.g., Ref.~\cite{Balkenhol:2025wms} for a recent joint analysis of Planck-BICEP-Keck-ACT-SPT data.
This is unfortunate, as although studying linear scalar fluctuations allowed us to discriminate between valid and invalid models of inflation, many possibilities remain open.
Remarkably, no lower bound on the amount of tensor modes can be reliably derived, leaving only wishful thinking in this direction.
But another aspect of fluctuations can be used to break model degeneracies and investigate the microphysical details of inflation, namely their non-linearities.

Non-linearities in the early universe are necessarily present due to the nature of gravity, but this gravitational floor may be complemented by direct interactions in the matter sector, making them a direct probe of high-energy physics beyond what we will ever access on Earth.
Indeed, they also have important phenomenological consequences, chief amongst which are primordial non-Gaussianities; however, other, more subtle effects might also appear, such as loop corrections.
The question of gauge issues at non-linear order was much less studied. 
Most applications generally pick a certain gauge and perform calculations within it, without formally proving the gauge-invariance of this procedure.
In his pioneering work, Maldacena~\cite{Maldacena:2002vr} streamlined two parallel derivations, one in the flat gauge, one in the comoving gauge.
In multi-field inflation, the flat gauge is usually chosen to study primordial non-Gaussianities~\cite{Langlois:2008mn,Gong_2011,Dias:2015rca}, though the comoving gauge is sometimes chosen after the adiabatic-entropic decomposition~\cite{Garcia_Saenz_2020,Pinol:2020kvw}.
Gauge fixing is also a common procedure in the Effective Field Theory of inflationary fluctuations~\cite{Creminelli:2006xe,Cheung:2007st} and its multi-field extensions~\cite{Senatore:2010wk, Noumi:2012vr, Pinol:2023oux, Pinol:2024arz}.
Although not necessarily pathologic, this procedure a priori requires a careful identification between the theoretical predictions and the frame in which the observations are made. 
A more robust approach would consist in directly using gauge-invariant non-linear variables.
A few remarkable works in this direction include the early Ref.~\cite{Bruni:1996im}, as well as Ref.~\cite{Malik:2003mv}, both identifying such combinations at the level of the fluctuations (see also~\cite{Nakamura:2004wr} for a more formal approach).
It is noticeable that no Hamiltonian approach was proposed at non-linear order, with the exception of~\cite{Dom_nech_2018} containing important consequences for scalar-induced tensor modes. 

The goal of this work is to assess whether the gauge-fixed procedure used in almost any non-linear applications truly removes all gauge redundancies.
To do so, we study the kinematics of fluctuations from covariant models of multifield inflation with curved field space, under perturbative gauge transformations.
In Sec.~\ref{sec: cosmo perturbations} we provide the perturbative expansion of the Hamiltonian of general relativity coupled to these general non-linear models, keeping all gauge degrees of freedom in full generality.
We also recall how one should use covariant phase-space fluctuations when the field space is curved.
With this Hamiltonian description at hand, in Sec.~\ref{sec: GI variables}, we first recover the usual Mukhanov-Sasaki variables at linear order and then apply our more general procedure to quadratic corrections.
We proceed by increasing order of complexity, starting from a single field to two fields and finally to any number of fields.
By working on super-Hubble scales where spatial gradients are negligible (assuming a standard attractor inflationary scenario), we explicitly construct the finite-dimensional basis of quadratic combinations that are truly gauge invariant.
We show that it is possible to organise these terms to recover both the flat-gauge and the comoving-gauge fluctuations, thus proving that they are indeed gauge invariant.
Strictly speaking, we can only confirm this at super-Hubble scales, but we argue that this is enough to connect phenomenological predictions to actual cosmological observations: our predictions below the horizon might depend on the frame used to make them, but they eventually collapse to unambiguous statements as the physical wavelengths of cosmological perturbations stretch above the horizon.
In Sec.~\ref{sec:gauge fixing}, we also devise a Hamiltonian flat-gauge-fixing procedure up to cubic interactions and prove that the result exactly matches the literature that usually relies on a Lagrangian approach, making for an additional robustness check.
Finally, after these concrete applications and explicit formulas, we provide in Sec.~\ref{ssec:GI 1 and 2} a formal construction of gauge-invariant variables at non-linear order, offering hints on the procedure to derive cubic-order gauge-invariant fluctuations, a priori relevant for applications including quartic interactions.
We finish this article with a conclusion and discussion in Sec.~\ref{sec: conclusion}.


\section{Cosmological perturbations at the cubic order}
\label{sec: cosmo perturbations}

\subsection{Non-linear sigma models in Hamiltonian}
We start by introducing the Hamiltonian of general relativity in the context of multifield inflation. We aim to study multifield models in a generic way. To this end we introduce scalar fields $\phi^I$ where $I$ runs from $1$ to $\mathcal N$ the number of fields. We also introduce $G\Tsr$ the coupling metric and $V(\phi^I)$ a potential, both depend on all fields. We also introduce $g\tsr$ the usual four-dimensional metric of the curved space time, and $g$ its determinant. We assume that the scalar fields are all minimally coupled to gravity, which leads to the following action
\begin{equation}
    S = \int \dd^4x \sqrt{-g} \left[\frac{\Mp^2}{2}  R - \frac 1 2 g\utsr \,G\Tsr(\phi^K)\, \partial_\mu \phi^I \partial_\nu \phi^J - V(\phi^I) \right]\,, 
\end{equation}
where $R$ is the Ricci scalar associated to the space time metric. The metric $g\tsr$ and its inverse $g\utsr$ are used to raise and lower spacetime indices. Similarly $G\Tsr$ and its inverse $G\uTsr$ lower and raise field space indices. This family of models is referred to as non-linear sigma models. To get the Hamiltonian description let us foliate our spacetime into three-dimensional hypersurfaces. To do so we introduce two Lagrange multipliers: the lapse $N(\tau,\vec x)$ and the shift $N^i(\tau,\vec x)$, where we have used $\vec x$ the coordinates on the hypersurfaces and $\tau$ the time coordinate orthogonal to the foliation. These quantities allow us to use the ADM formalism, writing the metric as
\begin{equation}
    \dd s^2 = -N^2(\tau, \vec x) \dd\tau^2 + \gamma\stsr(\tau,\vec x)\left[\dd x^i + N^i(\tau,\vec x)\dd\tau\right]\left[\dd x^j + N^j(\tau,\vec x)\dd\tau\right]\,,
\end{equation}
where $\gamma\stsr$ is the induced metric on the hypersurfaces and $\gamma$ its determinant. It can be used, alongside its inverse $\gamma\ustsr$ to raise and lower indices on these sheets. Let us also introduce the conjugate momenta of our variables, using for any tensor $\dot f \equiv \dd f / \dd \tau$, 
\begin{equation}
    \pi_I \equiv \frac{\delta L}{\delta \dot\phi^I},\quad \pi\ustsr \equiv \frac{\delta L}{\delta \gamma\stsr}.
\end{equation}
Since the lapse and the shift functions are Lagrange multipliers we do not need to define conjugate momenta. They are associated with the freedom of coordinate system\footnote{Fixing $N$ fixes the time coordinate, and fixing $N^i$ fixes the space coordinates on the hypersurfaces.}. Varying the action with respect to the functions themselves will lead to the scalar and diffeomorphism constraints $\mathcal C$ and $\mathcal D_i$. The action can now be written in the Hamiltonian form
\begin{equation}
    S=\int \dd \tau\int\dd^3x\left(\pi_I\dot{\phi}^I+\pi^{ij}\dot{\gamma}_{ij}-N\mathcal{C}-N^i\mathcal{D}_i\right). \label{eq:SHam}
\end{equation}
To express the constraints we introduce the trace of the gravitational momentum $\pi=\pi\ustsr\gamma\stsr$ and the three-dimensional Ricci scalar $R^{(3)}$ of the space-like hypersurfaces. We separate the scalar constraint in two parts for convenience $\mathcal C = \mathcal C^\phi + \mathcal  C^G$ and we find
\begin{align}
    \mathcal C^G =&\, \frac 2 {\Mp^2 \sqrt \gamma} \left(\pi\stsr\pi\ustsr - \frac 1 2 \pi^2\right) - \frac {\Mp^2\sqrt \gamma} 2 R^{(3)} \, , \\
    \mathcal C^\phi =&\, \frac{1}{2 \sqrt \gamma} G\uTsr \pi_I\pi_J  \frac { \sqrt \gamma }{2}\gamma^{ij}G\Tsr  \partial_i\phi^I\partial_j\phi^J + \sqrt{\gamma}V .
\end{align}
The first term contains the contribution from the gravitational sector, and the second from the matter sector. The diffeomorphism constraint reads $\mathcal D_i = \mathcal D_i^\phi + \mathcal D_i^G$ with
\begin{align}
    \mathcal D_i^G =&\, -2\partial_i(\gamma\stsr\pi^{jm}) + \pi^{mn}\partial_i\gamma_{mn} \,, \\
    \mathcal D_i^\phi =&\,\pi_I\partial_i\phi^I. \label{diff constraint}
\end{align}
We can read the Hamiltonian off of Eq. (\ref{eq:SHam}) 
\begin{equation}
    H[N,N^i, \phi^I,\pi_I,\gamma\stsr,\pi\ustsr] = \int \dd^3\vec x\left(N\mathcal C(\phi^I,\pi_J,\gamma\stsr,\pi\ustsr) + N^i\mathcal D_i(\phi^I,\pi_J,\gamma\stsr,\pi\ustsr)\right).
\end{equation}
We are working here with sets of canonical pairs, where the Poisson brackets read $\{\phi^I(\tau,\vec x),\pi_J(\tau,\vec y)\} = \delta^I_J\delta^3(\vec x - \vec y)$ and $\{\gamma\stsr(\tau,\vec x),\pi^{mn}(\tau\vec y)\} = \frac 1 2 (\delta^m_i\delta^n_j + \delta^n_i\delta^m_j)\delta^3(\vec x - \vec y)$. The time evolution of any given function $T$ of any of these variables is obtained from the Poisson bracket with the Hamiltonian
\begin{equation}
    \dot T(\phi^I,\pi_J;\gamma\stsr,\pi\ustsr) = \{T,H\}. \label{eq:EoM Ham}
\end{equation}
As previously mentioned, the lapse and shift functions are Lagrange multipliers and varying the action with respect to these functions imposes that the constraints vanish
\begin{equation}
    \mathcal C = 0, \quad \mathcal D_i = 0.
\end{equation}
These constraints are first class\footnote{This means that their Poisson brackets weakly vanish, \ie they verify $\{C_a,C_b\}=f^c_{ab}C_c$, where $C_a$ is either the scalar or the diffeomorphism constraint, and $f^c_{ab}$ is a function of background quantities.} and are preserved through the evolution of both the matter and gravitational fields. The constraints, and thus the Hamiltonian are explicitly covariant in the field phase-space, and this is a property we wish to keep throughout this work. To this effect we define the covariant derivatives of covariant and contravariant vectors $U^I$ and $W_I$
\begin{equation}
    D_\mu U^I = \partial_\mu U^I + \Gamma^I_{LK}\partial_\mu\phi^LU^K \quad \mathrm{and} \quad D_\mu W_I = \partial_\mu - \Gamma^K_{IL}\partial_\mu\phi^LW_K\,,
\end{equation}
where the Christoffel symbols are those associated to the coupling metric $G\Tsr$. We can finally write the equations of motions for the fields and their momenta, making use of Eq. (\ref{eq:EoM Ham})
\begin{align}
    \dot \phi^I =&\, \frac N{\sqrt\gamma}G\uTsr\pi_J + N^i\partial_i\phi^I \label{eq:dotphigen} \,,\\
    D_\tau   \pi_I =&\, - \sqrt \gamma N \frac{\partial V}{\partial\phi^I} + D_i (\sqrt \gamma N G\Tsr \gamma\ustsr \partial_j \phi^I) + D_i(N^i\pi_I)\,,\\
    \dot{\gamma}_{ij}=&\,\frac{2N}{\Mp^2\sqrt{\gamma}}\left[2\pi_{ij}-\gamma_{ij}\pi\right]+2\gamma_{mj}\partial_iN^m+N^m\partial_m\gamma_{ij}\,, \\
    \dot{\pi}^{ij}=&\,\frac{2N\gamma^{ij}}{\Mp^2\sqrt{\gamma}}\left(\pi_{mn}\pi^{mn}-\frac{1}{2}\pi^2\right)+\frac{2N}{\Mp^2\sqrt{\gamma}}\left(\pi\pi^{ij}-2\gamma_{mn}\pi^{im}\pi^{jn}\right)+\frac{N\Mp^2\sqrt{\gamma}}{2}\frac{\delta R^{(3)}}{\delta\gamma_{ij}} \nonumber \\
	&\,+\frac{N}{2 \sqrt \gamma} \gamma^{ij}G\uTsr \pi_I\pi_J +\frac{N}{2}\gamma^{im}\gamma^{jn}G\Tsr  \partial_m\phi^I\partial_n\phi^J-2\pi^{jm}\partial_mN^i-\partial_m\left(N^m\pi^{ij}\right).
\end{align}
Let us note that it is equivalent to use either the covariant derivative or the usual time derivative for the gravitational fields since they are scalars in the field space. It can be surprising however to see a covariant expression in Eq. (\ref{eq:dotphigen}) despite using a non covariant derivative. This is because the fields are the base vectors of the field space, hence the equality $\dot \phi^I = D_\tau  \phi^I$. In other words we are considering a collection of fields, not a vector composed of all the fields.

\subsection{Homogeneous and isotropic background}\label{ssec:background}
Let us assume homogeneity and isotropy for both the metric and the fields. The action is greatly simplified in this case since it leads to $N^i = 0$, $\mathcal D_i = 0$ and $R^{(3)} = 0$. The metric is assume to be a flat FLRW and reads
\begin{equation}
    \dd s^2=-N^2(\tau)\dd\tau^2+v^{2/3}(\tau)\tilde{\gamma}_{ij}\dd x^i\dd x^j\,,
\end{equation}
where $v\equiv a^3$, $a$ being the usual scale factor. We also introduce here $\theta$ the conjugate momenta to $v$. The induced metric and its conjugate momentum now read
\begin{equation}
    \gamma\stsr = v^{2/3}\delta\stsr, \quad \pi\ustsr = \frac{v^{1/3}\theta}{2}\delta\ustsr\,,
\end{equation}
where $\delta\stsr$ is the time independent three dimensional flat metric. Plugging this\footnote{In reality this corresponds to a canonical transform, but the derivative of the generating function vanishes, thus simply plugging the replacement is enough.} into Eq. (\ref{eq:SHam}) leads to the background\footnote{We have not yet separated the fields into a background and perturbations but the equations derived here will indeed be those of the background.} action and associated scalar constraint 
\begin{align}
    S^{(0)}=&\,\ds\int\dd \tau\left[\pi_I\dot{\phi}^I+\theta\dot{v}-N\mathcal{C}^{(0)}\right]\,,\\
	\mathcal{C}^{(0)}=&\,\frac{1}{2v}G^{IJ}\pi_I\pi_J+vV-\frac{3}{4\Mp^2}v\theta^2. \label{eq:C0}
\end{align}
And the equations of motion for the isotropic fields are
\begin{align}
    \dot\phi^I=&\,\frac{N}{v}G^{IJ}\pi_J,  \quad D_\tau  \pi_I=-NvV_{;I}\,, \label{eq:dotphi & dotpi} \\
	\dot{v}=&\,-\frac{3N}{2\Mp^2}v\theta, \quad \dot{\theta}=N\frac{G^{IJ}\pi_I\pi_J}{v^2}. \label{eq:dotpig & dotgamma}
\end{align}
Where we have used the notation $V_{;I}$ to indicate a covariant derivative with respect to the field $\phi^I$ \footnote{Since the potential is a scalar, this is equal to the usual derivative, however this does not remain true for tensors, for example $V_{;IJ} \equiv V_{,IJ} - \Gamma^K_{IJ}V_{;K}$.}. These equations of motion need to be solved under the constraint $\mathcal C^{(0)} = 0$. Let us also introduce the energy density and pressure of the matter sector 
\begin{align}
    \rho \equiv &\,\frac{1}{2v^2}G^{IJ}\pi_I\pi_J+V\,, \label{eq:rho} \\
	p \equiv &\,\frac{1}{2v^2}G^{IJ}\pi_I\pi_J-V\,, \label{eq:p}
\end{align}
where we recall that the potential can depend on all the matter fields $\phi^I(\tau)$.

\subsection{Cosmological perturbation theory}
To apply the Hamiltonian formalism to the context of inflationary space times we split the Lagrange multipliers and the fields into homogeneous and isotropic backgrounds, and cosmological perturbations. These perturbations are fluctuations around this background and represent deviations from homogeneity and isotropy. Mathematically, we define these perturbations as 
\begin{align}
    \delta\phi^I(\tau,\vec{x})=&\,\phi^I(\tau,\vec{x})-\phi^I(\tau)\,, \label{eq:deltaphi} \\
	\delta\pi_J(\tau,\vec{x})=&\,\pi_J(\tau,\vec{x})-\pi_J(\tau)\,, \label{eq:deltapi} \\
	\delta\gamma_{ij}(\tau,\vec{x})=&\,\gamma_{ij}(\tau,\vec{x})-v^{1/3}(\tau)\delta\stsr\,, \\
	\delta\pi^{ij}(\tau,\vec{x})=&\,\pi^{ij}(\tau,\vec{x})-\frac{1}{2}v^{1/3}(\tau)\theta(\tau)\delta\ustsr\,,\\
    \delta N(\tau,\vec{x})=&\,N(\tau,\vec{x})-N(\tau)\,, \\
	\delta N^i(\tau,\vec{x})=&\,N^i(\tau,\vec{x})-N^i(\tau).
\end{align}
These definitions are equivalent to performing a translation in the phase-space, as such it is a canonical transformation to go from the original fields to their perturbations, and the perturbations of a canonical pair are a canonical pair themselves. In this perturbed phase-space, the background quantities are treated as simple functions of time, an external parameter.

In all generality, these perturbations can be decomposed in scalar, vector and tensor perturbations degrees of freedom. The different degrees of freedom decouple at the leading order in perturbation and vector perturbations decay at all orders. In this work, we will focus solely on the scalar perturbations and leave the treatment of vector and tensor perturbations to future analysis. The perturbations of the scalar fields and the lapse function only contain scalar degrees of freedom. This is not the case for the gravitational perturbations or the shift function. We will thus set the vector and tensor components to 0 and we are left with two scalar components for the induced metric perturbation, and two for the conjugate momenta.
\begin{align}
    \delta\gamma_1 =&\, \frac{1}{\sqrt 3}\delta\ustsr \delta\gamma\stsr \,,\quad \delta\gamma_2 = D\ustsr \delta\gamma\stsr\,, \label{eq:gamma and pi 1}\\
    \delta\pi_1 =&\, \frac{1}{\sqrt 3}\delta\stsr \delta\pi\ustsr \,,\quad \delta\pi_2 = D\stsr \delta\pi\ustsr. \label{eq: gamma and pi 2}
\end{align}
Where we define $D\stsr = \sqrt{\frac 3 2}\left(\partial_i\partial_j\partial^{-2} - \frac 1 3 \delta\stsr\right)$ with $\partial^{-2}$ the inverse Laplacian operator. We can check that $\left(\delta\gamma_1(\tau,\vec x),\delta\pi_1(\tau,\vec x)\right)$ and $\left(\delta\gamma_2(\tau,\vec x),\delta\pi_2(\tau,\vec x)\right)$ also form canonical pairs. Similarly, the perturbation of the lapse vector can be expanded as a curl free part that derives from a scalar quantity, and a divergence-free vector. Neglecting the latter, the fluctuation reduces to $\delta N_i = \partial_i \delta N_1$. Let us study the dynamics of these perturbations.

\subsection{Covariant Hamiltonian}

\subsubsection*{Covariant variables}
The derivation of the second order action is usually done by naively expanding in the field perturbation Eqs. (\ref{eq:deltaphi}$\&\,$\ref{eq:deltapi}). However this leads to expressions that are not explicitly covariant, \ie some Christoffel symbols appear. This happens because the perturbations themselves are not covariant in the field space. To amend to this issue, we define the covariant perturbations $\left(Q^I,P_J\right)$ as \cite{Gong_2011, Pinol_2019, Grain_2026}
\begin{align}
    \delta\phi^I =&\,   \underbrace{Q^I}_{\delta_1 \phi^I} \underbrace{- \frac{1}{2}\Gamma^I_{JK}   Q^J   Q^K}_{\delta_2 \phi^I} \,, \label{eq:phi to Q}\\
    \delta\pi_I =&\,   \underbrace{P_I + \Gamma^K\Tsr \pi_K   Q^J}_{\delta_1 \pi_I} \label{eq:pi to P} \\ 
    \nonumber &\,+ \underbrace{ \Gamma^K\Tsr   P_K    Q^J + \frac 1 2 \left(\Gamma^S_{IJ,K} - \Gamma^S_{IR}\Gamma^R_{JK} + \Gamma^R\Tsr\Gamma^S_{RK}\right)\pi_S   Q^J   Q^K -\frac 1 6 R^S{}_{KLI}\pi_S Q^K Q^L}_{\delta_2 \pi_I}.
\end{align}
Where we have presented the relation up to the second order which is sufficient in this work since any replacement up to the third order will drop by least action principle. We also give here the generating function associated with the canonical transform.
\begin{align}
    F^{(3)}\left(\delta\phi^I\,P_I\,;\tau\right) =&\, \delta\phi^IP_I + \frac 1 2 \Gamma^K\Tsr\pi_K\delta\phi^I\delta\phi^J + \frac 1 2 \Gamma^K\Tsr P_K\delta\phi^I\delta\phi^J \label{eq:gen function}\\
    \nonumber &\,+\frac 1 6 \left(\Gamma^K_{IJ,L} + \Gamma^K_{LR}\Gamma^R\Tsr\right)\pi_K\delta\phi^I\delta\phi^J\delta\phi^L.
\end{align}
We call the Hamiltonian for these variables $H\cov$, and the action now reads
\begin{equation}
    S = \ds \int \dd^3 x\left( P_ID_\tau Q^I + \delta\pi\ustsr\dot{\delta\gamma}\stsr  \right) - H\cov(N,N^i,Q^I,P_I,\delta\gamma\stsr,\delta\pi\ustsr).
\end{equation}

\subsubsection*{Linearising the Hamiltonian}
At the leading order, the dynamics of these perturbation is obtained by linearising the action Eq. (\ref{eq:SHam}) around the homogeneous and isotropic background up to the second order in perturbations $S = S^{(0)}+S^{(1)}+S^{(2)}$. The background evolution is given by solving the equations of motions derived in \ref{ssec:background}, and assuming this is done, the linear action $S^{(1)}$ drops by least action principle. Thus the dynamics at leading order for the perturbation requires to go to the second order. This is translated to the Hamiltonian framework, with $\mathcal H^{(0)} = N(\tau)\mathcal C^{(0)}$ and $H^{(2)} = \int\dd^3x [\delta N \mathcal C^{(1)}+\delta N^i \mathcal D ^{(1)}_i+N\mathcal{C}^{(2)}]$. In this work we go one step further and compute the third order action as well. As just now presented, we wish to do so for the covariant variables $(Q^I,P_J)$. The Hamiltonian at second and third orders reads
\begin{align}
    H\cov^{(2),(3)}=\ds\int\dd^3x\left[\delta N \left(\tilde{\mathcal{C}}^{(1)}+\tilde{\mathcal{C}}^{(2)}\right)+\delta N^i\left(\mathcal{D}^{(1)}_{i,\text{cov}}+\mathcal{D}^{(2)}_{i,\text{cov}}\right)+N\left(\mathcal{C}\cov^{(2)}+\mathcal{C}\cov^{(3)}\right)\right], \label{eq:Ham 2 3}
\end{align}
Let us note that what we have derived is indeed a \textit{covariant} Hamiltonian, and that the linear equations of motion one derives for any field space tensor $T$ from the second order contributions to this Hamiltonian read 
\begin{equation}
    D_\tau  T = \left\{T,H\cov\right\}.
\end{equation}
We give here the constituents of the Hamiltonian density 
\begin{align}
    \tilde{\mathcal C}^{(n)} =&\, \ds\sum_{m=1}^n \mathcal C^{(m)}(\delta_{n+1-m}\phi^I,\delta_{n+1-m}\pi^I)\,,\\
    \mathcal C^{(n\ge2)}\cov =&\, \ds\sum_{m=2}^n \mathcal C^{(m)}(\delta_{n+1-m}\phi^I,\delta_{n+1-m}\pi^I) + \frac 1 N\frac{\partial F}{\partial t}^{(n)} - \mathcal M^{(n)}\,,\\
	\mathcal{D}^{(n)}_{i,\text{cov}} =&\, \ds\sum_{m=1}^n \mathcal D^{(m)}_i(\delta_{n+1-m}\phi^I,\delta_{n+1-m}\pi^I).
\end{align}
where we defined $\mathcal M^{(n)}$ as the total spatial-derivative terms of order $n$, and $\frac{\partial F}{\partial t}^{(n)}$ contributions to the $n$-th order of the derivative of the generating function. The expression $\tilde{\mathcal C}^{(n)}$ is the sum of all scalar constraint of order $m\leq n$, where we replace the naive perturbation at order $n+1-m$ in the covariant perturbations, whereas ${\mathcal{C}}^{(n)}\cov$ ignores the contribution from the first order and from total spatial-derivatives, but contains contributions from the derivative of the generating function. There are thus three clear differences between these two quantities despite both coming from the linearisation of the scalar constraint up to the $n-$th order. More details about this procedure are given in appendix \ref{app:full Hamiltonians}. Let us apply these definitions to the second and third order,  without assuming any gauge-choice nor any set of gauge-invariant variables.

\subsubsection{Second order Hamiltonian}
We present here the derivation up to the second order. As presented we need the contribution of $\tilde{\mathcal C}^{(1)}$, $\mathcal C^{(2)}\cov$ and $\mathcal D_{i,\text{cov}}^{(1)}$. The scalar constraint at first and second order reads
\begin{align}
    \tilde{\mathcal C}^{(1)} =&\, vV;_K Q^K + \frac{G^{KI}\pi_I}{v}P_K -\frac{\sqrt{3}}{\Mp^2}v^{2/3}\theta\delta\pi_1 \label{eq:scalar 1}\\
    \nonumber &\,+\left( -\frac{1}{\sqrt{3}}\frac{\pi_I\pi^I}{v^{5/3}} + \frac{1}{\sqrt{3}}v^{1/3}V -\frac{\Mp^2}{\sqrt{3}}\frac{\partial^2}{v^{1/3}} \right) \delta\gamma_1 + \frac{\Mp^2}{\sqrt{6}}\frac{\partial^2}{v^{1/3}} \delta\gamma_2 \,,\\
    \mathcal C^{(2),\phi}\cov =&\, \frac{1}{2v}P_IP^I - \frac{\sqrt{3}}{2}\frac{\pi^I}{v^{5/3}}P_I\delta\gamma_1 + \frac{1}{4v^{1/3}}\frac{\pi_I\pi^I}{v^2}\left(\frac{5}{4}\delta\gamma_1^2 + \frac{1}{2}\delta\gamma_2^2\right) - \frac{1}{2v}R_I{}^{KL}{}_J\pi_K\pi_LQ^IQ^J \label{eq:scalar 2 phi}\\
    \nonumber&\,+ \frac{v}{2}\left(V_{;IJ}-\frac{\partial_i\partial^i}{v^{2/3}}\right)Q^IQ^J+\frac{\sqrt 3}{2}v^{1/3}V_{;I}Q^I\delta\gamma_1 + \frac{1}{4} \frac{V}{v^{1/3}}\left(\frac{1}{2}\delta\gamma_1^2-\delta\gamma_2^2\right)\,,\\
    \mathcal C\cov^{(2),G} =&\, \frac 1 {\Mp^2} \left\{ v^{1/3}\left[-\delta\pi_1^2 + 2 \left(D_{mn}\delta\pi_2\right)\left(D^{mn}\delta\pi_2\right)\right] +\frac{\theta}2 \left[-\delta\gamma_1\delta\pi_1 + 2 \left(D_{mn}\delta\gamma_2\right)\left(D^{mn}\delta\pi_2\right)\right] \right. \label{eq:scalar 2 g}\\
    \nonumber &\,\left. +\frac 1 {32} \frac{\theta^2}{v^{1/3}}\left[\delta\gamma_1^2 + 10\left(D_{mn}\delta\gamma_2\right)\left(D^{mn}\delta\gamma_2\right)\right]\right\}\\
    \nonumber&\, +\frac{\Mp^2}{24v}\left\{-2\partial_m\delta\gamma_1\partial^m\delta\gamma_1 + 2\sqrt{2}\partial_m\delta\gamma_1\partial^m\delta\gamma_2 \right. \\
    \nonumber &\,\left. -  3\delta^{ij}\delta^{lk}\delta^{mn}\left(\partial_m D_{ik} + \partial_i D_{mk} - \partial_k D_{im}\right)\delta\gamma_2 \left( \partial_l D_{jn} + \partial_j D_{ln} - \partial_n D_{lj}\right) \delta\gamma_2\right\}.
\end{align}
Where we have once again split the purely gravitational part from the field contributions for better readability. The diffeomorphism constraint reads at the first order
\begin{align}
    \mathcal D_{i,\text{cov}}^{(1)} =&\, \pi_I\partial_i Q^I + \frac{1}{\sqrt{3}}v^{1/3}\theta\partial_i(\frac{1}{2}\delta\gamma_1 - \sqrt{2}\delta\gamma_2) - \frac{2}{\sqrt{3}}v^{2/3}\partial_i(\delta\pi_1 + \sqrt{2}\delta\pi_2). \label{eq:diff 1}
\end{align}
Let us now present the same derivation pushed to the third order.

\subsubsection{Third order Hamiltonian}

Here we provide the generic expression of the Hamiltonian up to the cubic order. As described in Eq. (\ref{eq:Ham 2 3}) we require contributions from $\tilde{\mathcal C}^{(2)}$, $\mathcal D_{i,\text{cov}}^{(2)}$ and $\mathcal C^{(3)}\cov$. We give here there full expressions, differentiating the gravitational and field contributions. The diffeomorphism constraint at the second order reads:
\begin{align}
    \mathcal D_i^{(2),G} =&\, \frac 1 3 \delta\pi_1 \partial_i\delta\gamma_1 - \frac 2 3 \delta\gamma_1 \partial_i\delta\pi_1 - \frac 2 {\sqrt3} \partial^m\delta\gamma_1D_{im}\delta\pi_2 - \frac{2\sqrt 2}{3}\delta\gamma_1\partial_i\delta\pi_2 \label{eq:diff 2 g}\\
    \nonumber &\, - \frac 2 {\sqrt3} \partial^m\delta\pi_1D_{im}\delta\gamma_2 - \frac{2\sqrt 2}{3}\delta\pi_1\partial_i\delta\gamma_2 - 2 \partial_mD\stsr \delta\gamma_2 D^{jm}\delta\pi_2 \\
    \nonumber &\, -\frac{2\sqrt 2}{\sqrt 3} D\stsr\delta\gamma_2\partial^j\delta\pi_2 + D^{mn}\delta\pi_2\partial_iD_{mn}\delta\gamma_2\,, \\
    D_{i,\text{cov}}^{(2),\phi} =&\, P_I\partial_iQ^I. \label{eq:diff 2 phi}\\
\end{align}
As previously explained, we need a second order contribution from the scalar constraint $\tilde{\mathcal C}^{(2)}$, which differs from the second order contribution of the scalar constraint at the second order Eqs. (\ref{eq:scalar 2 phi}$\&\,$\ref{eq:scalar 2 g})
\begin{align}
    \tilde{\mathcal C}^{(2),G} =&\, \mathcal C^{(2),G}\cov + \mathcal M^{(2)}\,, \label{eq:scalar tilde 2 g}\\
    \mathcal M^{(2)} =& - \frac{\Mp^2}{24v}\left\{ 8\partial^m\delta\gamma_1\partial_m\delta\gamma_1 - 2\sqrt{2}\partial^m\delta\gamma_1\partial_m\delta\gamma_2 - 8 \partial^m\delta\gamma_2\partial_m\delta\gamma_2  +8\delta\gamma_1\partial^2\delta\gamma_1  \right.\\
    \nonumber&\left.- 4\sqrt{2}\delta\gamma_1\partial^2\delta\gamma_2 + 2\sqrt{3}\partial_m\partial_n\delta\gamma_1 D^{mn}\delta\gamma_2  - 4\sqrt{6} \partial_m\partial_n\delta\gamma_2 D^{mn}\delta\gamma_2  \right.\\
    \nonumber&\left.+ 6 \partial^2D_{mn}\delta\gamma_2 D^{mn}\delta\gamma_2  + 6 \partial^iD^{mn}\delta\gamma_2\left(\partial_m D_{in} + \partial_i D_{mn} - \partial_n D_{im} \right)\delta\gamma_2 \right\} \, ,\\
    \tilde{\mathcal C}^{(2),\phi} =&\, \mathcal C^{(2),\phi}\cov + \frac 1 3 R_I{}^{KL}{}_J\pi_K\pi_LQ^IQ^J . \label{eq:scalar tilde 2 phi}
\end{align}
We have made the difference between $\tilde{\mathcal C}^{(2)}$ and $\mathcal C\cov^{(2)}$ explicit here. The difference for the gravitational contributions come from total derivatives, and the differences in the matter sector comes from the generating function on the one hand and the replacement at second order in covariant perturbations on the other. Finally, we write out here the third order contribution from the scalar constraint. Despite the complicated expression, we will see in Sec. \ref{sec:gauge fixing} that the entire gravitational contribution will drop with the adequate gauge choice.
\begin{align}
    \mathcal C^{(3),G}\cov =&\, \frac{2}{\Mp^2}\left\{ \theta^2 \left[\frac{\sqrt 3}{12}\delta\gamma_1^3 - \frac{5\sqrt 5}{12}\delta\gamma_1 (D\delta\gamma_2)^2 + \frac{1}{16}(D\delta\gamma_2)^3\right] \right. \label{eq:scalar 3 g} \\
    \nonumber&\,\left. + v^{1/3}\theta\left[ \frac{13}{16\sqrt 3}\delta\gamma_1^2\delta\pi_1 - \frac{1}{4\sqrt 3} \delta\gamma_1 D\delta\gamma_2 D\delta\pi_2 + \frac{11}{8\sqrt 3} (D\delta\gamma_2)^2\delta\pi_1 + (D\delta\gamma_2)^2\delta\pi_2 \right] \right. \\
    \nonumber&\,\left. +v^{2/3}\left[-\frac{1}{4\sqrt 3} \delta\gamma_1\delta\pi_1^2 + \frac{1}{2\sqrt 3}\delta\gamma_1 (D\delta\pi_2)^2 + \frac{1}{\sqrt 3} D\delta\gamma_2 D\delta\pi_2 \delta\pi_1  + 2 D\delta\gamma_2 (D\delta\pi_2)^2 \right]\right\}\\
    \nonumber&\,+ \frac{\Mp^2}{8v^{5/3}} \left\{ -26\sqrt 3 \delta\gamma_1^2 \partial^2\delta\gamma_1-\frac{34}{3\sqrt 3}\delta\gamma_1\partial_k\delta\gamma_1\partial^k\delta\gamma_1 -\frac{\sqrt 2}{3\sqrt 3}\delta\gamma_1\partial_k\delta\gamma_1\partial^k\delta\gamma_2 + \frac{113\sqrt 2}{3\sqrt 3}\delta\gamma_1^2\partial^2\delta\gamma_2 \right.\\
    \nonumber&\,\left. +3\delta\gamma_1\partial_i\partial_j\delta\gamma_1D^{ij}\delta\gamma_2 - \frac{5}{3}\partial_i\delta\gamma_1\partial_j\delta\gamma_1D^{ij}\delta\gamma_2 + \frac{68}{3\sqrt 3}\delta\gamma_1\partial_k\delta\gamma_2\partial^k\delta\gamma_2 +4\sqrt 2 \delta\gamma_1\partial_i\partial_j\delta\gamma_2 D^{ij}\delta\gamma_2 \right. \\
    \nonumber&\,\left. -\frac{\sqrt 2}{3}\partial_i\delta\gamma_1\partial_j\delta\gamma_2D^{ij}\delta\gamma_2 + \sqrt 3 \delta\gamma_1 D^{ij}\delta\gamma_2\partial^2D_{ij}\delta\gamma_2 - \frac{29}{2\sqrt 3} \delta\gamma_1 \partial^k  D^{ij}\delta\gamma_2 \partial_kD_{ij}\delta\gamma_2 \right. \\
    \nonumber&\,\left. +\frac{7}{\sqrt 3} \delta\gamma_1 \partial^k  D^{ij}\delta\gamma_2 \partial_i D_{kj}\delta\gamma_2 + \frac{2}{\sqrt 3} \partial^k \delta\gamma_1   D^{ij}\delta\gamma_2 \partial_kD_{ij}\delta\gamma_2 + 2\sqrt 3 \partial^k \delta\gamma_1   D^{ij}\delta\gamma_2 \partial_i D_{kj}\delta\gamma_2   \right.\\
    \nonumber&\,\left. + \frac{4}{\sqrt 3} \partial^2\delta\gamma_1(D\delta\gamma_2)^2 -2\sqrt 3 \partial_i\partial^j\delta\gamma_1D^{ik}\delta\gamma_2D_{kj}\delta\gamma_2+ \frac{4}{ 3}\partial_i\delta\gamma_2\partial_j\delta\gamma_2D^{ij}\delta\gamma_2 \right. \\
    \nonumber&\,\left.  + \frac{4\sqrt 2}{\sqrt 3} \partial^k\delta\gamma_2D^{ij}\delta\gamma_2\partial_kD_{ij}\delta\gamma_2 - \frac{4\sqrt 2}{\sqrt 3} \partial^k\delta\gamma_2D^{ij}\delta\gamma_2\partial_iD_{kj}\delta\gamma_2 + \frac{5 \sqrt 2}{\sqrt 3} \partial_i\partial^j\delta\gamma_2D^{ik}\delta\gamma_2D_{kj}\delta\gamma_2\right.\\
    \nonumber&\,\left.  - \frac{4\sqrt 2}{\sqrt 3}\partial^2\delta\gamma_2D^{ij}\delta\gamma_2D_{ij}\delta\gamma_2 - 4 D^{ij}\delta\gamma_2 \partial_iD_{kl}\delta\gamma_2\partial_j D^{kl}\delta\gamma_2 + 4 D^{ij}\delta\gamma_2\partial_iD^{kl}\delta\gamma_2\partial_kD_{jl}\delta\gamma_2  \right. \\
    \nonumber&\,\left. + 2D^{ij}\delta\gamma_2\partial^kD_{il}\delta\gamma_2\partial^lD_{kj}\delta\gamma_2 - 2 \delta^{kl}D^{ij}\delta\gamma_2\partial_mD_{ik}\delta\gamma_2\partial^mD_{jl}\delta\gamma_2 - 4 D^{ij}\delta\gamma_2D^{kl}\delta\gamma_2\partial_i\partial_j D_{kl}\delta\gamma_2 \right. \\
    \nonumber&\,\left. + 4D^{ij}\delta\gamma_2D^{kl}\delta\gamma_2\partial_i\partial_kD_{jl}\delta\gamma_2 \right\}\,, \\
    \mathcal C^{(3),\phi}\cov =&\,  \frac{7}{4\sqrt 3} \frac{1}{v^{1/3}} G_{IJ}\delta^{ij}\delta\gamma_1\partial_iQ^I\partial_jQ^J - \frac{1}{\sqrt 3}\frac{1}{v^{1/3}}G_{IJ}D^{ij}\delta\gamma_2\partial_iQ^I\partial_jQ^J \label{eq:scalar 3 phi} \\
    \nonumber&\,+ \frac{71}{48\sqrt 3}\frac 1 v \delta\gamma_1^3 V + \frac 1 v \delta\gamma_1(D\delta\gamma_2)^2 V + \frac 5 8 \frac{1}{v^{1/3}} \delta\gamma_1^2 V_{;I}Q^I +\frac 1 4 \frac{1}{v^{1/3}} (D\delta\gamma_2)^2 V_{;I}Q^I \\
    \nonumber&\,+ \frac{\sqrt 3}{2} v^{1/3}\delta\gamma_1 V_{;IJ}Q^IQ^J + \frac 1 6 V_{;IJK}Q^IQ^JQ^K \\
    \nonumber&\,-\frac{1}{6v} R^I{}_{KL}{}^{J}{}_{;M}\pi_I\pi_JQ^KQ^LQ^M -\frac{2}{3v} \pi_N G^{NJ} R^M{}_{KLJ} P_M  Q^L  Q^K - \frac{\sqrt 3}{4} \frac{1}{v^{5/3}} G^{IJ} \delta\gamma_1 P_IP_J \\
    \nonumber&\,-\frac{\sqrt 3}{12} \frac{1}{v^{5/3}} R^J{}_{LM}{}^{S} \delta\gamma_1 \pi_S\pi_J Q^LQ^M + \frac 1 8 \frac{1}{v^{7/3}}  \delta\gamma_1^2 \pi_I P_J - \frac 1 4 \frac{1}{v^{7/3}}G^{IJ} (D\delta\gamma_2)^2  \pi_I P_J  \\
    \nonumber&\,-\frac{5}{24\sqrt 3}\frac{1}{v^3} G^{IJ}\pi_I\pi_J \delta\gamma_1^3 + \frac{\sqrt 3}{8} \frac{1}{v^3} G^{IJ}\pi_I\pi_J\delta\gamma_1(D\delta\gamma_2)^2  - \frac{1}{12}\frac{1}{v^3} G^{IJ}\pi_I\pi_J(D\delta\gamma_2)^3.
\end{align}
Where we have used the shorthand notations for any quantities $f$ and $g$
\begin{align}
    (Df)(Dg) =&\, D\stsr f D\ustsr g \,,\\
    (Df)^3 =&\, D_{ij}f D_k{}^if D^{kj}f.
\end{align}
This result is novel in different aspects. First of all, all these computations have been done in the Hamiltonian formalism whereas most approaches to obtain higher order Hamiltonians first linearise the Lagrangian up to a given order then do a Legendre transform, making our approach more direct. Secondly we have not needed to assume a particular gauge in this work. We will see in the next sections how to compute gauge invariant quantities that are adequate to work with in these higher order dynamics, and how to reduce our full Hamiltonian to a gauge fixed Hamiltonian for said gauge invariant quantities.

\section{Gauge-invariant variables}
\label{sec: GI variables}

\subsection{Gauge transformation}\label{ssec:gauge trans}
In this section we provide the gauge transformation in the case of ADM variables, \ie in terms of the induced metric and extrinsic curvature. To build gauge invariant quantities, we first need to know how each configuration and momentum variables change under a gauge transformation. Let us call $\xi^\mu_{(m)}$ the family of gauge parameters, where $m$ is the order in perturbation theory. We follow here the procedure defined in \cite{Bruni_1997} to define the transformation of a field $T$ with respect to the gauge parameters. Defining the zeroth, first and second order perturbations of the tensor $T = \bar T + \delta T +\delta_2 T$, they change under the gauge transformation as
\begin{align}
    \tilde{\bar T} =&\, \bar T\,, \\
    \delta\tilde T =&\, \delta T + \mathcal L_{\xi_{(1)}}\bar T\,, \label{eq:GT 1 order} \\
    \delta_2\tilde T =&\, \delta_2 T +   \mathcal L_{\xi_{(1)}} \delta T + \frac 1 2 \mathcal L_{\xi_{(1)}}^2 \bar T +  \mathcal L_{\xi_{(2)}} \bar T. \label{eq:GT 2 order}
\end{align}
Where we define the Lie derivative of a tensor $T\tsr$ with respect to a vector field $\xi$ as
\begin{equation}
     \mathcal L_{\xi} T\tsr = \xi^\rho\partial_\rho T\tsr + T_{\mu\rho} \partial_\nu \xi^\rho + T_{\rho\nu} \partial_\mu \xi^\rho.
\end{equation}
This expression can easily be generalised to tensors of any rank. We wish to apply this method to find the first order transformation of all configuration and momentum parameters. We then find the second order transformations of the configurations and the perturbation of the conjugate momenta of fields. We do not treat the second order transform of the induced spacetime momenta perturbations since they are mathematically more involved and not useful in this work. Let us note however that the method is the same.

We present here the derivation up to the second order for the field covariant perturbations $Q^I$, and list the results for the other quantities. A full computation can be found in appendix \ref{app:gauge transform}, where the slightly more subtle case of the momenta perturbations $\delta\pi_1$ and $\delta\pi_2$ is discussed. Since the fields $\phi^I$ are scalars in the spacetime manifold, computing the first order gauge transform is straightforward. We need to introduce the first gauge parameter $\xi_{(1)}$. Let us find the gauge transformation of $\delta\phi^I$ as a first example. We start from the Lie derivative of the background field
\begin{align}
    \nonumber \mathcal L_{\xi_{(1)}}\bar\phi^I =&\, \xi_{(1)}^\rho\partial_\rho \bar\phi^I \\
    =&\, \xi_{(1)}^0\dot{\bar\phi}^I = \xi_{(1)}^0\frac N v G\uTsr \pi_J.
\end{align}
From this, and the definition of the gauge transformation of perturbations Eq. (\ref{eq:GT 1 order}), we get the gauge transform of the naive perturbation.
\begin{align}
    \delta\tilde\phi^I =&\, \delta\phi^I + \xi_{(1)}^0\frac N v G\uTsr \pi_J.
\end{align}
 Plugging in the definition of the covariant perturbations and keeping only first order terms, we get the gauge transform at the first order of the covariant perturbation. Using this same methods we can derive the following list of gauge transform at the first order.
\begin{align}
    \tilde Q^I =&\, Q^I + \xi_{(1)}^0\frac N v G\uTsr \pi_J\,, \label{eq:GT Q 1} \\
    \tilde{\delta\gamma_1} =&\,  \delta\gamma_1 + \frac{2 N v^{2/3}}{\Mp}\sqrt \rho\xi_{(1)}^0 + \frac{2}{\sqrt 3}v^{2/3}\partial_l \xi_{(1)}^l\,,\label{eq:GT gamma 1 1} \\ 
    \tilde{\delta\gamma_2} =&\, \delta\gamma_2 +2\sqrt{\frac 2 3} v^{2/3} \partial_l \xi_{(1)}^l\,,\label{eq:GT gamma 2 1} \\ 
    \tilde P_J =&\, P_J -  Nv V_{;J}\xi_{(1)}^0\,,\label{eq:GT P 1}\\ 
    \tilde{\delta\pi_1} =&\, \delta\pi_1 + \frac{ v^{1/3}\Mp}{ 3}\sqrt \rho \partial_l \xi_{(1)}^l + \frac{Nv^{1/3}}{\sqrt 3}\left[\frac{\pi^2}{v^2} - V\right] \xi_{(1)}^0\,,\label{eq:GT pi 1 1}\\ 
    \tilde{\delta\pi_2} =&\, \delta\pi_2 + \frac{2\sqrt 2v^{1/3}\Mp}{3}\sqrt \rho\partial_l \xi_{(1)}^l\,, \label{eq:GT pi 2 1}\\ 
    \tilde{N_i} =&\, N_i + v^{2/3}\delta_{ki}\partial_\tau  \xi_{(1)}^k - N^2 \partial_i \xi_{(1)}^0 \, ,\label{eq:GT shift 1}\\ 
    \tilde{\delta N} =&\, \delta N - 2 \dot N N \xi_{(1)}^0 - 2 N^2 \partial_\tau  \xi_{(1)}^0. \label{eq:GT lapse 1}
\end{align}
To go to the second order for $Q^I$ let us go back momentarily to the naive variables. As shown in Eq. (\ref{eq:GT 2 order}) we need to compute the Lie derivative of $\delta\phi^I$
\begin{align}
    \nonumber \mathcal L_{\xi_{(1)}}\delta\phi^I =&\, \xi_{(1)}^\rho\partial_\rho\delta\phi^I = \xi_{(1)}^\rho\partial_\rho\left(Q^I -\frac 1 2 \Gamma^I_{JK}Q^JQ^K\right) \\
    =&\, \xi_{(1)}^0\partial_\tau Q^I + \xi_{(1)}^i\partial_iQ^I. \label{eq:Lie derivative delta phi}
\end{align}
We also need 
\begin{align}
    \mathcal L_{\xi_{(1)}}^2 \bar\phi^I =&\, \xi_{(1)}^\nu\partial_\nu\left(\xi_{(1)}^\rho\partial_\rho\bar\phi^I\right) \nonumber \\
    =&\, (\xi_{(1)}^0)^2\partial_\tau \left(\frac N v \pi^I\right) + \xi_{(1)}^\nu\left(\partial_\nu\xi_{(1)}^0\right)\frac N v \pi^I. \label{eq:Lie derivative 2 background}
\end{align}
With the second order gauge parameter we define
\begin{align}
    \nonumber \mathcal L_{\xi_{(2)}} \bar\phi^I =&\, \xi_{(2)}^\rho\partial_\rho \bar\phi^I \\
    =&\, \xi_{(2)}^0\dot{\bar\phi}^I = \xi_{(2)}^0\frac N v G\uTsr \pi_J. \label{eq:lie derivative xi 2}
\end{align}
Combining these expressions, as expressed in Eq. (\ref{eq:GT 2 order}) we get the second order transform of $\delta\phi^I$ under the parameters $\xi_{(1)}$ and $\xi_{(2)}$. To get to the second order transform of $Q^I$ we recall the link between the naive and covariant perturbation and write
\begin{align}
    \delta_2\tilde\phi^{I} =&\, \tilde Q^{I(2)} - \frac 1 2 \Gamma^I_{JK} \tilde Q^{K(1)} \tilde Q^{J(1)} \label{eq:GT delta phi 2}\\ 
    \nonumber=&\, \delta_2\phi^I + \mathcal L_{\xi_{(1)}}\delta\phi^I + \frac 1 2 \mathcal L_{\xi_{(1)}}^2 \bar\phi^I + \mathcal L_{\xi_{(2)}}\bar\phi^I\\
    \nonumber=&\, - \frac 1 2 \Gamma^I_{JK} \tilde Q^{K(1)} \tilde Q^{J(1)}  + \mathcal L_{\xi_{(1)}}\delta\phi^I + \frac 1 2 \mathcal L_{\xi_{(1)}}^2 \bar\phi^I  + \mathcal L_{\xi_{(2)}}\bar\phi^I. 
\end{align}
Second order terms will either contain the gauge parameter squared, or the product of the gauge parameter with a perturbative quantity ($Q^I$ for example). Rearranging terms from the first and third line of the right hand side term in Eq. (\ref{eq:GT delta phi 2}), then injecting Eqs. (\ref{eq:GT Q 1}, \ref{eq:Lie derivative delta phi}, \ref{eq:Lie derivative 2 background} $\&\,$\ref{eq:lie derivative xi 2}) we get
\begin{align}
    \label{eq:GT Q 2} \tilde Q^{I(2)} =&\, (\xi_{(1)}^0)^2\left(\frac{\dot N}{2v}\pi^I -\frac{N^2 \sqrt 3}{2 v \Mp}\sqrt \rho  \pi^I - \frac{N^2} 2 G\uTsr V_{;J} \right) + \frac{N}{2v}\xi_{(1)}^\nu(\partial_\nu\xi_{(1)}^0) \\
    \nonumber &\,+\xi_{(1)}^0 D_\tau Q^I + \xi_{(1)}^iD_iQ^I + \xi_{(2)}^0\frac N v G\uTsr \pi_J\,,
\end{align}
where we have also used the background equations of motion. We have separated the two kinds of contributions to the second order gauge transform. On the first line are terms proportional to $\xi_{(1)}^2$, and the second line are terms where the gauge parameter $\xi_{(1)}$ multiplies a perturbation variable, as well as a term proportional to the second gauge parameter. This same procedure can be followed for all configuration and perturbation variables. We give here the second order gauge transform of the perturbation of the induced metric
\begin{align}
    \delta\tilde\gamma\stsr^{(2)} =&\, \frac{1}{\sqrt 3 }\delta\stsr\xi_{(1)}^\nu\partial_\nu \delta\gamma_1 + \xi_{(1)}^\nu\partial_\nu(D\stsr\delta\gamma_2) + \partial_{(i|}\xi_{(1)}^0\delta_{k|j)}N^k + \frac{1}{\sqrt 3}\partial_{(i|}\xi_{(1)}^k\delta_{k|j)}\delta\gamma_1 \label{eq:GT gamma 2} \\ \nonumber
    &\,+ \partial_{(i|}\xi_{(1)}^kD_{k|j)}\delta\gamma_2 + \frac{Nv^{2/3}}{\sqrt 3 \Mp}\sqrt \rho \delta\stsr\xi_{(1)}^\nu\partial_\nu\xi_{(1)}^0 + \left(\frac{\dot N v^{2/3}}{\sqrt 3 \Mp}\sqrt \rho + \frac{N^2v^{2/3}}{6\Mp^2}(\rho+3p)\right)\delta\stsr{\xi_{(1)}^0}^2 \\ \nonumber
    &\,+ \frac{4 N}{\sqrt 3 v^{1/3}\Mp}\sqrt \rho \xi_{(1)}^0\partial_{(i|}\xi_{(1)}^k\delta_{k|j)} + v^{2/3}\xi_{(1)}^\nu\partial_\nu\partial_{(i|}\xi_{(1)}^k\delta_{k|j)} + v^{2/3}\partial_{(i|}\xi_{(1)}^\nu\partial\nu\xi_{(1)}^k\delta_{k|j)} \\ \nonumber
    &\,- 2N^2\partial_i\xi_{(1)}^0\partial_j\xi_{(1)}^0 + 2v^{2/3}\partial_i\xi_{(1)}^k\partial_j\xi_{(1)}^l \delta_{kl} + \frac{2 N\sqrt\rho}{\sqrt3 \Mp} \xi_{(2)}^0 \delta\stsr + v^{2/3}\delta_{(i|k|}\partial_{j)}\xi_{(2)}^k\,,
\end{align}
where parenthesis around indices mean that the term needs to be symmetries for the indices that aren't being summed on (\ie on i and j in this case) with coefficient 1. From this expression we can easily deduce the second order gauge transform of the scalar variables 
\begin{equation}
    \delta\tilde\gamma_1^{(2)} = \frac{1}{\sqrt 3}\delta\ustsr \delta\tilde\gamma\stsr^{(2)} \,,\quad \delta\tilde\gamma_2^{(2)} = D\ustsr \delta\tilde\gamma\stsr^{(2)}. \label{eq:GT gamma 1 and 2 2}
\end{equation}

\subsection{Mukhanov-Sasaki variables}
Here we explicitly build the Mukhanov-Sasaki variables. We build these gauge invariant quantities and make sure that they reduce to the scalar-field fluctuations in the spatially-flat gauge. We do the same for the canonical momentum. We start by reminding how the linear Mukhanov-Sasaki variables are defined before demonstrating that we can also define variables that are gauge invariant up to the second order at large scales that match the scalar-field fluctuations in the flat gauge.

\subsubsection{First order gauge invariant construction}
Let us start be reconstructing the usual linear multifield Mukhanov-Sasaki variable. This quantity is gauge invariant even at small scales, but we will drop this requirement when constructing the second order gauge invariant quantity. As we have just shown there is a strong liberty when constructing gauge invariant quantities. From equations Eq. (\ref{eq:GT Q 1} - \ref{eq:GT lapse 1}) we can construct combinations such that all dependence on $\xi_{(1)}^\mu$ drops in the right hand side. The first solution we can build using Eqs. (\ref{eq:GT Q 1}, \ref{eq:GT gamma 1 1}$\&\,$ \ref{eq:GT gamma 2 1}) is 
\begin{equation}
    \mathcal Q^I = Q^I - \frac{\Mp G\uTsr \pi_J}{2\sqrt{2} v^{5/3} \sqrt\rho}\left(\sqrt 2 \delta\gamma_1 - \delta\gamma_2\right)\,, \label{eq:QMS}
\end{equation}
where we can verify that $\tilde{\mathcal Q}^I = \mathcal Q^I$. Similarly, a gauge invariant combination for the conjugate momenta of these variables reads
\begin{equation}
    \mathcal P_J = P_J + \frac{\Mp v^{1/3}}{2\sqrt 2\sqrt \rho} V_{;J}\left(\sqrt 2 \delta\gamma_1 - \delta\gamma_2\right). \label{eq:PMS}
\end{equation}
We can once again verify that $\tilde{\mathcal P}_J = \mathcal P_J$. We can also check that the variables we have defined here are a canonical pair.
\begin{equation}
    \{\mathcal Q^I,\mathcal P_J\} = \delta^I_J. \label{eq:MS poisson bracket}
\end{equation}
These canonical variables are particularly useful because not only are they linearly GI (since they were built as such), but they are also remarkably simple when one uses the flat gauge. In said gauge, the perturbations of the induced metric are set to $0$. In the ADM formalism, this is written as $\delta\gamma_1 = \delta\gamma_2 = 0$. Thus it is straightforward to see that the Mukhanov-Sasaki variables drop to $\mathcal Q^I \underset{\mathrm{SF}}{=} Q^I$ and $\mathcal P_J  \underset{\mathrm{SF}}{=} P_j$, where $ \underset{\mathrm{SF}}{=}$ indicates that this equality holds in the spatially-flat gauge.

We now wish to verify whether these variables are still gauge invariant at the second order. In order to proceed further and study gauge invariance at the next order, we make two simplifications. On the one hand, we go to the single field case, a condition we will later on relax to recover the multifield case. On the other hand we go to large scales, a condition that we will not relax.

\subsubsection{Second order gauge invariant at large scales construction}\label{sssec:MS GI second order}
In the single field case we can define the field and moment perturbation as $Q = \phi - \bar\phi$ and $P = \pi_\phi - \bar\pi_\phi$. From this we define the usual single field Mukhanov-Sasaki variable :
\begin{equation}
    \mathcal Q = Q -\frac{\Mp \pi_\phi}{2\sqrt{2} v^{5/3} \sqrt\rho} \left(\sqrt 2 \delta\gamma_1 - \delta\gamma_2\right)
\end{equation}
To check whether $\mathcal Q$ is gauge invariant up to the second order at large scales, we need the large-scale first and second order transformation of its constituents. We will also need to use the equations of motion of our variables, in particular of $Q$  and $P$, as derived from the second order Hamiltonian in Eq. (\ref{eq:CovDiffQP}). First, let us precise what we mean by large scales. We require every gradient term to drop and will define large scales in Fourier space as $\frac{k^2}{v^{2/3}H^2}\ll1$. In practice this means that any term with two spatial derivatives can be neglected. Since we are only considering scalars in this work, we take $\xi_i = \partial_i \xi$. Thus any spatial direction of the gauge parameter counts as a spatial derivative.

Let us take equations Eqs. (\ref{eq:GT Q 1}, \ref{eq:GT gamma 1 1}, \ref{eq:GT gamma 2 1} $\&\,$\ref{eq:GT gamma 2}) to large scales (and for a single field when necessary). We are also keeping notations from the covariant multifield description to stay consistent. For the single field case we have $D_\mu \equiv \partial_\mu$, and subsequently, $\left(V_{;\phi},V_{;\phi\phi}\right)=\left(V_{,\phi},V_{,\phi\phi}\right)$. We can remark from Eq. (\ref{eq:GT gamma 2 1}) that $\delta\gamma_2$ is gauge invariant at large scales at the first order since $\xi_{(1)}$ only appears with a gradient term. However, this is not the case at the second order, as we can see from Eq. (\ref{eq:GT gamma 1 and 2 2}), which in our case leads to
\begin{align}
    \delta\tilde\gamma_1^{(2)} =&\, \xi_{(1)}^0 \left( -\frac{\sqrt{3}}{\Mp^2}v^{2/3}\theta\delta N-\frac{N2v^{1/3}}{\Mp^2}\delta\pi_1 - \frac{N\theta}{2\Mp^2}\delta\gamma_1\right) +\frac{Nv^{2/3}}{\sqrt 3 \Mp}\sqrt \rho\xi_{(1)}^0\partial_\tau \xi_{(1)}^0 \label{eq:GT gamma 1 2 LS}\\ \nonumber
    &\, + (\xi_{(1)}^0)^2\left(\frac{\dot N v^{2/3}}{ \Mp}\sqrt \rho  + \frac{N^2v^{2/3}}{2\sqrt 3 \Mp^2}(\rho+3p)\right) +\frac{2 N v^{2/3}}{\Mp}\sqrt \rho\xi_{(2)}^0 \,,\\ \nonumber
    \delta\tilde\gamma_2^{(2)} =&\, D\ustsr(\xi_{(1)}^0\partial_\tau (D\stsr\delta\gamma_2)) .\label{eq:GT gamma 2 2 LS}
\end{align}
We can make another important remark here, even though $D\stsr$ looks like a local operator and could be expected to drop at large scales, it actually remains in the gauge transformation of $\delta\gamma_2$ at large scales. We have also made use of the equation of motion of $\delta\gamma_1$ Eq. (\ref{eq:CovDiff1}) obtained from the second order Hamiltonian. We are finally in the right position to compute the second order gauge transformation of the Mukhanov-Sasaki variable. By combining Eqs. (\ref{eq:GT Q 2}, \ref{eq:GT gamma 1 2 LS} $\&\,$\ref{eq:GT gamma 2 2 LS}) we get at large scales
\begin{align}
    \tilde{\mathcal Q}^{(2)} =&\, \xi_{(1)}^0 N \left(\frac 1 v P - \frac{2\pi_\phi}{\sqrt 3 v^{5/3}}\delta\gamma_1 + \frac{\pi_\phi}{v^{4/3}\Mp^2 \sqrt{\rho}}\delta\pi_1\right) + N\frac{\Mp \pi_\phi}{2\sqrt{2} v^{5/3} \sqrt\rho} D\ustsr(\xi_{(1)}^0\partial_\tau (D\stsr\delta\gamma_2)) \\
    \nonumber &\,+ (\xi_{(1)}^0)^2N^2\left(-vV_{,\phi} + \frac{\pi_\phi(\rho+3p)}{4\sqrt 3 v \Mp \sqrt \rho}\right).
\end{align}
It is in this step that we have explicitly used the linear equation of motion of $Q$. It is quite clear here that the second order contributions do not vanish in all generality at large scales, however all dependence on the second gauge parameter $\xi_{(2)}$ has vanished since it multiplies the same linear combination of background quantities as the first gauge parameter in the linear case, which is chosen to drop for gauge invariant quantities. The Mukhanov-Sasaki variable is thus not gauge invariant at this order, and needs to be corrected. It is also clear that small scale contributions are independent of the large scales one, thus our study is not affected by potential cancelling of terms that we would have missed.

We now set on the task of defining a new variable in which contributions of $\xi_{(1)}$ drop at the second order at large scales. We first focus on the remaining $\delta\gamma_2$ term. Since $\delta\gamma_2$ is itself GI at the first order at large scales, the necessary correction to get rid of its contribution can be decoupled from the rest. In practice we can add any term proportional to $\ D\ustsr(\delta z\partial_\tau (D\stsr\delta\gamma_2)) $ where $\delta \tilde z = \delta z + \beta \xi_{(1)}^0$, with $\beta$ a function of background quantities, to treat this correction. We choose $\delta z = \delta\gamma_1$ here, and the corrective term we need is thus $-\frac{\Mp^2\pi^I}{4\sqrt 2 v^{7/3}\rho} D^{mn}(\delta\gamma_1\partial_\tau (D_{mn}\delta\gamma_2))$.

We are now left with the task to cancel out the remaining terms by adding a linear combination of any term quadratic term of the form $\{Q,P,\delta\gamma_1,\delta\pi_1\}M\{Q,P,\delta\gamma_1,\delta\pi_1\}^{\mathrm T}$, with $M$ a $4\times4$ symmetric matrix of background quantities. We found the following particular solution
\begin{align}
    \mathcal Q_\mathrm{part \,sol} = &\,\mathcal Q  + \frac{1}{2 v^2 V_{,\phi}} P^2 + \frac{\Mp \pi_\phi }{2 \sqrt{3} \sqrt{\rho } v^{7/3}}\delta\gamma_1^2 - \frac{\sqrt{3} \pi_\phi }{\Mp \sqrt{\rho } v^{5/3} (3 p+\rho )}\delta\pi_1^2  \\
    \nonumber&\, - \frac{\Mp^2\pi^I}{4\sqrt 2 v^{7/3}\rho} D^{mn}(\delta\gamma_1\partial_\tau (D_{mn}\delta\gamma_2)).
\end{align}
This solution is not acceptable as such since we have lost one of the key aspects of the Mukhanov-Sasaki variable. It is no longer equal to $Q$ in the flat gauge. This is where our freedom to add any GI quantity up to the second order comes in. We have found six\footnote{And we show in \ref{ssec:GI 1 and 2} that there are no more linearly independent gauge invariant combinations at large scales.} linearly independent GI at the second order combinations.
\begin{align}
    v_1 =&\, \delta\gamma_1 \delta\pi_1-\frac{\Mp (3 p+\rho)}{8 \sqrt{3} \sqrt{\rho } v^{1/3}}\delta\gamma_1^2 -\frac{2 \sqrt{3}  \sqrt{\rho} v^{1/3}}{\Mp (3 p+\rho )} \delta\pi_1^2 \,, \label{eq:v1} \\
    v_2 =&\,  P  \delta\pi_1+\frac{(3 p+\rho)}{4 \sqrt{3} v^{2/3} V_{;\phi}} P^2 +\frac{\sqrt{3}  v^{2/3} V_{;\phi}}{3p+\rho }\delta\pi_1^2 \,, \label{eq:v2} \\
    v_3 =&\, P \delta\gamma_1+\frac{\sqrt{\rho }}{\Mp v^{1/3}V_{;\phi}}P^2+\frac{ \Mp v^{1/3} V_{;\phi}}{4\sqrt{\rho }}\delta\gamma_1^2\,, \label{eq:v3} \\
    v_4 =&\, Q \delta\pi_1 -\frac{\sqrt{3}  \pi_\phi}{v^{4/3} (3p+\rho )} \delta\pi_1^2-\frac{v^{4/3} (3 p+\rho )}{4 \sqrt{3}\pi_\phi}Q^2\,, \label{eq:v4}  \\
    v_5 =&\, QP +\frac{v V_{;\phi}}{2 \sqrt{p+\rho }}Q^2 +\frac{ \pi_\phi}{2 v^2V_{;\phi}} P^2\,,\label{eq:v5} \\
    v_6 =&\, Q\delta\gamma_1 - \frac{v^{2/3}\sqrt\rho}{\Mp\sqrt{p+\rho}}Q^2-\frac{\Mp\sqrt{p+\rho}}{4v^{2/3}\sqrt\rho}\delta\gamma_1^2. \label{eq:v6} 
\end{align}
Let us note that we have checked that these six independent solutions can be found by taking into account the redundant condition on the $(\xi_{(1)}^0)^2$ contribution or by discarding them, thus proving that it is indeed redundant\footnote{In practice we constructed both a $4\times10$ matrix and a $5\times10$ matrix combining all our gauge transformations at the second order of the cross products $\{Q,P,\delta\gamma_1,\delta\pi_1\}\cdot\{Q,P,\delta\gamma_1,\delta\pi_1\}^{\mathrm T}$, the former including the ${\xi_{(1)}^0}^2$ terms and the latter without. We then computed the kernel of both matrices and found the same dimension. The base vectors of the null space of the matrices are presented here.}. By combining our existent solution to $v_1$ and $v_3$ with the appropriate prefactors, we find that another GI up to the second order corrections is
\begin{align}
    \mathcal Q_\mathrm{flat} =&\, \mathcal Q_\mathrm{part \,sol} -\frac{\pi_\phi}{2v^2\rho} v_1  - \frac{\Mp}{2v^{5/3}\sqrt \rho} v_3 \label{eq:Q flat}\\
    \nonumber =&\,  Q - \frac{\Mp\pi_\phi}{2\sqrt 2 v^{5/3} \sqrt\rho}\left(\sqrt 2 \delta\gamma_1 - \delta \gamma_2\right) -\frac{\pi_\phi}{2v^2\rho}\delta\gamma_1\delta\pi_1 - \frac{\Mp}{2v^{5/3}\sqrt \rho}\delta\gamma_1 P \\
    \nonumber &\,+ \left[\frac{\Mp\pi_\phi}{2\sqrt 3 v^{7/3}\sqrt \rho}\left(1+\frac{3(\rho+p)}{8\rho} \right)-\frac{\Mp^2V_{;\phi}}{8v^{1/3}\rho}\right]\delta\gamma_1^2 - \frac{\Mp^2\pi^I}{4\sqrt 2 v^{7/3}\rho} D^{mn}(\delta\gamma_1\partial_\tau (D_{mn}\delta\gamma_2)). 
\end{align}
This corrected Mukhanov-Sasaki variable now verifies all the requirements we needed. It is GI both at the first and the second order at large scales. The corrective terms no longer contain prefactors that could be physically unacceptable\footnote{Except if our inflaton field has a vanishing pressure but this take us out of the inflationary dynamics context we are studying, or if the scale factor vanishes, which is outside of the context of this work.}. And finally, this expression simplifies to $\mathcal Q_\mathrm{flat} \underset{\mathrm{SF}}{=} Q$, which is a key characteristic of the Mukhanov-Sasaki variable.

We will now apply the exact same method to the associated momenta $\mathcal P$, where we still need to exhibit the second order transformation under a gauge parameter $\xi$ at large scales of the field momentum perturbation $P$
\begin{align} \label{eq:GT P 2 LS}
    \tilde P^{(2)} =&\, -(\xi_{(1)}^0)^2\left(\frac{N^2 \sqrt 3 v}{2\Mp}\sqrt \rho V_{;\phi} + \frac{N^2\pi_\phi}{2} V_{;\phi\phi} + \frac{\dot N v}{2}V_{;\phi}\right) - \frac{Nv}{2}V_{;\phi}\xi_{(1)}^0\partial_\tau \xi_{(1)}^0  \\ \nonumber 
    &\,+\xi_{(1)}^0 D_\tau  P -  Nv V_{;\phi}\xi_{(2)}^0 . 
\end{align}
Combining Eqs. (\ref{eq:GT gamma 2}, \ref{eq:GT gamma 1 2 LS} $\&\,$\ref{eq:GT P 2 LS}), and using the linear equation of motion for $P$, we get for the second order transformation under a parameter $\xi_{(1)}^\mu$ at large scale of the conjugate momentum to the Mukhanov-Sasaki variable. As it was the case for the configuration Mukhanov-Sasaki variable, this expression does not vanish in all generality. We need to correct this moment as we did previously. We do not present the intermediate steps but only the final result. We first found a particular solution then used the linearly independent gauge invariant combinations to find a solution where all second order terms contain at least one factor of $\delta\gamma_1$ or $\delta\gamma_2$ so that they drop in the flat gauge. Let us note that we treat once again the remaining $\delta\gamma_2$ dependence separately since it decouples from the rest of the problem thanks to the GI quality of $\delta\gamma_2$ at large scales. The final corrected momentum is
\begin{align}
    \mathcal P_\mathrm{flat} =&\, P + \frac{\Mp v^{1/3}}{2\sqrt 2\sqrt \rho} V_{;\phi}\left(\sqrt 2 \delta\gamma_1 - \delta\gamma_2\right) + \frac{\Mp^2}{4\sqrt 2 v^{1/3}\sqrt \rho} V_{;\phi} D^{mn}(\delta\gamma_1\partial_\tau (D_{mn}\delta\gamma_2)) \label{eq:P flat}\\
    &\,\left(\frac{\sqrt{3} \Mp  (\rho -p)}{16 \rho ^{3/2}v^{1/3}}V_{;\phi}-\frac{\Mp^2  \pi_\phi}{8 \rho v^{4/3}} V_{;\phi\phi}\right)\delta\gamma_1^2  +\frac{ \Mp v^{1/3} V_{;\phi\phi}}{2 \sqrt{\rho }}\delta\gamma_1 Q +\frac{V_{;\phi}}{2 \rho }\delta\gamma_1 \delta\pi_1.
\end{align}
Just like its conjugate configuration, this corrected momentum exhibits all the right properties. It is gauge invariant up to the second order at large scales, does not contain any potential problematic terms, and simplifies to $\mathcal P_\mathrm{flat} \underset{\mathrm{SF}}{=} P$ in the flat gauge. We however still need to check that $\mathcal Q_\mathrm{flat}$ and $\mathcal P_\mathrm{flat}$ are indeed still canonical variables. To this effect we compute their Poisson bracket, which will lead to three categories of terms, each at a different perturbation orders. Let us write $\mathcal Q_\mathrm{flat} = \mathcal Q + \mathcal Q_{\mathrm{flat}(2)}$ where we separated the second order corrections from the usual first order Mukhanov-Sasaki variable. Similarly we write $\mathcal P_\mathrm{flat} = \mathcal P + \mathcal P_{\mathrm{flat}(2)}$. We can now write the Poisson bracket as
\begin{equation}
    \left\{\mathcal Q_\mathrm{flat},\mathcal P_\mathrm{flat}\right\} = \left\{\mathcal Q,\mathcal P\right\} + \left\{\mathcal Q_{\mathrm{flat}(2)},\mathcal P\right\} + \left\{\mathcal Q,\mathcal P_{\mathrm{flat}(2)}\right\} + \left\{\mathcal Q_{\mathrm{flat}(2)},\mathcal P_{\mathrm{flat}(2)}\right\}.
\end{equation}
Where the first term of the right hand side automatically equals $1$ since the Mukhanov-Sasaki and its conjugate momenta are precisely a canonical pair. We have exhibited this behaviour in the multifield case Eq. (\ref{eq:MS poisson bracket}), and it is trivially also true in the single field case. We are left with the task of verifying that the rest of this expression vanishes. This will in fact not be the case exactly. We can verify that 
\begin{equation}
    \left\{\mathcal Q_{\mathrm{flat}(2)},\mathcal P\right\} + \left\{\mathcal Q,\mathcal P_{\mathrm{flat}(2)}\right\} = 0.
\end{equation}
However the last remaining term will not drop on its own. For it to vanish, we would need to introduce corrections up to the third order in both variables $\mathcal Q_{\mathrm{flat}(3)}$, and $\mathcal P_{\mathrm{flat}(3)}$, thus terms mixing the first and third order in the Poisson bracket like $\left\{\mathcal Q_{\mathrm{flat}(1)},\mathcal P_{\mathrm{flat}(3)}\right\}$ will be of the same order as $\left\{\mathcal Q_{\mathrm{flat}(2)},\mathcal P_{\mathrm{flat}(2)}\right\}$, and this total sum would then vanish. However it is sufficient in our case that the Poisson bracket equals one up to second order corrections. We can affirm that the variables $\left(\mathcal Q_\mathrm{flat},\mathcal P_\mathrm{flat}\right)$ form indeed a canonical pair. This is actually a remarkable result since we corrected the Mukhanov-Sasaki momenta variable in order to ensure that it remains gauge invariant and equal to the field momenta variable in the flat gauge and obtained this additional property.

\subsubsection{Extension to other gauge invariant quantities}
Let us point out how this method extends to other gauge invariant quantities. 
For example, we consider the linear curvature perturbation
\begin{align}
    \mathcal R \equiv \frac{\delta\gamma_1}{2\sqrt 3 v^{2/3}} - \frac{\theta v}{2\Mp^2\pi_\phi}Q\,,
\end{align}
proportional to $\delta \gamma_1$ in the comoving gauge where $Q$ vanishes.
At large scales, it becomes simply proportional to the Mukhanov-Sasaki variable,
\begin{equation}
    \mathcal R =  -\frac{\theta v}{2\Mp^2 \pi_\phi} \mathcal Q.
\end{equation}
This equality imposes that the second-order gauge-dependent part of $\mathcal{R}$, $\tilde{\mathcal{R}}^{(2)}$, is proportional to $\tilde{\mathcal{Q}}^{(2)}$ above
and, therefore, $\mathcal R_\text{part sol} =  -\theta v/(2\Mp^2 \pi_\phi) \mathcal Q_\text{part sol} $ is gauge invariant at second order on super-Hubble scales.
However, we also wish to impose that the gauge-invariant curvature perturbation remains proportional to $\delta\gamma_1$ in the comoving gauge. In order to  guaranty this property at the second order we will once again use the linearly independent GI quantities we previously defined, and we get
\begin{align}
    \mathcal R_\text{comoving} =& \mathcal R_\text{part sol} - \frac{1}{\sqrt 3 \Mp^2 v^{1/3}\sqrt{p+\rho}} v_4 - \frac{\sqrt \rho}{\sqrt 3 \Mp v (p+\rho)}v_5 + \frac{2\sqrt \rho}{3\Mp v^{2/3} \sqrt{p + \rho}}v_6 \\ \nonumber
    =& \frac{\delta\gamma_1}{2\sqrt 3 v^{2/3}} - \frac{\theta v}{2\Mp^2\pi_\phi}Q - \frac{\sqrt \rho}{\sqrt 3 \Mp v (p+\rho)}Q P - \frac{1}{\sqrt 3 \Mp^2 v^{1/3} \sqrt{p + \rho}}Q\delta\pi_1 \\ \nonumber
    &+ \frac{2\sqrt \rho}{3\Mp v^{2/3}\sqrt{p+\rho}}Q\delta\gamma_1 + \left(\frac{V_{;\phi}\sqrt\rho}{2\sqrt 3 \Mp^2 (p+\rho)^{3/2}} + \frac{2\rho}{3\Mp^2 (p+\rho)}-\frac{3p+\rho}{12\Mp^2 (p+\rho)}\right) Q^2.
\end{align}
We can now check explicitly that in the comoving gauge, imposing $Q=0$ (and $\delta\gamma_2 = 0$ but this condition disappears at large scales) leads to 
\begin{equation}
    \mathcal R_\text{comoving} \underset{\mathrm{CG}}{=} \frac{\delta\gamma_1}{2\sqrt 3 v^{2/3}},
\end{equation}
even at the second order, and we used $\underset{\mathrm{CG}}{=} $ to indicate an equality in the comoving gauge. This method can be applied to any other first order gauge invariant quantity, and additional requirements can be imposed thanks to the linearly independent GI quantities we have defined.

\subsection{Extension to multiple fields}\label{ssec:multifield}
Let us now extend the construction of corrected Mukhanov-Sasaki variables to the multifield case. We will however not relax the assumption that we are working at large scales, all gradient terms are taken to be vanishing. We will not be deriving the corrections systematically as we did in the single field case. We will however be presenting the covariant generalisation of Eqs. (\ref{eq:Q flat} $\&\,$\ref{eq:P flat}), and checking that these terms are indeed GI at the second order and form a canonical pair\footnote{Let us note that the generalisation could not have been guessed from the single field case, we systematically computed the corrections in the two field case with a covariant coupling metric, and then extended it to the $\mathcal N$ field case. The two field case is treated explicitly in appendix \ref{app:two field}.}.

We start once again from the known multifield Mukhanov-Sasaki Eq. (\ref{eq:QMS}) and their conjugate momenta Eq. (\ref{eq:PMS}). We can then compute their second order transformation under gauge parameter $\xi_{(1)}^\mu$ and $\xi_{(2)}^\mu$ at large scales, which read
\begin{align}
    \tilde Q^{I(2)} =&\, (\xi_{(1)}^0)^2\left(\frac{\dot N}{2v}\pi^I -\frac{N^2 \sqrt 3}{2 v \Mp}\sqrt \rho  \pi^I - \frac{N^2} 2 G\uTsr V_{;J} \right) + \frac{N}{2v}\xi_{(1)}^0(\partial_\tau \xi_{(1)}^0)+\xi_{(1)}^0 D_\tau Q^I + \xi_{(2)}^0\frac N v \pi^I \,,\\
    \tilde P_J ^{(2)}=&\,  -(\xi_{(1)}^0)^2\left(\frac{N^2 \sqrt 3 v}{2\Mp}\sqrt \rho V_{;J} + \frac{N^2\pi^I}{2} V_{;IJ} + \frac{\dot N v}{2}V_{;J} + \frac 1 2 R_{KIJL}\pi^I\pi^K\pi^L\right) - \frac{Nv}{2}V_{;J}\xi_{(1)}^0\partial_\tau \xi_{(1)}^0  \\
    \nonumber &\,-\xi_{(1)}^0\left[\delta N v V_{;J} + \frac{\sqrt 3 Nv^{1/3}}{2}V_{;J} \delta\gamma_1 + \left(\frac{N}{2v}R_{KIJL}\pi^K\pi^L - NvV_{;IJ}\right) Q^I \right] - Nv V_{;J}\xi^0_{(2)}.
\end{align}
Where we have already used the equation of motion for $P_J$ in the last line. Similarly as previously we can now combine these expressions to Eqs. (\ref{eq:GT gamma 1 2 LS} $\&\,$\ref{eq:GT gamma 2 2 LS}) to get the transform of the Mukhanov-Sasaki variables and their momenta. We will not write them out since they are similar to the single field case and we are mostly interested by the corrected variables, which we shall now present
\begin{align}
    \mathcal Q^I_\mathrm{flat} =&\,  Q^I - \frac{\Mp  \pi^I}{2\sqrt{2} v^{5/3} \sqrt\rho}\left(\sqrt{2} \delta\gamma_1 - \delta \gamma_2\right) +\frac{\pi^I}{2v^2\rho}\delta\gamma_1\delta\pi_1 + \frac{G^{IJ}\Mp}{2v^{1/3}\sqrt \rho}\delta\gamma_1P_J \\
    \nonumber &\,+ \left[\frac{G^{IJ}\Mp^2V_{;J}}{8v^{1/3}\rho} - \frac{\Mp\pi^I}{2\sqrt 3 v^{7/3}\sqrt \rho}\left(1+\frac{3(\rho+p)}{8\rho} \right)\right]\delta\gamma_1^2 + \frac{\Mp^2\pi^I}{4\sqrt 2 v^{7/3}\rho} D^{mn}(\delta\gamma_1\partial_\tau (D_{mn}\delta\gamma_2))\,,\\
    \mathcal P_{J,\mathrm{flat}} =&\, P_J + \frac{\Mp v^{1/3}}{2\sqrt 2\sqrt \rho} V_{;J}\left(\sqrt 2 \delta\gamma_1 - \delta\gamma_2\right)  + \frac{\Mp}{4v^{5/3}\sqrt \rho}\left(\mathcal R_{IJ} - 2 v^2 V_{;IJ}\right)\delta\gamma_1Q^I -\frac{V_{;J}}{2\rho}\delta\gamma_1\delta\pi_1 \\
    \nonumber &\,+\left[\frac{\Mp V_{;J}}{4\sqrt 3 v^{1/3}\sqrt\rho}\left(\frac{\rho+3p}{8\rho}-1\right)-\frac{\Mp^2}{16v^{10/3}\rho}\left(\mathcal R_{IJ} - 2 v^2 V_{;IJ})\pi^I \right)\right] \delta\gamma_1^2 \\
    \nonumber &\,-\frac{\Mp^2}{4\sqrt 2 v^{1/3}\sqrt \rho} V_{;J} D^{mn}(\delta\gamma_1\partial_\tau (D_{mn}\delta\gamma_2)).
\end{align}
Where we have introduced the notation $\mathcal R\Tsr = R_{KIJL}\pi^K\pi^L$. These sets of variables once again exhibit all the expected behaviours. They are GI up to the second order and simplify to the covariant perturbations of the fields $\left(Q^I,P_J\right)$ in the flat gauge. Similarly to the single field case, we can check that
\begin{equation}
    \left\{\mathcal Q^I_\mathrm{flat},\mathcal P_{J,\mathrm{flat}}\right\} = \delta^I_J + \mathcal O(\text{second order corrections})\,,
\end{equation}
insuring that we have built canonical pairs. 

\section{Gauge fixed Hamiltonian}
\label{sec:gauge fixing}

Now that we have found quantities that are gauge invariant at large scales and reduce to the covariant field perturbations in the flat gauge we shall look into reducing the gauge independent Hamiltonian to the flat gauge Hamiltonian for these quantities. Even though our corrected Mukhanov-Sasaki variables are only gauge invariant at large scales we can still compute the Hamiltonian for these quantities in the flat gauge at all scales. Let us present the method we followed to achieve this. This method can be used for any well defined gauge. In our example we take the spatial metric to have no fluctuations, \ie $\delta\gamma_1 = \delta\gamma_2 = 0$. This already greatly simplifies the obtained Hamiltonian since all terms containing any of those two fluctuations vanish
\begin{align}
    C^{(1)}\cov \underset{\mathrm{SF}}{=}&\, vV;_K Q^K + \frac{G^{KI}\pi_I}{v}P_K -\frac{\sqrt{3}}{\Mp^2}v^{2/3}\theta\delta\pi_1 \, ,\\
    C^{(2)}\cov \underset{\mathrm{SF}}{=}&\, \frac{1}{2v}P_IP^I +\frac{v}{2}\left(V;_{IJ} - \frac 1 {v^2}R_I{}^{KL}{}_J\pi_K\pi_L -\frac{\partial_i\partial^i}{v^{2/3}}\right)Q^IQ^J - \frac{v^{1/3}}{\Mp^2}\delta\pi_1^2 + \frac{2v^{1/3}}{\Mp^2}D_{mn}\delta\pi_2D^{mn}\delta\pi_2\,,\\
    \tilde{\mathcal C}^{(2)} \underset{\mathrm{SF}}{=}&\, \frac{1}{2v}P_IP^I +\frac{v}{2}\left(V;_{IJ} - \frac 1 {3v^2}R_I{}^{KL}{}_J\pi_K\pi_L -\frac{\partial_i\partial^i}{v^{2/3}}\right)Q^IQ^J - \frac{v^{1/3}}{\Mp^2}\delta\pi_1^2 + \frac{2v^{1/3}}{\Mp^2}D_{mn}\delta\pi_2D^{mn}\delta\pi_2 \, ,\\
    C^{(3)}\cov \underset{\mathrm{SF}}{=}&\,  \frac 1 6 V_{;IJK}Q^IQ^JQ^K -\frac{1}{6v} R^I{}_{KL}{}^{J}{}_{;M}\pi_I\pi_JQ^KQ^LQ^M -\frac{2}{3v} \pi_N G^{NJ} R^M{}_{KLJ} P_M  Q^L  Q^K .
\end{align}
Since we have imposed the constraints to vanish at the background and at first order by least action principle, we can use those to express the gravitational momenta 
\begin{align}
    \delta\pi_1 =&\, \frac{\Mp^2}{\sqrt{3}v^{5/3}\theta}\pi_I P_I+\frac{\Mp^2v^{1/3}}{\sqrt{3}\theta}V_{;I}Q^I\,, \\
    \delta\pi_2 =&\, \frac{\sqrt 3 \pi_I}{2\sqrt 2 v^{2/3}}Q^I - \frac{\delta\pi_1}{\sqrt 2} .
\end{align}
We wish to stay in the flat gauge during the evolution of our fluctuations. Thus we impose that the gauge condition hold through time : $D_\tau \delta\gamma_1 = D_\tau \delta\gamma_2 = 0$. Reinjecting this into the covariant equations of motion Eqs. (\ref{eq:EoM gamma 1}$\&\,$\ref{eq:EoM gamma 2}), and using both the gauge conditions and the expressions for the gravitational momenta we have just derived, we can also express the perturbation of the lapse and shift functions in terms of covariant field perturbations $\left(Q^I,P_J\right)$ only. This leads to 
\begin{align}
    \delta N =&\, -\frac 1 {v\theta}\pi_IQ^I \,,\\
    \partial_i \delta N^i =&\, \frac 1 {v^2\theta}\pi_IP^I + \frac 1 \theta V_{;I}Q^I - \frac 3 {2v\Mp^2}\pi_IQ^I.
\end{align}
We can now combine all of these steps to express the second and third order Hamiltonian Eq. (\ref{eq:Ham 2 3}) in terms of the field perturbations only. We can now use the fact that the Mukhanov Sasaki variables reduce to the field perturbations in the flat gauge to claim that this Hamiltonian is \textit{at large scales} the Hamiltonian describing the evolution of the variables $\left(\mathcal Q^I_\mathrm{flat},\mathcal P_{J,\mathrm{flat}}\right)$. Before presenting the complete result let us proceed to a canonical redefinition of the field momenta in order to simplify further our expression and drop the cross terms in the second order Hamiltonian. 
\begin{align}
    P_J \rightarrow \tilde P_J =&\, P_J -\frac{1}{v\theta}\pi_I\pi_JQ^J\,, \\
    Q^I \rightarrow \tilde Q^I =&\, Q^I\,,
\end{align}
with the associated type two generating function 
\begin{equation}
    G(\tilde Q^I,P_J,\tau) = -P_I\tilde Q^I- \frac{1}{2v\theta}\pi_I\pi_J\tilde Q^I \tilde Q^J.
\end{equation}
One could ask whether the two operations we have done here commute. In our case, we would get the same result if we had done the canonical transform before reducing the Hamiltonian since the canonical transform and the gauge condition do not affect the same variables, however this is not true in all generality. We can see that thanks to the field redefinition, the cross term in the second order Hamiltonian drops. Going to Fourier space, we finally get the following second and third order contributions in the flat gauge
\begin{align}
    H^{(2),(3)} = \int \dd^3k &\,\left[ D\uTsr \tilde P_I\tilde P_J + M\Tsr Q^I Q^J + A_{IJK} Q^IQ^JQ^K \right.\\
    &\,\left.+ B\Tsr^KQ^IQ^J\tilde P_K + C\uTsr_K \tilde P_I\tilde P_JQ^K\right].
\end{align}
With
\begin{align}
    D\uTsr =&\, \frac{1}{2v} G\Tsr \,,\\
    M\Tsr =&\, \frac{v}{2} V;_{IJ} - \frac 1 {2v}R_I{}^{KL}{}_J\pi_K\pi_L +\frac{v^{1/3} k^2}{2}G_{IJ} -\frac{2}{\theta} V_{;I}\pi_J + \frac{(\rho-p)}{v\theta^2}\pi_I\pi_J\,,\\
    A_{IJK} =&\, \frac 1 6 V_{;IJK} -\frac{1}{6v} R_{MIJL;K}\pi^L\pi^M - \frac{2}{3v^2\theta} \pi^L \pi^M \pi_I R_{LJKM} \\
    \nonumber &\,-\frac{V;_{JK}}{2\theta}\pi_I + \frac 1 {2\theta v^2}R_{JLMK}\pi_I\pi^M\pi_L -\frac{k^2}{2\theta v^{2/3}}G_{JK}\pi_I \\
    \nonumber &\,+\frac{1}{v\theta^2} V_{;I}\pi_J\pi_K - \frac{(\rho-p)}{2v^2\theta^3}\pi_I\pi_J\pi_K +\left(\frac{1}{v\theta^2}V_{;I}\pi_J\pi_K - \frac{3}{2v^2\theta\Mp}\pi_I\pi_J\pi_K\right)\frac{k_1\cdot k_3}{k_1^2}\\
    \nonumber &\,\left(\frac{\Mp^2}{2\theta^3}\pi_IV_{;J}V_{;K} - \frac{\Mp^2}{v\theta^4}(\rho-p)\pi_I\pi_JV_{;K} - \frac{3}{8 v^2\theta^3}(\rho-p)^2\pi_I\pi_J\pi_K\right) \left(1-\frac{(k_2.k_3)^2}{k_2^2k_3^2}\right) \,,\\
    \nonumber B\Tsr^K =&\, -\frac{1}{v} \pi_M G^{LM} R^K{}_{IJL} + \left(\frac{\Mp^2}{v^2\theta^3} \pi_IV_{;J}\pi^K + \frac{\Mp(p-\rho)}{v^3\theta}\pi_I\pi_J\pi^K\right)\left(1-\frac{(k_2.k_3)^2}{k_2^2k_3^2}\right) \\
    &\,+\frac{1}{v^3\theta^2}\pi_I\pi_J\pi^K \frac{k_1\cdot k_3}{k_3^2} + \left(\frac 1 \theta V_{;I}G^{KL}G_{JL} - \frac 3 {2v\Mp^2}\pi_I G^{KL}G_{JL}\right)\frac{k_1\cdot k_2}{k_1^2} \,,\\
    C\uTsr_K =&\, -\frac{1}{2v^2\theta} G\uTsr \pi^K + \frac \Mp{2v^4\theta^3} \pi^I\pi^J\pi_K \left(1-\frac{(k_1.k_2)^2}{k_1^2k_2^2}\right) + \frac{1}{v^2\theta}\pi_I G^{JL}G_{LK} \frac{k_1\cdot k_3}{k_1^2}.
\end{align}
Where the indices $_1,\,_2$ and $_3$ correspond to the wave-number of the  quantity with index $I,J$ and $K$ respectively. We have opted to write our results in such a way that comparing to the literature is quite direct. We have thus shown how to construct the Hamiltonian in the flat gauge for the Mukhanov-Sasaki gauge invariant variables. The method we followed can be extended to any gauge, and is better suited once the adequate gauge-invariant variables have been identified for said gauge. We proved in this work that this Hamiltonian converges at large scales to the gauge fixed Hamiltonian for the gauge invariant at second order Mukhanov Sasaki variables $(\mathcal Q^I_\mathrm{flat},P_{J,\mathrm{flat}})$ we have constructed. One thus needs to be careful when computing results with this Hamiltonian and these quantities at small scales since we have not shown that there exists a correction to the Mukhanov-Sasaki variable that is gauge invariant at all scales \textit{and} that coincides with the field fluctuations in the flat gauge.

\section{Formal construction of gauge invariant variables}
\label{ssec:GI 1 and 2}

A general process to build gauge-invariant variables at the second order is defined by extending the phase-space analysis done in \cite{Artigas:2023kyo}, which is limited to the linear theory. Here, we choose to keep considering large scales as it greatly simplifies the construction of gauge-invariant variables by neglecting gradients. In practice, it means that we can work in the separate-universe picture where perturbative degrees of freedom enjoy the same symmetries as the background \cite{article,PhysRevD.42.3936,LIFSHITZ1992493,PhysRevD.49.2759,Wands_2000,I_M_Khalatnikov_2002,PhysRevD.68.103515,PhysRevD.68.123518,Tanaka_2021,Artigas_2022,Grain_2026}. As a result, the diffeomorphism constraint is large-scale suppressed at any order in perturbation and the anisotropic degrees of freedom can be ignored. Hence, we are left with a $(2\N+2)$-dimensional phase-space (instead of a $2\N+4$), and made of the scalar-field fluctuations $(\delta\phi^I,\delta\pi^I_\phi)$ and the {\it isotropic} fluctuations of the metric $(\delta\gamma_1,\delta\pi_1)$, as well as one constraint $\mathcal{C}$ whose associated Lagrange multiplier is the lapse function $\delta N$. 

We summarize the phase-space analysis of \cite{Artigas:2023kyo} in the framework of the separate-universe approximation before turning to second order. We present this extension in details considering one single scalar field for simplicity. Its generalizations to the multifield context and to the third order are then briefly outlined.

\subsection{Phase-space parametrization}
\subsubsection*{Dynamics at first order}
In the separate-universe picture, the background remains unchanged and we organize the phase-space variables in a 4-dimensional vector, \ie
\bea   
    \boldsymbol{z}^\mathrm{T}\equiv(\phi,\gamma_1,\pi_\phi,\pi_1),
\eea
where we perform a slight phase-space reparametrization of the gravitational sector, $\gamma_1=\sqrt{3}v^{2/3}$ and $\pi_1=(\sqrt{3}/2)v^{1/3}\theta$ (note that it is easily checked to correspond to a canonical transformation). In what follows, boldface variables will denote a vector in the phase-space and we will use $a,b,\cdots\in\{1,2,3,4\}$ as indices to label their coordinates. The superscript T means the transpose. Similarly, we arrange the perturbative degrees of freedom in the vector 
\bea   
    \boldsymbol{\delta z}^\mathrm{T}\equiv\left(\delta\phi,\delta\gamma_1,\delta\pi_\phi,\delta\pi_1\right),
\eea
where the entries are all isotropic and now considered as homogeneous. Finally, the lapse function reads $N+\delta N$ where its fluctuations are homogeneous too. 

The Hamiltonian at third order reads
\bea
    \mathcal{H}=N\mathcal{C}^{(0)}+\delta N\left(\mathcal{C}^{(1)}+\mathcal{C}^{(2)}\right)+N\left(\mathcal{C}^{(2)}+\mathcal{C}^{(3)}\right).
\eea
At the background level, $\mathcal{C}^{(0)}$ is unchanged (up to the phase-space reparametrization). For perturbations, the constraints $\mathcal{C}^{(n\geq1)}$ are obtained from their full expressions where gradients and anisotropic degrees of freedom are discarded. Using such vector notations, the Hamilton equations for the background are recasted as
\bea
    \dot{\boldsymbol{z}}_a=N\boldsymbol{\Omega}_{ab}\,\partial_b\mathcal{C}^{(0)}, \label{eq:Ham0vect}
\eea
where $\partial_b\equiv\partial/\partial\boldsymbol{z}_b$ and where $\boldsymbol{\Omega}$ is the totally antisymmetric matrix that reads as
\bea    
    \boldsymbol{\Omega}=\left(\begin{array}{cc}
        0 &\, \boldsymbol{I} \\
        -\boldsymbol{I} &\, 0
    \end{array}\right). \label{eq:Omega}
\eea
In the above, $\boldsymbol{I}$ stands for $2\times2$ identity and the inverse of $\boldsymbol{\Omega}$ is $\boldsymbol{\Omega}^{-1}=\boldsymbol{\Omega}^\mathrm{T}=-\boldsymbol{\Omega}$.

For perturbations, the constraint at first order is written as $\mathcal{C}^{(1)}=\boldsymbol{C}^{(1)}_a\boldsymbol{\delta z}_a$ where the vector $\boldsymbol{C}^{(1)}$ is made of functions of the background variables derived by expanding the full constraint around $\mathcal{C}^{(0)}$ up to linear order. In the separate-universe picture where gradients are negligible, one easily sees that $\boldsymbol{C}^{(1)}_a=\partial_a\mathcal{C}^{(0)}$. Similarly, the scalar constraint at second order can be written as a quadratic form in the phase-space $\mathcal{C}^{(2)}=\frac{1}{2}\boldsymbol{\delta z}_a\left(\partial_{a}\partial_b\mathcal{C}^{(0)}\right)\boldsymbol{\delta z}_b$. At linear order then, the Hamilton equations are
\bea
    \dot{\boldsymbol{\delta z}}=\delta N\,\boldsymbol{\Omega}\boldsymbol{C}^{(1)}+N\,\boldsymbol{\Omega}\boldsymbol{C}^{(2)}\boldsymbol{\delta z} \label{eq:Ham1vect},
\eea
where $\boldsymbol{C}^{(2)}_{ab}\equiv\partial_a\partial_b\mathcal{C}^{(0)}$. The dynamics should be solved for on the surface of constraints, \ie on the surface defined by $\mathcal{C}^{(1)}=0$, since the action is minimized irrespectively of the choice of $\delta N$. As shown in \cite{Artigas:2023kyo}, the algebra of constraints of general relativity imposes that $\dot{\boldsymbol{C}}^{(1)}=N\boldsymbol{C}^{(2)}\boldsymbol{\Omega}\boldsymbol{C}^{(1)}$ where the contribution from the diffeomorphism constraint has been dropped here since it is large-scale suppressed. This ensures that the evolution remains on the surface of constraints providing that one initially starts on it. Finally, the constraint at first order generates gauge transformations of the perturbative degrees of freedom through $\mathcal{L}_{\xi_{(1)}^0}(\boldsymbol{\delta z})=\left\{\boldsymbol{\delta z},\xi_{(1)}^0 N\mathcal{C}^{(1)}\right\}$. Using vector notations, it explictly reads
\bea
    \mathcal{L}_{\xi_{(1)}^0}(\boldsymbol{\delta z})=\xi_{(1)}^0\,N\boldsymbol{\Omega C}^{(1)}.
\eea
It results that gauge transformations are given by translations along the phase-space direction given by the vector $(\boldsymbol{\Omega C}^{(1)})$.

\subsubsection*{Phase-space reparametrization}
Following \cite{Artigas:2023kyo}, it points toward a convenient parametrization of the phase-space of perturbations when it comes to identifying physical degrees of freedom from non-physical ones. To that end, let us first introduce the vector $\boldsymbol{e}^\mathcal{C}\equiv\boldsymbol{C}^{(1)}/\left|\boldsymbol{C}^{(1)}\right|$ which projects on the surface of constraint, which is 1-dimensional in the separate-universe picture. As a result, the variable $\delta C\equiv\boldsymbol{e}^\mathcal{C}_a\,\boldsymbol{\delta z}_a$ is proportional to the linear constraint $\mathcal{C}^{(1)}$. It can be interpreted as one configuration variable in the phase-space whose dynamics reads $\delta C=0$ and $\dot{\delta C}=f(\tau) \delta C$ where $f$ depends on the background variables and whose exact expression is not needed here. Thus, we simply recover that the constraint is vanishing and that it is preserved through evolution. 

Second, we introduce the vector $\bar{\boldsymbol{e}}^\mathcal{C}\equiv-\boldsymbol{\Omega e}^\mathcal{C}$. This vector generates gauge transformations since it is aligned with $\boldsymbol{\Omega C}^{(1)}$, and it is orthogonal to the vector $\boldsymbol{e}^\mathcal{C}$ which projects on the surface of constraints. Hence, one can build a second variable in the phase-space through $\delta G\equiv \bar{\boldsymbol{e}}^\mathcal{C}_a\,\boldsymbol{\delta z}_a$ which is linearly independent of $\delta C$. The variable $\delta G$ stands for the gauge degree of freedom since $\mathcal{L}_{\xi_{(1)}^0}(\delta g)=\xi_{(1)}^0\left|\boldsymbol{C}^{(1)}\right|$ (note that as expected, $\delta C$ is recovered to be gauge invariant since ${\boldsymbol{e}}^\mathcal{C}$ is orthogonal to $\bar{\boldsymbol{e}}^\mathcal{C}$). Finally, one can easily check that $\left\{\delta C,\delta G\right\}=1$ and $\delta G$ is the momentum canonically conjugated to the constraint. One can further check from Eq. (\ref{eq:Ham1vect}) that $\dot{\delta G}$ is sourced by $\delta N$, hence dependent from our freedom in choosing the lapse function.

Finally, we introduce two vectors to complete our basis of the phase-space  defining them as being orthogonal to both ${\boldsymbol{e}}^\mathcal{C}$ and $\bar{\boldsymbol{e}}^\mathcal{C}$. In practice, one can use for instance the Gram-Schmidt procedure to built a first vector of that sort and that we dub $\boldsymbol{e}^\mathcal{P}$. Then the second easily follows as $\bar{\boldsymbol{e}}^\mathcal{P}\equiv-\boldsymbol{\Omega e}^\mathcal{P}$ and because of that, these two last vectors are also orthogonal one to each other. This allows us to define two new variables in the phase-space as $\delta Q\equiv{\boldsymbol{e}}^\mathcal{P}_a\,\boldsymbol{\delta z}_a$ and $\delta P\equiv\bar{\boldsymbol{e}}^\mathcal{P}_a\,\boldsymbol{\delta z}_a$. By construction, these variables are gauge-invariant which further justifies that $\delta G$ is the sole variable bearing the gauge redundancy. Moreover, it is easy to verify that $\left\{\delta Q,\delta P\right\}=1$ and these two variables are no more than the canonical pair of physical degrees of freedom. It is seen from Eq. (\ref{eq:Ham1vect}) that the time evolution of $\delta Q$ and $\delta P$ has no contribution from $\delta N$ and it was shown in \cite{Artigas:2023kyo} that they are not coupled to the gauge degree of freedom.

To summarize, the set of basis vector that we have just built provides us with an orthonormal basis of the phase-space which neatly separates the physical degrees of freedom $(\delta Q,\delta P)$ from the unphysical ones $(\delta C,\delta G)$, and which organizes them by pairs of canonical variables. From now on, we organize the perturbative phase-space using the following basis
\bea
    \left\{\boldsymbol{e}^A\right\}_{A\in\{1,2,3,4\}}\equiv\left\{\boldsymbol{e}^\mathcal{P},\boldsymbol{e}^\mathcal{C},\bar{\boldsymbol{e}}^\mathcal{P},\bar{\boldsymbol{e}}^\mathcal{C}\right\}.
\eea
Here, the two first entries stand for the configuration variables and the two last ones for their conjugated momenta. In each subspace (either configuration or momentum), the first entry stands for the physical degree of freedom and the second for the unphysical ones (\ie projection on the constraint for configuration and projection on the gauge-transformation generator for the momentum). Hence, we name it the {\it physical basis} since it separates the physical degrees of freedom from the constraint and the gauge degree of freedom at linear order. Components of phase-space vectors and matrices projected in the above basis will be denoted with $A,B,\cdots$ indices, \ie any vector is decomposed as 
\bea
    \boldsymbol{V}=\displaystyle\sum_{A=1}^4V_A\,\boldsymbol{e}^A.
\eea
Because the basis is orthonormal, the components are obtained by taking the inner-dot product with the basis vectors, for instance ${\delta z}_A\equiv\boldsymbol{e}^A\cdot\boldsymbol{\delta z}$. In practice, it corresponds to implementing the canonical transformation from the set of {\it natural variables} $(\delta\phi,\delta\gamma_1,\delta\pi_\phi,\delta\pi_1)$ to the {\it physical variables} $(\delta Q, \delta C,\delta P, \delta G)$. Because it is a canonical transformation, the matrix $\boldsymbol{\Omega}$ admits the very same expression in this new basis, Eq. (\ref{eq:Omega}). Finally, let us note that the above construction is mostly inspired by first-order considerations. Yet, the phase-space reparametrization just introduced can be implemented at any order in principle. It does not mix the different orders because the canonical transformation which generates the new variables is a mere time-dependent rotation in the phase-space. However, interpreting $(\delta Q,\delta P)$ as physical degrees of freedom and $(\delta C,\delta G)$ as the redundant ones is tied to the linear order. In the rest of this section, it will prove convenient to make explicit the summations over $A,B,\cdots$ indices.

\subsection{Gauge-invariant canonical variables}
\subsubsection*{General construction}
Using our vector notations, gauge-transformations at first and second order are recast as
\bea
    \widetilde{\boldsymbol{\delta z}}^{(1)}&=&\,{\boldsymbol{\delta z}}^{(1)}+\lambda N{\xi}^{0}_{(1)}\boldsymbol{e}^4. \label{eq:GT1vect} \\
    \widetilde{\boldsymbol{\delta z}}^{(2)}&=&\,{\boldsymbol{\delta z}}^{(2)}+\lambda N{\xi}^{0}_{(2)}\boldsymbol{e}^4+\xi^0_{(1)}\partial_\tau{\boldsymbol{\delta z}^{(1)}}+\frac{1}{2}\xi^0_{(1)}\partial_\tau\left(\xi^0_{(1)}\partial_\tau\boldsymbol{z}\right) \label{eq:GT2vect}
\eea
where we made explicit that $\boldsymbol{\Omega C}^{(1)}$ is aligned with $\boldsymbol{e}^4$ and $\lambda\equiv-\left|\boldsymbol{C}^{(1)}\right|$.\footnote{The minus sign comes from $\boldsymbol{e}^2=\boldsymbol{C}^{(1)}/\left|\boldsymbol{C}^{(1)}\right|$ and $\boldsymbol{e}^4$ is defined as $\boldsymbol{e}^4\equiv-\boldsymbol{\Omega e}^2$.} At linear order, the construction of a pair of canonically conjugated variables was proposed in \cite{Artigas:2023kyo} and it works as follows. We first pick up a vector which is orthogonal to $\boldsymbol{e}^4$, \ie $\boldsymbol{Q}=\sum_{A=1}^3Q_A\,\boldsymbol{e}^A$. Hence the variable $Q^{(1)}_{\mathrm{GI}}\equiv \boldsymbol{Q}\cdot\boldsymbol{\delta z}$ is automatically gauge-invariant. The physical momentum associated to $Q^{(1)}_{\mathrm{GI}}$ is obtained introducing a second vector orthogonal to the gauge direction, \ie $\boldsymbol{P}=\sum_{A=1}^3P_A\,\boldsymbol{e}^A$. Then, $P^{(1)}_{\mathrm{GI}}\equiv \boldsymbol{P}\cdot\boldsymbol{\delta z}$ is gauge-invariant too. It is further constrained requiring that $\left\{Q^{(1)}_{\mathrm{GI}},P^{(1)}_{\mathrm{GI}}\right\}=1$. It boils down to imposing $Q_1P_3-Q_3P_1=1$ while $P_2$ remains arbitrary since it corresponds to the projection on the constraint. Note that at least $Q_1$ and/or $Q_3$ should be nonzero in order to  define a {\it pair} of canonical variables which are gauge-invariant. It automatically ensures that $Q^{(1)}_{\mathrm{GI}}$ and $P^{(1)}_{\mathrm{GI}}$ have non-vanishing projection in the plane of physical degrees of freedom.

Let us now turn to the second order. Here, we first need to rearrange Eq. (\ref{eq:GT2vect}) to make the different contribution of the constraint direction, the gauge direction, and the physical directions more explicit. To that end, we first make use of the Hamilton equation for $\boldsymbol{\delta z}^{(1)}$, Eq. (\ref{eq:Ham1vect}), to obtain
\bea
    \xi^0_{(1)}\partial_\tau{\boldsymbol{\delta z}^{(1)}}=\xi^0_{(1)}\left[\lambda\delta N \boldsymbol{e}^4+N\boldsymbol{\Omega C}^{(2)}\boldsymbol{\delta z}^{(1)}\right].
\eea
Similarly, we uses the Hamilton equation for the background evolution, Eq. (\ref{eq:Ham0vect}), as well as $\partial_a\mathcal{C}^{(0)}=\boldsymbol{C}^{(1)}$, to rewrite
\bea
    \xi^0_{(1)}\partial_\tau\left(\xi^0_{(1)}\partial_\tau\boldsymbol{z}\right)&=&\,\xi^0_{(1)}\partial_\tau\left(N\xi^0_{(1)}\boldsymbol{\Omega C}^{(1)}\right).
\eea
Then, we expand the time derivative and use  $\dot{\boldsymbol{C}}^{(1)}=N\boldsymbol{C}^{(2)}\boldsymbol{\Omega}\boldsymbol{C}^{(1)}$ \cite{Artigas:2023kyo}. By making explicit that $\boldsymbol{\Omega}\boldsymbol{C}^{(1)}$ is aligned with the gauge direction $\boldsymbol{e}^4$, it gives
\bea
    \xi^0_{(1)}\partial_\tau\left(\xi^0_{(1)}\partial_\tau\boldsymbol{z}\right)&=&\,\lambda\left[\xi^0_{(1)}\partial_\tau\left(N{\xi}^0_{(1)}\right)\boldsymbol{e}^4+\left(N\xi^0_{(1)}\right)^2\boldsymbol{\Omega C^{(2)}}\boldsymbol{e}^4\right].
\eea
Collecting everything, Eq. (\ref{eq:GT2vect}) boils down to
\bea
    \widetilde{\boldsymbol{\delta z}}^{(2)}&=&\,{\boldsymbol{\delta z}}^{(2)}+\lambda\left[N\xi^0_{(2)}+\xi^0_{(1)}\delta N+\frac{1}{2}\xi^0_{(1)}\partial_\tau\left(N{\xi}^0_{(1)}\right)\right]\boldsymbol{e}^4+\frac{\lambda}{2} \left(N\xi^0_{(1)}\right)^2\boldsymbol{\Omega C}^{(2)}\boldsymbol{e}^4 \nonumber \\
    &\,&\,+N\xi^0_{(1)}\boldsymbol{\Omega C}^{(2)}\boldsymbol{\delta z}^{(1)}. \label{eq:GT2bisvect}
\eea
We note that one recognizes in the above the gauge-transformed Lagrange multiplier since at linear order, fluctuations in the lapse function transform as $\widetilde{\delta N}=\delta N+\partial_\tau(N\xi^0_{(1)})$ [see Eq. (\ref{eq:GT lapse 1})]. To interpret our expression of the gauge-transformed variables at second order, we decompose the vector $\boldsymbol{\delta z}$ and the matrix $\boldsymbol{C}^{(2)}$ on the basis $\left\{\boldsymbol{e}^A\right\}_{A\in\{1,2,3,4\}}$, \ie $\boldsymbol{\delta z}=\sum_{A=1}^4\delta z_A \,\boldsymbol{e}^A$ and $\boldsymbol{C}^{(2)}=\sum_{A,B=1}^4C_{AB}\,(\boldsymbol{e}^A\otimes\boldsymbol{e}^B)$. It leads to
\bea
    \widetilde{\boldsymbol{\delta z}}^{(2)}&=&\,{\boldsymbol{\delta z}}^{(2)}+\lambda\left[N\xi^0_{(2)}+\xi^0_{(1)}\delta N+\frac{1}{2}\xi^0_{(1)}\partial_\tau\left(N{\xi}^0_{(1)}\right)\right]\boldsymbol{e}^4+\frac{\lambda}{2} \left(N\xi^0_{(1)}\right)^2\ds\sum_{A=1}^4C_{A4}\,\left(\boldsymbol{\Omega e}^A\right) \nonumber \\
    &\,&\,+N\xi^0_{(1)} \ds\sum_{A,B=1}^4\left(C_{AB}\,\delta z_B\right)\,\left(\boldsymbol{\Omega e}^A\right).
\eea
At second order then, a gauge transformation can be viewed as the contribution of three translations. The first one is along $\boldsymbol{e}^4$ and its amplitude is set by the background evolution through $\lambda$ and $N$, as well as by perturbations in the lapse function gauge-transformed at linear order through $\delta N+\mathcal{L}_{\xi_{(1)}^0}(\delta N)$. The second is aligned with a linear combination of the four vectors $\left\{\boldsymbol{e}^A\right\}_{A\in\{1,2,3,4\}}$ through $\boldsymbol{\Omega C}^{(2)}\boldsymbol{e}^4$ (note that by construction the vector $\boldsymbol{\Omega e}^A$ is one of the four basis vector). Their relative amplitudes are solely set by the background evolution. The last one is also aligned with a linear combination of the four vectors $\left\{\boldsymbol{e}^A\right\}$ through $\boldsymbol{\Omega C}^{(2)}\boldsymbol{\delta z}^{(1)}$. However, their relative amplitudes are now set by the dynamics of the perturbations at first order. Let us also stress that Eq. (\ref{eq:GT2bisvect}) is equivalently derived starting directly from the expression of the Lie derivative as Poisson brackets with the constraints, rather than from a direct calculation of Lie derivatives of the ADM variables. This is done in App. \ref{app:GTHam} and a similar strategy was adopted in \cite{Dom_nech_2018}. However, our expression slightly differs from the one we would obtain starting from the Hamiltonian definition of gauge transformations proposed in \cite{Dom_nech_2018}. Indeed, the contribution of $\widetilde{\delta N}$ to the translation along $\boldsymbol{e}^4$ is absent in the latter, while all the rest perfectly match. This mismatch is detailed in App. \ref{app:GTHam} and as it will be shown hereafter, this difference does not impact on the construction of gauge-invariant variables.

We now define the most generic variable at second order by
\bea
    Q^{(2)}_\mathrm{GI}=\boldsymbol{Q}\cdot\boldsymbol{\delta z}^{(2)}+\frac{1}{2}\boldsymbol{\delta z}^{(1)\,\mathrm{T}}\boldsymbol{M}\boldsymbol{\delta z}^{(1)}, \label{eq:GIzgen}
\eea
where $\boldsymbol{Q}$ is  a 4-dimensional vector and $\boldsymbol{M}$ a $4\times4$ symmetric matrix, which can both depend on time, either explicitly or implicitly through functions of the background variables. Combining Eqs. (\ref{eq:GT1vect}) \&\, (\ref{eq:GT2bisvect}), it is gauge-transformed as 
\bea
    \widetilde{Q}^{(2)}_\mathrm{GI}-Q^{(2)}_\mathrm{GI}&=&\,\lambda\left[N\xi^0_{(2)}+\xi^0_{(1)}\delta N+\frac{1}{2}\xi^0_{(1)}\partial_\tau\left(N{\xi}^0_{(1)}\right)\right]\left(\boldsymbol{Q}\cdot\boldsymbol{e}^4\right) \\
    &\,&\,+\frac{\lambda}{2} \left(N\xi^0_{(1)}\right)^2\left(\boldsymbol{Q}^\mathrm{T}\boldsymbol{\Omega C}^{(2)}\boldsymbol{e}^4+\lambda \boldsymbol{e}^{4\,\mathrm{T}}\boldsymbol{M}\boldsymbol{e}^4\right) \nonumber \\
    &\,&\,+N\xi^0_{(1)}\left(\boldsymbol{Q}^\mathrm{T}\boldsymbol{\Omega C}^{(2)}+\lambda\boldsymbol{e}^{4\,\mathrm{T}}\boldsymbol{M}\right)\boldsymbol{\delta z}^{(1)}. \nonumber \label{eq:GT GI 2}
\eea
Gauge-invariance is obtained setting the right-hand-side to zero. To do so, we show that we can systematically design $\boldsymbol{Q}$ and $\boldsymbol{M}$ such that each lines in the above are cancelled independently. The first line in the right-hand-side is easily set to zero by imposing $\boldsymbol{Q}$ to be orthogonal to $\boldsymbol{e}^4$. Hence, we write it as $\boldsymbol{Q}=\sum_{A=1}^3 Q_A\,\boldsymbol{e}^A$. Then, we decompose the vectors and matrices of the second and third lines in the right-hand-side on the basis $\left\{\boldsymbol{e}^A\right\}_{A\in\{1,2,3,4\}}$. Imposing the second line proportional to $(\xi^0_{(1)})^2$ to vanish yields
\bea
    \ds\sum_{A=1}^3Q_A\left[\boldsymbol{\Omega C}^{(2)}\right]_{A4}+\lambda \left[\boldsymbol{M}\right]_{44}=0,
\eea
which fixes one of the entries of $\boldsymbol{M}$ given the choice of $\boldsymbol{Q}$. Finally, requiring the last line proprotional to $\xi^0_{(1)}$ to equal zero for any $\boldsymbol{\delta z}_{(1)}$ yields four constraints. They read
\bea
    \ds\sum_{A=1}^3Q_A\left[\boldsymbol{\Omega C}^{(2)}\right]_{AB}+\lambda \left[\boldsymbol{M}\right]_{4B}=0, \label{eq:Mcond}
\eea
for each $B\in\{1,2,3,4\}$. The case $B=4$ leads to the very same constraint as demanding the term proportional to $(\xi^0_{(1)})^2$ to vanish, as it should. Hence, the above fixes three additional entries in $\boldsymbol{M}$. 

To summarize, gauge-invariant variables at second order are constructed starting from Eq. (\ref{eq:GIzgen}). First, one constrains the vector $\boldsymbol{Q}$ to be orthogonal to $\boldsymbol{e}^4$. It fixes one of its components to vanish while the three remaining ones are arbitrarily chosen (up to ensuring that it has a nonzero projection on the plane of physical degrees of freedom). Second, one sets the matrix $\boldsymbol{M}$ from the four equations in Eq. (\ref{eq:Mcond}) and given the choice of $\boldsymbol{Q}$. Since $\boldsymbol{M}$ is symmetric, it fixes four independent entries among the ten independent ones. This is because the remaining six degrees of freedom in $\boldsymbol{M}$ generates quadratic variables in $\boldsymbol{\delta z}^{(1)}$ which are automatically gauge invariant. Indeed, one can split this matrix as $\boldsymbol{M}=\boldsymbol{M}_\mathrm{GI}+\boldsymbol{M}_\mathrm{GT}$ with
\bea
    \boldsymbol{M}_\mathrm{GI}&=&\,\ds\sum_{A,B=1}^3M_{AB}\,\left(\boldsymbol{e}^A\otimes\boldsymbol{e}^B\right), \label{eq:MGI} \\
    \boldsymbol{M}_\mathrm{GT}&=&\,M_{44}\,\left(\boldsymbol{e}^4\otimes\boldsymbol{e}^4\right)+\ds\sum_{A=1}^3M_{A4}\,\left(\boldsymbol{e}^A\otimes\boldsymbol{e}^4+\boldsymbol{e}^4\otimes\boldsymbol{e}^A\right) \label{eq:MGT}
\eea
where we remind that $M_{AB}=M_{BA}$. Hence, the first matrix has 6 independent degrees of freedom (note that the summations stop at 3 in $\boldsymbol{M}_\mathrm{GI}$) and the second has four. Because gauge transformations at the linear order are translations along $\boldsymbol{e}^4$ which is in the null space of $\boldsymbol{M}_\mathrm{GI}$, one straightforwardly obtains 
\bea
\mathcal{L}_{\xi^0_{(1)}}\left(\boldsymbol{\delta z}^{(1)\,\mathrm{T}}\boldsymbol{M}_\mathrm{GI}\boldsymbol{\delta z}^{(1)}\right)=0.
\eea
Hence, it is gauge-invariant. On the contrary, a gauge transformation yields  
\bea
\mathcal{L}_{\xi^0_{(1)}}\left(\boldsymbol{\delta z}^{(1)\,\mathrm{T}}\boldsymbol{M}_\mathrm{GT}\boldsymbol{\delta z}^{(1)}\right)=\left(\lambda N\xi^0_{(1)}\right)^2M_{44}+2\left(\lambda N\xi^0_{(1)}\right)\sum_{A=1}^4M_{A4}\delta z_A.
\eea
It results that gauge-invariance is obtained by solely fixing the matrix $\boldsymbol{M}_\mathrm{GT}$ given a choice of $\boldsymbol{Q}$. However, one can then add any quadratic contributions in the form of $\boldsymbol{\delta z}^{(1)\,\mathrm{T}}\boldsymbol{M}_\mathrm{GI}\boldsymbol{\delta z}^{(1)}$ to the constructed gauge-invariant variable to define another gauge-invariant variable. These remaining degrees of freedom can thus be used to design gauge-invariant variables with some desired properties.

Let us note that in this construction, the vector $\boldsymbol{Q}$ is constrained requiring that it is orthogonal to the gauge direction; hence independently of the time-dependent function in front of $\boldsymbol{e}^4$ in Eq. (\ref{eq:GT2bisvect}). The matrix $\boldsymbol{M}_\mathrm{GT}$ is independent of that function too since it is only set by $\boldsymbol{Q}$. It results that building gauge-invariant variables is insensitive to the exact shape of that function, \ie it is done irrespectively of the amplitude of the translation along the gauge direction. Since our expression of gauge transformation differs from the one in \cite{Dom_nech_2018} by $\widetilde{\delta N}\boldsymbol{e}^4$, this difference is harmless to the construction of gauge-invariant variables.

In Sec. \ref{sssec:MS GI second order} where we constructed the gauge-invariant variable $\mathcal{Q}_\mathrm{flat}$, we found an expression of $\boldsymbol{Q}$ and $\boldsymbol{M}$ for which the above requirements are met but working in the natural basis of cosmological perturbations instead, \ie $(\delta\phi,\delta\gamma_1,\delta\pi_\phi,\delta\pi_1)$. The case of $\boldsymbol{Q}$ was already well-known as it directly follows from the construction of gauge-invariant variables at the linear order. For the matrix $\boldsymbol{M}$, we derived it as follows. Its ten independent entries are organized into a 10-dimensional vector, that we named $\vec{M}$, and which is subject to four constrains. It can be recast as a linear algebraic system reading $C\vec{M}=\vec{S}$. Here, $\vec{S}$ is a non-vanishing right-hand-side which is 4-dimensional and given by the four components of the vector $\boldsymbol{\Omega C}^{(2)}\boldsymbol{Q}$ in the natural basis. Then, $C$ is a $4\times10$ matrix which is built by rewriting $\boldsymbol{M}_{ab}\left[\boldsymbol{\Omega C}^{(1)}\right]_b$ as $C\vec{M}$. Hence the entries of $C$ are solely made of the components of the vector $(\boldsymbol{\Omega C}^{(1)})$ in the natural basis. As a result, the kernel of $C$ is 4-dimensional and it provides us with the basis for $\boldsymbol{M}_\mathrm{GT}$, while its null space is 6-dimensional and provides us with the basis of $\boldsymbol{M}_\mathrm{GI}$.\footnote{The set of vectors $\{v_\alpha\}_{\alpha\in\{1,\cdots,6\}}$ given from Eq. (\ref{eq:v1}) to Eq. (\ref{eq:v6}) are no more than the set $\left\{\boldsymbol{e}^A\otimes\boldsymbol{e}^B\right\}_{A,\,B\in\{1,2,3\}}$ expressed in the natural basis, \ie $v_\alpha=\boldsymbol{\delta z}^\mathrm{T}\boldsymbol{M}^\alpha_\mathrm{GI}\boldsymbol{\delta z}$ where $\alpha$ labels the basis $\boldsymbol{e}^A\otimes\boldsymbol{e}^B$ with $A,\,B\in\{1,2,3\}$.} Finally, we exploited the freedom in $\boldsymbol{M}_\mathrm{GI}$ to ensure that $\boldsymbol{\delta z}^{(1)\,\mathrm{T}}\boldsymbol{M}\boldsymbol{\delta z}^{(1)}=0$ in the spatially-flat gauge where $\delta\gamma_1=0$. It results that $\mathcal{Q}_\mathrm{flat}^{(2)}=\delta\phi^{(2)}$ in that gauge with no quadratic contributions from $(\delta\phi)^2$, $(\delta\pi_\phi)^2$, and $(\delta\phi\delta\pi_\phi)$, as well as $(\delta\pi_1)^2$, $(\delta\phi\delta\pi_1)$ and $(\delta\pi_\phi\delta\pi_1)$. This amounts to removing six contribution and we have just the right number of degrees of freedom in $\boldsymbol{M}_\mathrm{GI}$ to do so. However in practice, we found that $\boldsymbol{M}_\mathrm{GT}$ brings contributions from $(\delta\pi_\phi)^2$ and $(\delta\pi_1)^2$ only. Hence, we just need to use two degrees of freedom in $\boldsymbol{M}_\mathrm{GI}$ to remove such unwanted contribution.

\subsubsection*{Conjugate momentum}\label{sssec:conjugate gi}
Having established the procedure to build any gauge-invariant variable $Q^{(2)}_\mathrm{GI}$, we now turn to the construction of its conjugated momentum. We define it as
\bea   
    P^{(2)}_\mathrm{GI}=\boldsymbol{P}\cdot\boldsymbol{\delta z}^{(2)}+\frac{1}{2}\boldsymbol{\delta z}^{(1)\,\mathrm{T}}\boldsymbol{N}\boldsymbol{\delta z}^{(1)}, \label{eq:GI2P}
\eea
where the vector $\boldsymbol{P}$ and the matrix $\boldsymbol{N}$ needs to satisfy the same conditions as $\boldsymbol{Q}$ and $\boldsymbol{M}$ for the variable to be gauge-invariant. Then, they have to be further constrained to ensure that $\left\{Q^{(2)}_\mathrm{GI},P^{(2)}_\mathrm{GI}\right\}=1$. An explicit caluclation of the Poisson bracket leads to
\bea
    \left\{Q^{(2)}_\mathrm{GI},P^{(2)}_\mathrm{GI}\right\}&=&\,\boldsymbol{Q}^\mathrm{T}\boldsymbol{\Omega P}+\boldsymbol{\delta z}^{(1)\,\mathrm{T}}\left(\boldsymbol{M\Omega}\boldsymbol{P}-\boldsymbol{N\Omega}\boldsymbol{Q}\right)+\cdots,
\eea
where higher orders are not needed since they would contribute to the dynamics at the quartic order at least. The right-hand-side has to equal one independently of the perturbative degrees of freedom $\boldsymbol{\delta z}^{(1)}$. It imposes $\boldsymbol{Q}^\mathrm{T}\boldsymbol{\Omega P}=1$ and $\boldsymbol{M\Omega}\boldsymbol{P}-\boldsymbol{N\Omega}\boldsymbol{Q}=\boldsymbol{0}$. Let us stress that the second requirement yields four conditions since it is an equality on phase-space vectors. 

The first condition is identical to the one at first order. It results that the vectors $\boldsymbol{Q}$ and $\boldsymbol{P}$ read
\bea
    \boldsymbol{Q}&=&\,Q_2\,\boldsymbol{e}^2+\ds\sum_{s=0}^1Q_{2s+1}\,\boldsymbol{e}^{2s+1}, \\
    \boldsymbol{P}&=&\,P_2\,\boldsymbol{e}^2+\frac{1}{Q_1^2+Q_3^2}\ds\sum_{s=0}^1Q_{2s+1}\,\left(\boldsymbol{\Omega e}^{2s+1}\right),
\eea
with $Q_1^2+Q_3^2\neq0$ to capture physical degrees of freedom. We note that $Q_2$ and $P_2$ are arbitrary since they have vanishing contributions to the Poisson bracket. 

For the second condition, we first split the matrices into their gauge-invariant part and their non-gauge-invariant part. First, we note that the vectors $\boldsymbol{M}_\mathrm{GI}\boldsymbol{\Omega P}$ and $\boldsymbol{N}_\mathrm{GI}\boldsymbol{\Omega Q}$ have zero projection on the gauge direction $\boldsymbol{e}^4$ since the gauge-invariant part of the matrices $\boldsymbol{M}$ and $\boldsymbol{N}$ are orthogonal to that direction by construction [see Eq. (\ref{eq:MGI})]. Given that $\boldsymbol{\Omega e}^2=-\boldsymbol{e}^4$ which is in the null space of $\boldsymbol{M}_\mathrm{GI}$ and $\boldsymbol{N}_\mathrm{GI}$, one further arrives at 
\bea
    \boldsymbol{M}_\mathrm{GI}\boldsymbol{\Omega P}=\boldsymbol{M}_\mathrm{GI}\boldsymbol{\Omega P}^\mathcal{P} &\,\quad\mathrm{and}\quad &\,\boldsymbol{N}_\mathrm{GI}\boldsymbol{\Omega Q}=\boldsymbol{N}_\mathrm{GI}\boldsymbol{\Omega Q}^\mathcal{P},
\eea
where $\boldsymbol{Q}^\mathcal{P}\equiv \boldsymbol{Q}-Q_2\boldsymbol{e}^2$ is the projection of $\boldsymbol{Q}$ on the plane of physical degrees of freedom, and similarly for $\boldsymbol{P}$. (Note that in such a plane, one easily shows that $\boldsymbol{\Omega P}^\mathcal{P}=-\left|\boldsymbol{Q}^\mathcal{P}\right|^{-2}\boldsymbol{Q}^\mathcal{P}$.) It results $\boldsymbol{M}_\mathrm{GI}\boldsymbol{\Omega P}$ and $\boldsymbol{N}_\mathrm{GI}\boldsymbol{\Omega Q}$ are also independent of $Q_2$ and $P_2$. Let us now consider the part of the matrices which is gauge-transformed. From Eq. (\ref{eq:Mcond}), one can view ${M}_{A4}$ as the components of the column vector $\lambda^{-1}\left[\boldsymbol{C}^{(2)}\boldsymbol{\Omega Q}\right]$ in the basis $\left\{\boldsymbol{e}^A\right\}_{A\in\{1,2,3,4\}}$, and similarly for $N_{A4}$ replacing $\boldsymbol{Q}$ by $\boldsymbol{P}$. The peculiar structure of the matrices $\boldsymbol{M}_\mathrm{GT}$ and $\boldsymbol{N}_\mathrm{GT}$, Eq. (\ref{eq:MGT}), then gives 
\bea
    \boldsymbol{M}_\mathrm{GT}\boldsymbol{\Omega P}&=&\,\frac{-1}{\lambda}\left(\boldsymbol{Q}^\mathrm{T}\boldsymbol{\Omega C}^{(2)}\boldsymbol{\Omega P}\right)\,\boldsymbol{e}^4-\frac{P_2}{\lambda}\sum_{A=1}^3\left[\boldsymbol{C}^{(2)}\boldsymbol{\Omega Q}\right]_A\,\boldsymbol{e}^A, \\
    \boldsymbol{N}_\mathrm{GT}\boldsymbol{\Omega P}&=&\,\frac{-1}{\lambda}\left(\boldsymbol{P}^\mathrm{T}\boldsymbol{\Omega C}^{(2)}\boldsymbol{\Omega Q}\right)\,\boldsymbol{e}^4-\frac{Q_2}{\lambda}\sum_{A=1}^3\left[\boldsymbol{C}^{(2)}\boldsymbol{\Omega P}\right]_A\,\boldsymbol{e}^A.
\eea
The first term in the right-hand-side of the above two equations are equal since $\boldsymbol{C}^{(2)}$ is symmetric and $\boldsymbol{\Omega}^\mathrm{T}=-\boldsymbol{\Omega}$, thus cancelling each other in $\boldsymbol{M\Omega}\boldsymbol{P}-\boldsymbol{N\Omega}\boldsymbol{Q}=\boldsymbol{0}$. The remaining terms are vectors with vanishing projection along the gauge direction $\boldsymbol{e}^4$. Hence, they lie in the same subspace as $\boldsymbol{M}_\mathrm{GI}\boldsymbol{\Omega P}$ and $\boldsymbol{N}_\mathrm{GI}\boldsymbol{\Omega Q}$. Collecting everything and demanding $\boldsymbol{M\Omega}\boldsymbol{P}-\boldsymbol{N\Omega}\boldsymbol{Q}=\boldsymbol{0}$ boils down to three constraints reading
\bea
    \left[\left(\boldsymbol{M}_\mathrm{GI}-\frac{Q_2}{\lambda}\boldsymbol{C}^{(2)}\right)\boldsymbol{\Omega P}\right]_A&=&\,\left[\left(\boldsymbol{N}_\mathrm{GI}-\frac{P_2}{\lambda}\boldsymbol{C}^{(2)}\right)\boldsymbol{\Omega Q}\right]_A \, , \label{eq:NGIvsMGI}
\eea
for $A=1,\,2$ and 3. They come from the projection of $\boldsymbol{M\Omega}\boldsymbol{P}-\boldsymbol{N\Omega}\boldsymbol{Q}=\boldsymbol{0}$ in the plane of physical degrees of freedom and in the constraint directions, while the projection of it on the gauge direction automatically vanishes. Once $\boldsymbol{Q},\,\boldsymbol{P}$ and $\boldsymbol{M}_\mathrm{GI}$ has been done, one can easily tune $\boldsymbol{N}_\mathrm{GI}$ for the three above constraints to be fulfilled since this matrix has six independent degrees of freedom. A concrete example of this is when $Q_2=0=P_2$ which gives $\boldsymbol{N}_\mathrm{GI}=\left|\boldsymbol{Q}^\mathcal{P}\right|^{-2}\boldsymbol{M}_\mathrm{GI}\boldsymbol{\Omega}$ as a sufficient solution. 

This finalizes to prove that one can systematically built a gauge-invariant momentum conjugated to any gauge-invariant variable at second order. In this construction, the freedom that is offered by the gauge-invariant subspace of matrices plays a key role to obtain a Poisson bracket equal to one up to second order. In practice, a gauge-invariant variable is defined through $\boldsymbol{Q}$ and $\boldsymbol{M}_\mathrm{GI}$, and we remind that $\boldsymbol{M}_\mathrm{GT}$ is fixed by $\boldsymbol{Q}$. Then, its conjugate momentum is defined by $\boldsymbol{P}$ which is partly fixed by $\boldsymbol{Q}$ and also fixes $\boldsymbol{N}_\mathrm{GT}$, and by $\boldsymbol{N}_\mathrm{GI}$ which is partly fixed by $\boldsymbol{Q}$, $\boldsymbol{P}$ and $\boldsymbol{M}_\mathrm{GI}$. 

We note that in Sec. \ref{sssec:MS GI second order}, the matrix $\boldsymbol{N}_\mathrm{GI}$ is solely designed to ensure that $\mathcal{P}_\mathrm{flat}^{(2)}=\delta\pi_\phi^{(2)}$ in the spatially-flat gauge. Yet, it was remarkable that it resulted in a matrix which also satisfies the condition given in Eq. (\ref{eq:NGIvsMGI}). From a naive counting of degrees of freedom, imposing $\mathcal{P}_\mathrm{flat}^{(2)}=\delta\pi_\phi^{(2)}$ in the spatially-flat gauge yields at most 6 conditions on $\boldsymbol{N}_\mathrm{GI}$ which are its entries corresponding to $(\delta\phi)^2$, $(\delta\pi_\phi)^2$, and $(\delta\phi\delta\pi_\phi)$, as well as $(\delta\pi_1)^2$, $(\delta\phi\delta\pi_1)$ and $(\delta\pi_\phi,\delta\pi_1)$. Given that $\boldsymbol{N}_\mathrm{GI}$ has 6 independent degrees of freedom, one might run out of freedom to further constrain this matrix to satisfy Eq. (\ref{eq:NGIvsMGI}) which brings three additional conditions. However, $\boldsymbol{N}_\mathrm{GT}$ yields quadratic contributions from $(\delta\phi)^2$ and $(\delta\pi_1)^2$ only. Hence, one just needs to use two degrees of freedom in $\boldsymbol{N}_\mathrm{GI}$ to derive $\mathcal{P}_\mathrm{flat}^{(2)}$ which leaves four of them to meet Eq. (\ref{eq:NGIvsMGI}). This explains why a momentum conjugated to $\mathcal{Q}_\mathrm{flat}$ and satisfying $\mathcal{P}_\mathrm{flat}^{(2)}=\delta\pi_\phi^{(2)}$ in the spatially-flat gauge can be found.

Finally, let us stress that once a canonical pair of gauge-invariant variables has been given, it is possible to build any other ones. For instance, suppose that we start from $(\mathcal{Q}_\mathrm{flat},\mathcal{P}_\mathrm{flat})$ as obtained in Sec. \ref{sssec:MS GI second order} and let us introduce 
\bea
    \left(\begin{array}{c}
        Q_\mathrm{GI} \\
        P_\mathrm{GI}
    \end{array}\right)
    =\boldsymbol{S}\left(\begin{array}{c}
    \mathcal{Q}_\mathrm{flat} \\
    \mathcal{P}_\mathrm{flat}\end{array}\right)
\eea
where $\boldsymbol{S}$ is a $2\times2$ matrix potentially time-dependent  (either explicitly or implicitly via functions of the background variables) and belonging to the symplectic group $\mathrm{Sp}(2,\mathbb{R})$. This is a canonical transformation (thus $\left\{Q_\mathrm{GI},P_\mathrm{GI}\right\}=1$) and the new variables are gauge-invariant too. It amounts to picking up the vectors $\boldsymbol{S}_{11}\boldsymbol{Q}+\boldsymbol{S}_{12}\boldsymbol{P}$ and $\boldsymbol{S}_{21}\boldsymbol{Q}+\boldsymbol{S}_{22}\boldsymbol{P}$ (and similarly for $\boldsymbol{M}$ and $\boldsymbol{N}$) to define the new configuration and the new momentum from $\boldsymbol{\delta z}$. This new pair of canonical and gauge-invariant variables can be further extended adding $\boldsymbol{\delta z}^\mathrm{T}\boldsymbol{\mathcal{M}}_\mathrm{GI}\boldsymbol{\delta z}$ and $\boldsymbol{\delta z}^\mathrm{T}\boldsymbol{\mathcal{N}}_\mathrm{GI}\boldsymbol{\delta z}$ to $Q_\mathrm{GI}$ and $P_\mathrm{GI}$ respectively. These new variables remain gauge-invariant and they form a canonical pair providing that Eq. (\ref{eq:NGIvsMGI}) holds. It results that the whole set of gauge-invariant variables at large scales is generated from the knowledge of $(\mathcal{Q}_\mathrm{flat},\mathcal{P}_\mathrm{flat})$ and of the basis vectors of $\boldsymbol{M}_\mathrm{GI}$, which are all given in the natural basis from Eq. (\ref{eq:v1}) to Eq. (\ref{eq:v6}).

\subsection{Extension to multifield and third order}
\subsubsection*{Multiple fields}
The above construction can be extended to theories with multiple field. Using covariant variables for the fluctuations in the scalar fields, the background and perturbative phase-space are
\bea
    \boldsymbol{z}^\mathrm{T}&=&\,\left(\phi^1,\cdots,\phi^{\N},\gamma_1,\pi_1,\cdots,\pi_\N,\pi_1\right) \\
    \boldsymbol{\delta z}^\mathrm{T}&=&\,\left(Q^1,\cdots,Q^\N,\delta\gamma_1,P_1,\cdots,P_\N,\delta\pi_1\right).
\eea
The full phase-space is now $(2\N+2)$-dimensional. However, there is still one constraint and one gauge degree of freedom. It means that by adding new scalar fields we are simply extending the space of physical degrees of freedom and we write the physical basis as
\bea
    \left\{\boldsymbol{e}^A\right\}_{A\in\{1,\cdots,2\N+2\}}\equiv\left\{\boldsymbol{e}^\mathcal{P}_I,\boldsymbol{e}^\mathcal{C},\bar{\boldsymbol{e}}^\mathcal{P}_I,\bar{\boldsymbol{e}}^\mathcal{C}\right\},
\eea
where $I$ runs from $1$ to $\N$. Each basis vector is $(2\N+2)$-dimensional. The vectors $\boldsymbol{e}^{\N+1}$ and $\boldsymbol{e}^{2\N+2}$ are canonically conjugated and they provide us with the subspace of unphysical degrees of freedom since they generate the constraint and its associated gauge degree of freedom respectively. Working in the field basis, the subspace generating physical degrees of freedom can be constructed following the Gran-Schmidt algorithm. Alternatively, one can project the covariant scalar-fields fluctuations on the adiabatic/entropic basis. In practice, the adiabatic mode is defined as $Q_\sigma\equiv\pi_IQ^I/(\sqrt{G^{IJ}\pi_I\pi_J}$ and similarly for $P_\sigma$. These are not gauge-invariant and they behave as the scalar field fluctuations in the single-field context, replacing derivatives of background functions with respect to $\phi$ by covariant derivatives with respect to $\sigma$. This defines a first pair of basis vector in the physical subspace $(\boldsymbol{e}^\mathcal{P}_\sigma,\bar{\boldsymbol{e}}^\mathcal{P}_\sigma)$. Then, entropic modes are dealt with by using a Frenet basis \cite{Kaiser_2013,Achucarro:2018ngj,Pinol:2020kvw}. In this construction, the configuration of the first entropic mode, $Q_{s=1}$, is automatically gauge invariant hence it defines $\boldsymbol{e}^\mathcal{P}_{s=1}$ by projecting out the component of $Q_{s=1}$. However, its associated momentum is not gauge-invariant \cite{Grain:2020wro} and the vector $\bar{\boldsymbol{e}}^\mathcal{P}_{s=1}$ is obtained by deprojecting from $P_{s=1}$ its components on the constraint and on the gauge degrees of freedom. Finally, all the remaining entropic modes $(Q_{s\geq2},P_{s\geq2})$ are gauge-invariant and they can be directly used to complete the basis of the physical degrees of freedom.

Having established the physical basis, the whole process to construct gauge-invariant follows exactly as the single-field case. It is easy to show that gauge transformations at the linear order are still given by translations along the gauge vector which is now $\boldsymbol{e}^{2\N+2}$. At that order, gauge-invariant variables are $\boldsymbol{Q}_a\boldsymbol{\delta z}_a$ where $\boldsymbol{Q}$ is a $(2\N+2)$-dimensional vector constrained to satisfy $Q_{2\N+2}=0$. Similarly, gauge-invariant variables at second order are defined by Eq. (\ref{eq:GIzgen}) where $\boldsymbol{M}$ is $(2\N+2)\times(2\N+2)$-symmetric matrix. In the physical basis, this matrix is constrained to satisfy $\lambda\boldsymbol{M}_{(2\N+2)\,B}=\sum_{A=1}^{2\N+1}\left[\boldsymbol{\Omega C}^{(2)}\right]_{AB}$. Here again, the matrix can be expressed as the sum of a gauge-invariant part which has $(2\N+1)(2\N+2)/2$ independent degrees of freedom, and a part which is not gauge-invariant and has $(2\N+2)$ independent entries that are subsequently fixed requiring gauge-invariance. Considering two fields for instance, the gauge-invariant part of $\boldsymbol{M}$ has 15 degrees of freedom and the gauge-transformed one has 6 degrees of freedom to fix, in agreement with the explicit calculations done in App. \ref{app:two field}.

\subsubsection*{Third order}
FInally, let us show how such an approach can be implemented at higher orders. For instance, it is shown in App. \ref{app:GTHam} that gauge-transformations at the cubic order reads
\bea
    \widetilde{\boldsymbol{\delta z}}_a^{(3)}-{\boldsymbol{\delta z}}_a^{(3)}&=&\,F(\xi^0)\boldsymbol{K}^{(1)}_a+\boldsymbol{K}^{(2)}_{ab}\left[N\xi^0_{(1)}\boldsymbol{\delta z}^{(2)}_b+G(\xi^0)\left(\boldsymbol{\delta z}_b^{(1)}+N\xi^0_{(1)}\boldsymbol{K}^{(1)}_b\right)\right] \nonumber \\
    &\,&\,+\frac{1}{2}\left(N\xi^0_{(1)}\right)^2\left(\boldsymbol{K}^{(2)}_{ab}\boldsymbol{K}^{(2)}_{bc}+\boldsymbol{K}^{(3)}_{abc}\boldsymbol{K}^{(1)}_b\right)\left(\boldsymbol{\delta z}_c^{(1)}+\frac{1}{3}N\xi^0_{(1)}\boldsymbol{K}^{(1)}_c\right) \label{eq:GT3} \\
    &\,&\,+\frac{1}{2}N\xi^0_{(1)}\boldsymbol{K}^{(3)}_{abc}\boldsymbol{\delta z}_b^{(1)}\boldsymbol{\delta z}_c^{(1)} \nonumber 
\eea
where we introduce $\boldsymbol{K}^{(1)}_a\equiv\left[\boldsymbol{\Omega C}^{(1)}\right]_a$, $\boldsymbol{K}^{(2)}_{ab}\equiv\left[\boldsymbol{\Omega C}^{(2)}\right]_{ab}$, and $\boldsymbol{K}^{(3)}_{abc}\equiv\boldsymbol{\Omega}_{ab}\partial_a\partial_b\partial_c\mathcal{C}^{(0)}$ to lighten the expression\footnote{Since the expression are already quite heavy we do not go into extensive detail here, but in general non linear sigma models this should read $\boldsymbol{K}^{(3)}_{abc}\equiv\boldsymbol{\Omega}_{ab}D_aD_bD_c\mathcal{C}^{(0)}$, with covariant derivatives when they are applicable. If we consider multifield models with flat field space metric this expression is perfectly valid. In any case, we are focusing in the end on the single field case, so taking $\N=1$ in this whole discussion, in which case this difference is irrelevant. \label{fn:third order constraint multifield}}. The function $F$ and $G$ are cubic and quadratic in the gauge parameters and perturbations of the lapse function, respectively. They read
\bea
    F(\xi^0)&=&\,\xi^0_{(3)}N+\xi^0_{(2)}\left[\delta N^{(1)}+\frac{1}{2}\partial_\tau\left(N\xi^0_{(1)}\right)\right]+\xi^0_{(1)}\left[\delta N^{(2)}+\frac{1}{2}\partial_\tau\left(N\xi^0_{(2)}\right)\right], \\
    G(\xi^0)&=&\,\xi^0_{(2)}N+\xi^0_{(1)}\left[\delta N^{(1)}+\frac{1}{2}\partial_\tau\left(N\xi^0_{(1)}\right)\right].
\eea
We note that with such shorthand notations, gauge transformations at first and second order are $\widetilde{\boldsymbol{\delta z}}^{(1)}_a=\boldsymbol{\delta z}^{(1)}_a+N\xi^0_{(1)}\boldsymbol{K}^{(1)}$ and $\widetilde{\boldsymbol{\delta z}}^{(2)}_a=\boldsymbol{\delta z}^{(2)}_a+G(\xi^0)\boldsymbol{K}^{(1)}+N\xi^0_{(1)}\boldsymbol{K}^{(2)}_{ab}\left(\boldsymbol{\delta z}^{(1)}_b+\frac{1}{2}N\xi^0_{(1)}\boldsymbol{K}^{(1)}\right)$.

To build gauge-invariant variables at third order, we introduce
\bea
    Q_\mathrm{GI}^{(3)}=\boldsymbol{Q}_a\boldsymbol{\delta z}^{(3)}_a+\boldsymbol{M}_{ab}\boldsymbol{\delta z}_a^{(2)}\boldsymbol{\delta z}^{(1)}_b+\frac{1}{6}\boldsymbol{T}_{abc}\boldsymbol{\delta z}^{(1)}_a\boldsymbol{\delta z}^{(1)}_b\boldsymbol{\delta z}^{(1)}_c,
\eea
where $\boldsymbol{T}$ is totally symmetric. Gauge-transforming the above and imposing the variables to be gauge invariant yields the following constraints
\bea
   &\,&\,\boldsymbol{Q}_a\boldsymbol{K}^{(1)}_a=0, \\
   &\,&\,\boldsymbol{M}_{ab}\boldsymbol{K}^{(1)}_b+\boldsymbol{K}^{(2)}_{ab}\boldsymbol{Q}_b=0, \\
   &\,&\,\boldsymbol{T}_{abc}\boldsymbol{K}^{(1)}_a\boldsymbol{K}^{(1)}_b+\boldsymbol{M}_{ab}\boldsymbol{K}^{(2)}_{bc}\boldsymbol{K}^{(1)}_a+\frac{1}{2}\boldsymbol{M}_{cb}\boldsymbol{K}^{(2)}_{ba}\boldsymbol{K}^{(1)}_a \\ \nonumber
   &\,&\,\quad\quad\quad\quad\quad\quad\quad\quad\quad\quad\quad\quad\quad\quad+\frac{1}{2}\left(\boldsymbol{K}^{(2)}_{ab}\boldsymbol{K}^{(2)}_{bc}+\boldsymbol{K}^{(3)}_{abc}\boldsymbol{K}^{(1)}_b\right)\boldsymbol{Q}_a=0 \label{eq:T1}\\
   &\,&\,\boldsymbol{T}_{abc}\boldsymbol{K}^{(1)}_a+\left(\boldsymbol{M}_{ab}\boldsymbol{K}^{(2)}_{ac}+\boldsymbol{M}_{ac}\boldsymbol{K}^{(2)}_{ab}\right)+\boldsymbol{K}^{(3)}_{abc}\boldsymbol{Q}_a=0 \label{eq:T2}
\eea
The first one comes from cancelling $F(\xi^0)\boldsymbol{\mathcal{C}}^{(1)}$ in Eq. (\ref{eq:GT3}), and it fixes the vector $\boldsymbol{Q}_a$. The second is needed to cancel the second term in the first line of Eq. (\ref{eq:GT3}). It is an equality between two vectors and given $\boldsymbol{Q}_a$, it sets $2\N+2$ elements of $\boldsymbol{M}_{ab}$ among its $(2\N+2)(2\N+3)/2$ independent degrees of freedom. These two conditions are identical to the ones needed at second order. Working in the natural basis, one recovers $Q_{2\N+2}=0$ and $\boldsymbol{M}_{A\,(2\N+2)}=-\lambda^{-1}\sum_{B=1}^{2\N+1}\boldsymbol{K}^{(2)}_{AB}\boldsymbol{Q}_B$ for $A\in\{1,\cdots,2\N+2\}$. The third constraints is needed to cancel the second line in Eq. (\ref{eq:GT3}) which is linear in $\boldsymbol{\delta z}^{(1)}$. It is also an equality between vectors and it sets $2\N+2$ elements of $\boldsymbol{T}$ knowing $\boldsymbol{M}$ and $\boldsymbol{Q}$. In the physical basis, one easily sees that $\boldsymbol{T}_{abc}\boldsymbol{K}^{(1)}_a\boldsymbol{K}^{(1)}_b=\lambda^2\boldsymbol{T}_{A\,(2\N+2)\,(2\N+2)}$. The last conditions is required to cancel the last line Eq. (\ref{eq:GT3}) which is quadratic in $\boldsymbol{\delta z}^{(1)}$. It is an equality between symmetric matrices and it fixes $(2\N+2)(2\N+3)/2$ entries of $\boldsymbol{T}$ (note that in the physical basis, one has $\boldsymbol{T}_{abc}\boldsymbol{K}^{(1)}_a=\lambda \boldsymbol{T}_{(2\N+2)\,B\,C}$) with potential redundancies with the previous one. The tensor $\boldsymbol{T}$ being totally symmetric, it has $(2\N+2)(2\N+3)(2\N+4)/6$ independent degrees of freedom \cite{axler2023linear}, which leaves enough freedom to fulfill the last two conditions. Let us consider single-field inflation for example. Then $\N =1$ and $\boldsymbol{T}$ has 20 independent degrees of freedom. The two last conditions in the above each leads to 4 and 10 constraints on it. However in this case, 4 of the 10 constraints coming from Eq. (\ref{eq:T2}) are redundant with the ones from Eq. (\ref{eq:T1}). It results that they fix only 10 degrees of freedom in $\boldsymbol{T}$ among the 20 independent ones. The ten unfixed degrees of freedom stands for the gauge-invariant part of $\boldsymbol{T}$ which reads $\boldsymbol{T}_\mathrm{GI}=\sum_{A,B,C=1}^3T_{ABC}\boldsymbol{e}^A\otimes\boldsymbol{e}^B\otimes\boldsymbol{e}^C$. It has the same number of independent degrees of freedom as a totally symmetric tensor of order 3 in a $(2\N+1)$-dimensional space, hence 10 for $\N=1$.

\section{Conclusion}
\label{sec: conclusion}

In this work we explored the Hamiltonian formalism for cosmological perturbation theory for non-linear sigma models as a proxy for general multifield inflation. We derived the third order Hamiltonian without assuming any gauge, thus extending previous studies of perturbations in several directions. Not only did we consider multiple fields in a covariant way, extending the single field work done in  Refs. \cite{Maldacena:2002vr,Braglia:2024zsl}, we also went up to the cubic Hamiltonian in perturbation theory, thus going further than in Ref. \cite{Grain_2026}, and finally we also conducted this analysis without gauge fixing, completing the analysis done in Refs. \cite{Gong_2011,Butchers:2018hds}. We showed how to track and identify gauge transformations of scalar fields order by order. At the second order, and on super horizon scales, this program led to a set of gauge invariant variables generalizing the multifield Mukhanov-Sasaki variables. The usual Mukhanov-Sasaki variables are corrected by quadratic terms involving both field and gravitational perturbations. We found that we can simultaneously correct the configuration variables and also their conjugate momenta, while keeping their key properties. Not only do the corrected variables drop to the field and momenta perturbations in the flat gauge, but they also remain canonical pairs. 
We then showed how to impose the flat gauge in the gauge independent Hamiltonian we previously
derived.
We found that the large-scale limit of this procedure reproduces the cubic Hamiltonian governing the evolution of the gauge-invariant quantities we just exhibited. Finally, we gave an algebraic proof of the existence of these variables that are gauge invariant at the second order at large scales in the single field case, as well as a proof for the existence of their conjugate momenta. We also hinted at how this proof should extend to higher orders, and to the multifield case.

Several directions naturally follow from this construction. One could naturally ask whether our method can be extended to other gauge invariant quantities and other gauges. As exhibited we can apply the same method to the comoving curvature perturbation and guaranty that the gauge invariant at the second order and at large scale corrected quantity drops to the right form in the comoving gauge. 
Moreover, we have shown that any second order gauge invariant quantity at large scales can be constructed from the ones we have already built. In order to make observational predictions, the gauge independent Hamiltonian should then be gauge fixed to the comoving gauge for simplicity. This can be done by following the same procedure, first imposing the gauge conditions and then ensuring that they hold through the evolution.

The second, and quite immediate application of the formalism we developed is stochastic inflation. In this coarse-grained effective theory, the dynamics of the long wavelength modes is formulated at the leading order (so with a quadratic Hamiltonian), with the noise term being sourced by the linear mode functions evaluated at horizon crossing. Having the third order Hamiltonian means the noise can in principle be extended beyond the Gaussian and Markovian approximation.  
Cubic interactions correct both the amplitude and the correlation structure of the stochastic kicks, which in turn source a non trivial drift. Since we have already defined quantities that are gauge invariant at large scales, they may be the right variables to coarse grain and to which we should apply the stochastic formalism, thus discarding any gauge dependent artifacts that could remain. Deriving the Langevin equation sourced by the Hamiltonian we constructed, and it's consequences for the statistics of the curvature perturbation at the end of inflation is a direct extension of this work.

A third direction concerns loop corrections to the primordial power spectrum and bispectrum. Having the gauge independent cubic Hamiltonian explicitly is precisely the ingredient needed to compute one-loop contributions to the two-point correlation function of the curvature perturbation consistently, and part of those involved in the three-point correlation function. A fourth order Hamiltonian would be necessary to complete this discussion. Since our construction is independent of any gauge choice, the resulting loop integrals are free of gauge effects that have recently plagued loop computations with the in-in formalism, even in single-field inflation \cite{Inomata:2022yte,Kristiano:2022maq}.
Additionally, a Hamiltonian treatment frees us from the dangers of omitting contributing boundary terms and total derivatives \cite{Braglia:2024zsl}, which are indeed present in the cubic Lagrangian \cite{Maldacena:2002vr,Arroja_2011}. 
Using our cubic Hamiltonian to compute the renormalized one-loop power spectrum, in particular in the presence of a non-trivial field space curvature, would provide a natural extension to the recent proof of super-horizon conservation of the curvature perturbation in single-field scenarios~\cite{Braglia:2025qrb,Braglia:2025cee} including beyond scale-invariance~\cite{Braglia:2026fle}, and hint at observational targets if we find enhanced loop effects in multifield scenarios.

Finally, the algebraic structure underlying our proof of existence for the gauge invariant variables suggests a more systematic route via canonical transformations. Rather than constructing gauge invariant variables order by order by hand, we could look for a canonical transform that maps the gauge independent Hamiltonian to the gauge fixed one with the appropriate set of gauge invariant quantities. In this construction, the gauge degrees of freedom would appear as pure gauge canonical pairs and decouple from the physical degrees of freedom. This would provide a clear picture of how our large scale, second order construction fits in the fully non-linear multifield structure of the primordial universe. We leave these directions to future work.

\appendix

\section{Second and third order Hamiltonian}\label{app:full Hamiltonians}
We will present in this appendix the computation leading to the covariant Hamiltonian at any order, and give the results up to the third order. Let us start by writing the action as  
\begin{align}
	S=S^{(0)}+S^{(1)}+\ds\sum_{n=2}^\infty S^{(n)}\,,
\end{align}
and the relation between the naive variables and the covariant ones up to any order 
\begin{align}
	\delta\phi^I&=\ds\sum_{n=1}^\infty{\delta_n\phi}^I(Q^I,P_I)\,, \\
	\delta\pi_I&=\ds\sum_{n=1}^\infty\delta_n \pi_I(Q^I,P_I)\,,
\end{align}
where the expression $\delta_n$ means that the expression (of $\delta\phi^I$ or $\delta\pi_I$) contains added powers of $\left(Q^I,P_J\right)$ to the $n$ only. The tilde means that we are writing out our expression in terms of the covariant quantities. The associated type 2 generating function is $G = -Q^IP_I + F(\delta\phi^I,P_I,t)$
\begin{align}
	F(\delta\phi^I,P_I,t)=\sum_{n=2}F_n(\delta\phi^I,P_I,t)\,,
\end{align}
where $F_n$ contains terms of power $n$ in the covariant variables only. We can now make use of the fact that two actions have the same extremum as long as they differ from a total derivative. In our case we will call $\mathcal H = N\mathcal C + N^i\mathcal D_i$ the Hamiltonian density of the naive variables and $\mathcal K$ that of the new covariant variables.
\begin{align}
	\ds\int\dd^3x\left[-\dot{\pi}_I\delta\phi^I+\dot{\phi}^I\delta\pi_I+\dot{\delta\phi^I}\delta\pi_I-\mathcal{H}\left(\delta\phi^I,\delta\pi_I\right)\right]=&\,\ds\int\dd^3x\left[-\dot{\pi}_IQ^I+\dot{\phi}^IP_I+\dot{Q^I}P_I\right. \nonumber \\
	&\,\left.-\mathcal{K}\left(Q^I,P_I\right)+\frac{\dd G}{\dd t}\right]\,, \label{eq:actionF}
\end{align}
In this expression the background quantities have already been removed since we naturally have $\mathcal H^{(0)} = \mathcal K^{(0)}$. Plugging the replacements and the generating function gives the new Hamiltonian from the second order up as
\begin{align}
	\mathcal{K}^{(n\geq2)}=&\,\sum_{n=1}\left[\delta N\tilde{\mathcal{C}}^{(n)}+\delta N^i\mathcal{D}^{(n)}_{i,\text{cov}}\right]+N\sum_{n=2}\tilde{\mathcal{C}}^{(n)}+\frac{\partial \tilde{F}_n}{\partial t}.
\end{align}
Where we have defined $\forall n \geq 1$
\begin{align}
    \tilde{\mathcal C}^{(n)} =&\, \ds\sum_{m=1}^n \mathcal C^{(m)}(\delta_{n+1-m}\phi^I,\delta_{n+1-m}\pi^I)\,,\\
	\mathcal{D}^{(n)}_{i,\text{cov}} =&\, \ds\sum_{m=1}^n \mathcal D^{(m)}_i(\delta_{n+1-m}\phi^I,\delta_{n+1-m}\pi^I).
\end{align}
Where the contribution of $\mathcal C^{(1)}(\delta_n\phi^I,\delta_n\pi_I)$ will systematically drop in the $N\tilde{\mathcal C}^{(n)}$ by least action principle, but not in the $\delta N\tilde{\mathcal C}^{(n)}$ term. Let us write $\tilde{\mathcal C}^{(n)}_{-1}$ the term without the contribution from the first order. Finally, to go from $\mathcal K$ to $\mathcal H\cov$, the covariant Hamiltonian density, we need to recall that the action reads at the quadratic order as $S = \int \dd \tau\int\dd^3\vec x P_I\dot Q^I - \ds\sum_{n=2}^\infty \mathcal K^{(n)}$ which we can rewrite as
\begin{align}
    S =&\, \int \dd \tau\int\dd^3\vec x P_ID_\tau  Q^I - \ds\sum_{n+2}^\infty \mathcal H\cov^{(n)}\,, \\
    \mathcal H^{(n\geq2)}\cov =&\, \delta N\tilde{\mathcal{C}}^{(n-1)}+\delta N^i\mathcal{D}^{(n-1)}_{i,\text{cov}}+ \underbrace{N\tilde{\mathcal{C}}^{(n)}_{-1}+\frac{\partial \tilde{F}_n}{\partial t}}_{N{\mathcal C^{(n)}\cov}_\text{, full}}.
\end{align}
All three contributions to $\mathcal H^{(n\geq2)}\cov$ are explicitly covariant in the field space, thus their sum also is. Let us render explicit one last simplification. The term ${\mathcal C^{(n)}\cov}_\text{, full}$ may contain total spatial-derivatives coming from $\tilde{\mathcal{C}}^{(n)}_{-1}$ that automatically drop in the Hamiltonian, let us write these terms $\mathcal M^{(n)}$. We define
\begin{equation}
    \mathcal C^{(n)}\cov = {\mathcal C^{(n)}\cov}_\text{, full}-\mathcal M^{(n)}.
\end{equation}
We can now finally write the covariant Hamiltonian up to the third order as
\begin{align}
    H\cov^{(2),(3)}=\ds\int\dd^3x\left[\delta N \left(\tilde{\mathcal{C}}^{(1)}+\tilde{\mathcal{C}}^{(2)}\right)+\delta N^i\left(\mathcal{D}^{(1)}_{i,\text{cov}}+\mathcal{D}^{(2)}_{i,\text{cov}}\right)+N\left(\mathcal{C}\cov^{(2)}+\mathcal{C}\cov^{(3)}\right)\right].
\end{align}
Let us apply the method in our case making use of Eqs. (\ref{eq:phi to Q}, \ref{eq:pi to P} $\&\,$\ref{eq:gen function}). We also cut the contributions to the scalar constraints in a field part and a purely gravitational part. For $n\ge2$ we have
\begin{align}
    \tilde{\mathcal C}^{(n)} =&\, \tilde{\mathcal C}^{(n),\phi}+\tilde{\mathcal C}^{(n),G}\,, \\
    {\mathcal C}^{(n)}\cov =&\, \tilde{\mathcal C}^{(n),\phi}\cov+\tilde{\mathcal C}^{(n),G}\cov \,, \\
    {\mathcal D_{i,\text{cov}}^{(n)}} =&\, {\mathcal D_{i,\text{cov}}^{(n),\phi}}+{\mathcal D_{i,\text{cov}}^{(n)nG}}.
\end{align}
We can apply this method up to any order. The second order Hamiltonian is given in the main text in equations Eqs. (\ref{eq:scalar 1}, \ref{eq:scalar 2 phi}, \ref{eq:scalar 2 g} $\&\,$\ref{eq:diff 1}.) From this Hamiltonian, we can also write the equation of motions for the perturbation variables
\bea
	&\,&\,\left\{\begin{array}{rcl}
	D_\tau  \delta\gamma_1&=&\,\ds\frac{-2}{\sqrt{3}}v^{2/3}k\delta N_1-\frac{\sqrt{3}}{\Mp^2}v^{2/3}\theta\delta N-\frac{N}{\Mp^2}\left(2v^{1/3}\delta\pi_1+\frac{\theta}{2}\delta\gamma_1\right) \label{eq:EoM gamma 1}\\
	D_\tau  \delta\pi_1&=&\,\ds\frac{-v^{1/3}\theta}{2\sqrt{3}}k\delta N_1+\frac{v^{1/3}}{\sqrt{3}}\left[\frac{1}{2}\left(\rho+3p\right)+\Mp^2\frac{k^2}{v^{2/3}}\right]\delta N \\
	&\,&\,\ds-N\left[\frac{1}{6v^{1/3}}\left(5\rho+3p-\Mp^2\frac{k^2}{v^{2/3}}\right)\delta\gamma_1-\frac{\theta}{2\Mp^2}\delta\pi_1+\frac{\Mp^2\sqrt{2}}{12v}k^2\delta\gamma_2\right]  \\
	&\,&\,\ds-\frac{\sqrt{3}}{2}v^{1/3}N\left(V_{;I}Q^I-\frac{G^{IJ}\pi_J}{v^2}P_I\right),
	\end{array}\right. \label{eq:CovDiff1} \\
	&\,&\,\left\{\begin{array}{rcl}
	D_\tau  \delta\gamma_2&=&\,\ds-2\sqrt{\frac{2}{3}}v^{2/3}k\delta N_1+\frac{N}{\Mp^2}\left(4v^{1/3}\delta\pi_2+{\theta}\delta\gamma_2\right) \label{eq:EoM gamma 2}\\
	D_\tau  \delta\pi_2&=&\,\ds\sqrt{\frac{2}{3}}v^{1/3}k\delta N_1-\frac{\Mp^2}{v^{1/3}\sqrt{6}}k^2\delta N \\
	&\,&\,\ds-N\left[\frac{1}{6v^{1/3}}\left(5\rho+3p-\Mp^2\frac{k^2}{2v^{2/3}}\right)\delta\gamma_2+\frac{\theta}{\Mp^2}\delta\pi_2+\frac{\Mp^2\sqrt{2}}{12v}k^2\delta\gamma_1\right], 
	\end{array}\right. \label{eq:CovDiff2} \\
	&\,&\,\left\{\begin{array}{rcl}
	D_\tau  Q^I&=&\,\ds\frac{1}{v}G^{IJ}\pi_J \delta N + N\left(\frac{1}{v}G^{IJ}P_J-\frac{\sqrt{3}}{2v^{5/3}}G^{IJ}\pi_J\delta\gamma_1\right) \\
	D_\tau  P_I&=&\,\ds-\pi_I k\delta N_1-vV_{;I}\delta N \\
	&\,&\,\ds-N\left[v\left(\frac{k^2}{v^{2/3}}+V_{;IJ}-\frac{1}{v^2}R_I{}^{KL}{}_J\pi_K\pi_L\right)Q^J+\frac{\sqrt{3}}{2}v^{1/3}V_{;I}\delta\gamma_1\right].
	 \end{array}\right. \label{eq:CovDiffQP}
\eea
And finally, the third order is given in the core of the text in Eqs. (\ref{eq:diff 2 g}, \ref{eq:diff 2 phi}, \ref{eq:scalar tilde 2 g}, \ref{eq:scalar tilde 2 phi}, \ref{eq:scalar 3 g} $\&\,$\ref{eq:scalar 3 phi}). We however recover here the explicit expression of Eqs. (\ref{eq:scalar tilde 2 g} $\&\,$\ref{eq:scalar tilde 2 phi}).
\begin{align}
    \tilde{\mathcal C}^{(2),G} =&\, \frac 1 {\Mp^2} \left\{ v^{1/3}\left[-\delta\pi_1^2 + 2 \left(D_{mn}\delta\pi_2\right)\left(D^{mn}\delta\pi_2\right)\right] +\frac{\theta}2 \left[-\delta\gamma_1\delta\pi_1 + 2 \left(D_{mn}\delta\gamma_2\right)\left(D^{mn}\delta\pi_2\right)\right] \right.\\
    \nonumber &\,\left. +\frac 1 {32} \frac{\theta^2}{v^{1/3}}\left[\delta\gamma_1^2 + 10\left(D_{mn}\delta\gamma_2\right)\left(D^{mn}\delta\gamma_2\right)\right]\right\}\\
    \nonumber &\, + \frac{\Mp^2}{12v}\left\{-4\partial_m\delta\gamma_1\partial^m\delta\gamma_1 + 2\sqrt{2}\partial_m\delta\gamma_1\partial^m\delta\gamma_2 + 4 \partial^m\delta\gamma_2\partial_m\delta\gamma_2 \right. \\
    \nonumber&\, -2\delta\gamma_1\partial^2\delta\gamma_1  + \sqrt{2}\delta\gamma_1\partial^2\delta\gamma_2 - 2\sqrt{3}\partial_m\partial_n\delta\gamma_1 D^{mn}\delta\gamma_2 \\
    \nonumber&\,+ 4\sqrt{6} \partial_m\partial_n\delta\gamma_2 D^{mn}\delta\gamma_2 - 6 \partial^2D_{mn}\delta\gamma_2 D^{mn}\delta\gamma_2 \\
    \nonumber&\,\left. - \frac{3}{2}\left(\partial^m D^{in} + \partial^i D^{mn} - \partial^n D^{im}\right)\delta\gamma_2 \left( \partial_m D_{in} + \partial_i D_{mn} - \partial_n D_{im}\right) \delta\gamma_2\right\}\,, \\
    \tilde{\mathcal C}^{(2),\phi} =&\, \frac{1}{2v}P_IP^I - \frac{\sqrt{3}}{2}\frac{\pi^I}{v^{5/3}}P_I\delta\gamma_1 + \frac{1}{4v^1/3}\frac{\pi_I\pi^I}{v^2}\left(\frac{5}{4}\delta\gamma_1^2 + \frac{1}{2}\delta\gamma_2^2\right) - \frac{1}{6v}R_I{}^{KL}{}_J\pi_K\pi_LQ^IQ^J \\
    \nonumber &\,+ \frac{v}{2}\left(V_{;IJ}-\frac{\partial_i\partial^i}{v^{2/3}}\right)Q^IQ^J+\frac{\sqrt 3}{2}v^{1/3}V_{;I}Q^I\delta\gamma_1 + \frac{1}{4} \frac{V}{v^{1/3}}\left(\frac{1}{2}\delta\gamma_1^2-\delta\gamma_2^2\right).
\end{align}

\section{Gauge transformations for ADM variables}\label{app:gauge transform}
In this appendix we will present the subtle case of computing the gauge transformation of the induced metric scalar perturbation and their momenta. In order to do so let us introduce a few quantities, notably the extended space time metric, a normal vector and the extended intrinsic curvature
\begin{align}
     n_\mu =&\, \left(-N, 0, 0, 0\right)\,,\quad n^\mu = \left(\frac 1 N, -\frac{N^i}{N}\right)\,, \\
    \gamma_{\mu\nu} =&\, \begin{pmatrix}
        -N_iN^i &\, N_i\\
        N_j &\, \gamma\stsr
    \end{pmatrix}\,,\\
    K_{\mu\nu} =&\, -\frac 1 2 \mathcal{L}_n \gamma_{\mu\nu} = \begin{pmatrix}
        0 &\,  \frac{N_i v^{2/3}\theta}{2\Mp^2} \\
        \frac{N_j v^{2/3}\theta}{2\Mp^2} &\, K\stsr
        \end{pmatrix}\,, \\
    K\stsr =&\, \frac{1}{\sqrt \gamma}\left(\frac 1 2 \gamma_{kl}\pi^{kl}\gamma\stsr  - \gamma_{il}\gamma_{jk}\pi^{kl}\right)\, ,
\end{align}
where $\gamma\stsr$ is the induced metric and $K\stsr$ its extrinsic curvature. We can easily define background quantities and perturbations (up to the second order for $\gamma\tsr$ since we needed the second order gauge transform of $\delta\gamma_1$ and $\delta\gamma_2$, but only up to the first order for $K\tsr$ since we needed the gauge transform of $\delta\pi_1$ and $\delta\pi_2$ up to the first order only in this work)
\begin{align}
    \gamma\tsr =&\, \bar\gamma\tsr + \delta\gamma\tsr + \delta_2\gamma\tsr = \begin{pmatrix}
        0 &\, 0\\
        0 &\, v^{2/3} \delta\stsr
    \end{pmatrix} +\begin{pmatrix}
        0 &\, N_i\\
        N_j &\, \delta\gamma\stsr
    \end{pmatrix} + \begin{pmatrix}
        -N_iN^i &\, 0\\
        0 &\, 0
    \end{pmatrix}\,, \\
    K_{\mu\nu} =&\, \bar K_{\mu\nu} + \delta K_{\mu\nu} = \begin{pmatrix}
        0 &\,  0 \\
        0 &\, \bar K\stsr
        \end{pmatrix} + \begin{pmatrix}
        0 &\,  \frac{N_i v^{2/3}\theta}{2\Mp^2} \\
        \frac{N_j v^{2/3}\theta}{2\Mp^2} &\, \delta K\stsr.
        \end{pmatrix}\,, \\
    \delta K\stsr =&\, \frac{\theta}{8\sqrt 3}\delta\gamma_1\delta\stsr + \frac{1}{2\sqrt 3}v^{1/3}\delta\pi_1\delta\stsr - \frac \theta 4 D\stsr \delta\gamma_2 -v^{1/3} D\stsr \delta\pi_2.
\end{align}
It is now quite direct to apply the method explained in \ref{ssec:gauge trans} and derived from Eqs. (\ref{eq:GT 1 order}$\&\,$\ref{eq:GT 2 order}) to these tensors. We will start by applying to $\gamma\tsr$. We find 
\begin{align}
    \left(\mathcal L_{\xi_{(1)}}\bar\gamma\tsr\right)\stsr =&\, \frac 2 3 \frac{\dot v}{v^{1/3}}\xi_{(1)}^0\delta\stsr + v^{2/3}\delta_{(ik}\partial_{j)}\xi_{(1)}^k\nonumber \\
    =&\, \frac{2 N\sqrt\rho}{\sqrt3 \Mp} \xi_{(1)}^0 \delta\stsr + v^{2/3}\delta_{(ik}\partial_{j)}\xi_{(1)}^k.
\end{align}
Where we have used the equation of motion for $v$ and we define the parenthesis as symmetrising over the indices $i$ and $j$. This allows us to compute $\delta\tilde\gamma\stsr$, and thus $\delta\tilde\gamma_1 $ and $\delta\tilde\gamma_2$, through the relations Eqs. (\ref{eq:gamma and pi 1} $\&\,$\ref{eq: gamma and pi 2}). For the latter, the following equality is useful
\begin{equation}
    D\ustsr\partial_j = \sqrt{\frac 2 3 }\partial^i.
\end{equation}
This finally leads to Eqs. (\ref{eq:GT gamma 1 1} $\&\,$\ref{eq:GT gamma 2 1}). The same procedure is applied to the extrinsic curvature tensor
\begin{equation}
    \left(\mathcal L_{\xi_{(1)}}\bar K\tsr\right)\stsr = \frac 1 4 \partial_\tau (v^{2/3}\theta)\xi_{(1)}^0\delta\stsr + \frac 1 4 v^{2/3}\theta\delta_{k(i}\partial_j)\xi_{(1)}^k.
\end{equation}
Now using Eq. (\ref{eq:GT 1 order} and reinjecting the results we just obtained for $\delta\tilde\gamma_1$ and $\delta\tilde\gamma_2$, we can compute $\delta\tilde\pi_1$ and $\delta\tilde\pi_2$.

We will not detail the mathematics here but this procedure is used to compute $\delta\tilde N$, $\delta\tilde N_i$ and $\tilde P_J$. We can also extend this procedure up to the second order on $\gamma\tsr$ in order to arrive at Eq. (\ref{eq:GT gamma 2}).

\section{Two field gauge invariant quantities}\label{app:two field}
Let us extend the results derived in \ref{sssec:MS GI second order} to a generic two field model. We call $(\phi,\pi_\phi)$ and $(\chi,\pi_\chi)$ our fields and their conjugate momenta. Their covariant perturbations are $(Q^\phi,P_\phi)$ and $(Q^\chi,P_\chi)$. Keeping the same notations as previously for the gravitational sector we can build the two Mukhanov-Sasaki variables and their conjugate momenta
\begin{align}
    \mathcal Q^\phi =&\, Q^\phi - \frac{\Mp G^{\phi \alpha} \pi_\alpha}{2\sqrt{2} v^{5/3} \sqrt\rho}\left(\sqrt 2 \delta\gamma_1 - \delta\gamma_2\right)\,, \\
    \mathcal P_\phi =&\, P_\phi + \frac{\Mp v^{1/3}}{2\sqrt 2\sqrt \rho} V_{;\phi}\left(\sqrt 2 \delta\gamma_1 - \delta\gamma_2\right)\,, \\
    \mathcal Q^\chi =&\, Q^\chi - \frac{\Mp G^{\chi\alpha} \pi_J}{2\sqrt{2} v^{5/3} \sqrt\rho}\left(\sqrt 2 \delta\gamma_1 - \delta\gamma_2\right)\,, \\
    \mathcal P_\chi =&\, P_\chi + \frac{\Mp v^{1/3}}{2\sqrt 2\sqrt \rho} V_{;\chi}\left(\sqrt 2 \delta\gamma_1 - \delta\gamma_2\right).
\end{align}
Where we use $\{\alpha,\beta,\zeta,..\}$ to indicate either $\phi$ or $\chi$. We then proceed as we did for the single field case and compute the second order gauge transform at large scales of these Mukhanov-Sasaki variables. Let us focus on $\chi$ since the second case will be symmetrical in $\phi$ and $\chi$ indices. We get for the configuration and momenta variables
\begin{align}
    \tilde Q^{\phi(2)} =&\, (\xi_{(1)}^0)^2\left(\frac{\dot N}{2v}\pi^\phi -\frac{N^2 \sqrt 3}{2 v \Mp}\sqrt \rho  \pi^\phi - \frac{N^2} 2 G^{\phi\alpha} V_{;\alpha} \right) + \frac{N}{2v}\xi_{(1)}^0(\partial_\tau \xi_{(1)}^0)+\xi_{(1)}^0 D_\tau Q^\phi + \xi_{(2)}^0\frac N v \pi^\phi\,, \\
    \tilde P_\phi ^{(2)}=&\,  -(\xi_{(1)}^0)^2\left(\frac{N^2 \sqrt 3 v}{2\Mp}\sqrt \rho V_{;\phi} + \frac{N^2\pi^\alpha}{2} V_{;\phi\alpha} + \frac{\dot N v}{2}V_{;\phi} + \frac 1 2 R_{\alpha\beta\phi\zeta}\pi^\alpha\pi^\beta\pi^\zeta\right) - \frac{Nv}{2}V_{;\phi}\xi_{(1)}^0\partial_\tau \xi_{(1)}^0  \\
    \nonumber &\,-\xi_{(1)}^0\left[\delta N v V_{;\phi} + \frac{\sqrt 3 Nv^{1/3}}{2}V_{;\phi} \delta\gamma_1 + \left(\frac{N}{2v}R_{\alpha\beta\phi \zeta}\pi^\alpha\pi^\zeta - NvV_{;\phi\beta}\right) Q^\beta \right] -NvV_{;\phi}\xi^0_{(2)}.
\end{align}
Where we have made use of the equations of motion Eq. (\ref{eq:CovDiffQP}). Combining these to equations Eqs. (\ref{eq:GT gamma 1 2 LS} $\&\,$\ref{eq:GT gamma 2 2 LS}) we get for the Mukhanov-Sasaki variables
\begin{align}
    \tilde{\mathcal Q}^{\phi(2)} =&\, \xi_{(1)}^0 N \left(\frac 1 v G^{\phi\alpha}P_\alpha - \frac{2\pi^\phi}{\sqrt 3 v^{5/3}}\delta\gamma_1 + \frac{\pi^\phi}{v^{4/3}\Mp^2 \sqrt{\rho}}\delta\pi_1\right) + \frac{\Mp \pi^\phi}{2\sqrt{2} v^{5/3} \sqrt\rho} D\ustsr(\xi_{(1)}^0\partial_\tau (D\stsr\delta\gamma_2)) \\
    \nonumber &\,+ (\xi_{(1)}^0)^2N^2\left(-vG^{\phi\alpha}V_{,\alpha} + \frac{\pi^\phi(\rho+3p)}{4\sqrt 3 v \Mp \sqrt \rho}\right)\,,\\
    \tilde{\mathcal P}^{(2)}_\phi =&\, -(\xi_{(1)}^0)^2N^2\left(\frac{\sqrt 3 v V_{;\phi}\left(p + \frac{5\rho}{3}\right)}{4\Mp \sqrt \rho} + \frac{\pi^\alpha}{2} V_{;\phi\alpha} + \frac 1 2 R_{\alpha\beta\phi\zeta}\pi^\alpha\pi^\beta\pi^\zeta\right) \\
    \nonumber &\,-\xi_{(1)}^0\left[+ \frac{v^{1/3}}{\sqrt 3} V_{;\phi}\delta\gamma_1  + \frac{v^{2/3}}{\Mp\sqrt \rho}V_{;\phi}\delta\pi_1+ \left(\frac{N}{2v}R_{\alpha\beta\phi \zeta}\pi^\alpha\pi^\zeta - NvV_{;\phi\beta}\right) Q^\beta \right] 
\end{align}
From this we first computed a particular solution for the correction to both configuration and the momentum Mukhanov-Sasaki variables. We also computed a basis of fifteen linearly independent gauge invariant combinations of the second order quantities $\boldsymbol{\delta z}^{(1)\,\mathrm{T}}\boldsymbol{M}_\mathrm{GI}\boldsymbol{\delta z}^{(1)}$, with $\boldsymbol{\delta z} = (Q^\phi,Q^\chi,\delta\gamma_1,P_\phi,P_\chi,\delta\pi_1)$ in this case. We give this basis here
\begin{align}
    b_1 =&\,\delta\gamma_1\delta\pi_1-\frac{2\sqrt{3}v^{1/3}\delta\pi_1^2\sqrt{\rho}}{\Mp(3p+\rho)}-\frac{\Mp\delta\gamma_1^2(3p+\rho)}{8\sqrt{3}v^{1/3}\sqrt{\rho}} \,, \\
    b_2=&\,\delta\pi_1P_\chi +\frac{\sqrt{3}v^{2/3}V_{;\chi}\delta\pi_1^2}{3p+\rho}+\frac{P_\chi ^2(3p+\rho)}{4\sqrt{3}v^{2/3}V_{;\chi}} \,, \\
    b_3 =&\, \delta\gamma_1P_\chi +\frac{\Mp v^{1/3}V_{;\chi}\delta\gamma_1^2}{4\sqrt{\rho}}+\frac{P_\chi ^2\sqrt{\rho}}{\Mp v^{1/3}V_{;\chi}} \,, \\
    b_4 =&\, \delta\pi_1P_\phi +\frac{\sqrt{3}v^{2/3}V_{;\phi}\delta\pi_1^2}{3p+\rho}+\frac{P_\phi ^2(3p+\rho)}{4\sqrt{3}v^{2/3}V_{;\phi}} \,, \\
    b_5 =&\, P_\phi P_\chi -\frac{V_{;\chi}P_\phi ^2}{2V_{;\phi}} -\frac{V_{;\phi}P_\chi ^2}{2V_{;\chi}} \,, \\
    b_6=&\,\delta\gamma_1P_\phi +\frac{\Mp v^{1/3}V_{;\phi}\delta\gamma_1^2}{4\sqrt{\rho}}+\frac{P_\phi ^2\sqrt{\rho}}{\Mp v^{1/3}V_{;\phi}} \,,\\
    b_7 =&\, \delta\pi_1Q^\chi-\frac{\sqrt{3}\delta\pi_1^2\pi^{\chi}}{v^{4/3}(3p+\rho)}-\frac{v^{4/3}{Q^\chi}^2(3p+\rho)}{4\sqrt{3}\pi^{\chi}} \,,\\
    b_8 =&\,  P_\chi Q^\chi+\frac{v^2V_{;\chi}{Q^\chi}^2}{2\pi^{\chi}}+\frac{P_\chi ^2\pi^{\chi}}{2v^2V_{;\chi}} \,, \\
    b_9 =&\, P_\phi Q^\chi +\frac{v^2V_{;\phi}{Q^\chi}^2}{2\pi^{\chi}} +\frac{P_\phi ^2\pi^{\chi}}{2v^2V_{;\phi}} \,, \\
    b_{10} =&\, \delta\gamma_1Q^\chi-\frac{\Mp\delta\gamma_1^2\pi^{\chi}}{4v^{5/3}\sqrt{\rho}}-\frac{v^{5/3}{Q^\chi}^2\sqrt{\rho}}{\Mp\pi^{\chi}} \,, \\
    b_{11}=&\,\delta\pi_1Q^\phi-\frac{\sqrt{3}\delta\pi_1^2\pi^{\phi}}{v^{4/3}(3p+\rho)}-\frac{v^{4/3}{Q^\phi}^2(3p+\rho)}{4\sqrt{3}\pi^{\phi}} \,, \\
    b_{12} =&\,P_\chi Q^\phi+\frac{v^2V_{;\chi}{Q^\phi}^2}{2\pi^{\phi}}+\frac{P_\chi ^2\pi^{\phi}}{2v^2V_{;\chi}} \,, \\
    b_{13} =&\, P_\phi Q^\phi+\frac{v^2V_{;\phi}{Q^\phi}^2}{2\pi^{\phi}}+\frac{P_\phi ^2\pi^{\phi}}{2v^2V_{;\phi}}\, , \\
    b_{14} =&\,\delta\gamma_1Q^\phi-\frac{\Mp \delta\gamma_1^2\pi^{\phi}}{4v^{5/3}\sqrt{\rho}}-\frac{v^{5/3}{Q^\phi}^2\sqrt{\rho}}{\Mp\pi^{\phi}} \,, \\
    b_{15} =&\, Q^\phi Q^\chi -\frac{{Q^\chi}^2\pi^{\phi}}{2\pi^{\chi}} -\frac{{Q^\phi}^2\pi^{\chi}}{2\pi^{\phi}}.
\end{align}
We then find a particular solution, and combine it adequately with linear combinations of the expressions we just gave, to write out our final solution for the two corrected Mukhanov-Sasaki variables and their momenta. They read
 \begin{align}
    \mathcal Q^\phi_\mathrm{flat} =&\,  Q^\phi - \frac{\Mp  \pi^\phi}{2\sqrt{2} v^{5/3} \sqrt\rho}\left(\sqrt{2} \delta\gamma_1 - \delta \gamma_2\right) +\frac{\pi^\phi}{2v^2\rho}\delta\gamma_1\delta\pi_1 + \frac{G^{\phi\alpha}\Mp}{2v^{1/3}\sqrt \rho}\delta\gamma_1P_\alpha \\
    \nonumber &\,+ \left[\frac{G^{\phi\alpha}\Mp^2V_{;\alpha}}{8v^{1/3}\rho} - \frac{\Mp\pi^\phi}{2\sqrt 3 v^{7/3}\sqrt \rho}\left(1+\frac{3(\rho+p)}{8\rho} \right)\right]\delta\gamma_1^2 + \frac{\Mp^2\pi^\phi}{4\sqrt 2 v^{7/3}\rho} D^{mn}(\delta\gamma_1\partial_\tau (D_{mn}\delta\gamma_2))\,,\\
    \mathcal Q^\chi_\mathrm{flat} =&\,  Q^\chi - \frac{\Mp  \pi^\phi}{2\sqrt{2} v^{5/3} \sqrt\rho}\left(\sqrt{2} \delta\gamma_1 - \delta \gamma_2\right) +\frac{\pi^\chi}{2v^2\rho}\delta\gamma_1\delta\pi_1 + \frac{G^{\chi\alpha}\Mp}{2v^{1/3}\sqrt \rho}\delta\gamma_1P_\alpha \\
    \nonumber &\,+ \left[\frac{G^{\chi\alpha}\Mp^2V_{;\alpha}}{8v^{1/3}\rho} - \frac{\Mp\pi^\chi}{2\sqrt 3 v^{7/3}\sqrt \rho}\left(1+\frac{3(\rho+p)}{8\rho} \right)\right]\delta\gamma_1^2 + \frac{\Mp^2\pi^\chi}{4\sqrt 2 v^{7/3}\rho} D^{mn}(\delta\gamma_1\partial_\tau (D_{mn}\delta\gamma_2))\,,\\
    \mathcal P_{\phi\mathrm{flat}} =&\, P_\phi + \frac{\Mp v^{1/3}}{2\sqrt \rho} V_{;\phi}\delta\gamma_1 + \frac{\Mp}{4v^{3/3}\sqrt \rho}\left(\mathcal R_{\alpha\phi} - 2 v^2 V_{;\alpha\phi}\right)\delta\gamma_1Q^\alpha -\frac{V_{;\phi}}{2\rho}\delta\gamma_1\delta\pi_1 \\
    \nonumber &\,+\left[\frac{\Mp V_{;\phi}}{4\sqrt 3 v^{1/3}\sqrt\rho}\left(\frac{\rho+3p}{8\rho}-1\right)-\frac{\Mp^2}{16v^{10/3}\rho}\left(\mathcal R_{IJ} - 2 v^2 V_{;\alpha\phi})\pi^\alpha \right)\right] \delta\gamma_1^2 \\
    \nonumber &\,-\frac{\Mp^2}{4\sqrt 2 v^{1/3}\sqrt \rho} V_{;\phi} D^{mn}(\delta\gamma_1\partial_\tau (D_{mn}\delta\gamma_2))\,, \\
    \mathcal P_{\chi\mathrm{flat}} =&\, P_\chi + \frac{\Mp v^{1/3}}{2\sqrt \rho} V_{;\chi}\delta\gamma_1 + \frac{\Mp}{4v^{3/3}\sqrt \rho}\left(\mathcal R_{\alpha\chi} - 2 v^2 V_{;\alpha\chi}\right)\delta\gamma_1Q^\alpha -\frac{V_{;\chi}}{2\rho}\delta\gamma_1\delta\pi_1 \\
    \nonumber &\,+\left[\frac{\Mp V_{;\chi}}{4\sqrt 3 v^{1/3}\sqrt\rho}\left(\frac{\rho+3p}{8\rho}-1\right)-\frac{\Mp^2}{16v^{10/3}\rho}\left(\mathcal R_{IJ} - 2 v^2 V_{;\alpha\chi})\pi^\alpha \right)\right] \delta\gamma_1^2 \\
    \nonumber &\,-\frac{\Mp^2}{4\sqrt 2 v^{1/3}\sqrt \rho} V_{;\chi} D^{mn}(\delta\gamma_1\partial_\tau (D_{mn}\delta\gamma_2)).
\end{align}
As we did in the single field case, we can check that these variables are indeed canonical conjugates. Finally this allows us to extend our result to any number of fields since the multifield aspect of the theory is rendered explicit via the coupling metric and its Riemann tensor in this two field case. The more general case was not done via this method, we simply extended this result in a covariant manner, and this is what is presented in \ref{ssec:multifield}.

\section{Gauge transformations in the phase-space} \label{app:GTHam}
We consider the following  transformation of the 4-dimensional coordinates system 
\bea
    x^\mu\to\widetilde{x}^\mu=x^\mu+\xi^\mu.
\eea
Following \cite{Bruni_1997,Dom_nech_2018}, phase-space variables describing perturbative degrees of freedom are transformed as $\boldsymbol{\delta z}\to\widetilde{\boldsymbol{\delta z}}=e^{\mathcal{L}_{\xi^\mu}}\boldsymbol{\delta z}$ with $\mathcal{L}_{\xi^\mu}$ the Lie derivative. In order to implement it in the Hamiltonian language, the infinitesimal vector generating the 4-dimensional transformation in the embedding space-time should be expressed in terms of hypersurface deformation \cite{Hojman:1976vp,thieman_book}. In practice, it means that one needs to extract from $\xi^\mu$ the infinitesimal lapse function $\delta N_{\xi^\mu}$ and the infinitesimal shift vector $\delta N^i_{\xi^\mu}$ which deform the point $x^i$ sitting on the hypersuface at $\tau$, to the point $x^i+\xi^i$ sitting on the hypersurface at $\tau+\xi^0$. Then, a gauge transformation is an infinitesimal deformation of the hypersurfaces which can be decomposed into a pure deformation given by $\delta N_{\xi^\mu}$, that is a displacement along the vector normal to the hypersurfaces and whose generator is the scalar constraint, followed by a pure stretching given by $\delta N^i_{\xi^\mu}$, that is a displacement within the hypersurface and whose generator is the diffeomorphism constraint (see \cite{Hojman:1976vp,thieman_book} for details). The infinitesimal lapse corresponds to the proper time of the geodesic connecting the two hypersurfaces and starting along $n^\mu$. It is given by $\delta N_{\xi^\mu}=n_\mu\xi^\mu=N\xi^0$. Performing the pure deformation given by $\delta N_{\xi^\mu}$ from $(\tau,x^i)$ sends to $(\tau+\xi^0,x^i-N^i\xi^0)$. (Note that one does not end at $x^i$ under pure deformation since the geodesic normal to a hypersurface at $x^i$ has no reason to be aligned with constant-$x^i$ curves and this is why one also needs a shift vector.) Hence to arrive at $(\tau+\xi^0,x^i+\xi^i)$, one should perform a pure stretching given by $\delta N^i_{\xi^\mu}=\xi^i+N^i\xi^0$. (Note that these expressions obtained from hypersurface deformation are in agreement with the one given in \cite{Salisbury:1982ez,Pons:1995su,PhysRevD.55.658,Pons:1998ht} for gauge transformations to be properly projected in the phase-space under the Legendre map.)

Let us now consider any function of the phase-space, $F(\phi,\gamma_{ij},\pi_\phi,\pi^{ij})$, that is displaced along $\xi^\mu$. Its infinitesimal variation is given by the Lie derivative which can be decomposed as a pure deformation generated by $(\delta N_{\xi^\mu}\mathcal{C})$ followed by a pure stretching generated by $(\delta N^i_{\xi^\mu}\mathcal{D}_i)$. Hence, Lie derivatives of phase-space variables are recast using Poisson brackets as follows
\bea
    \mathcal{L}_{\xi^\mu}\left(F\right)=\left\{F,\ds\int\dd^3x\left[\xi^0N\mathcal{C}+\left(\xi^i+\xi^0N^i\right)\mathcal{D}_i\right]\right\},
\eea
where it is worth noting the presence of the Lagrange multiplier in the right-hand-side. The above defines gauge transformation at linear order which is then solely expressed using Poisson brackets. However at higher orders, it means that gauge transformations are given by Poisson brackets of phase-space variables {\it and} Lie derivatives of the Lagrange multipliers because of terms like $\mathcal{L}^n_{\xi^\mu}\left(\boldsymbol{\delta z}\right)$ with $n\geq2$.

\subsubsection*{Gauge transformation at second order}
To illustrate this, we consider the simpler case of the separate-universe picture. In this case, anisotropic degrees of freedom and the diffeomorphism constraint all vanish because of isotropy, and all gradients are removed because of homogeneity. This last condition leads to $\xi^i\equiv\partial_i\xi$ to equal zero and $\xi^\mu=(\xi^0,0,0,0)$. At second order in perturbations, phase-space variables are transformed as follows
\bea
    \boldsymbol{\delta z}\to\widetilde{\boldsymbol{\delta z}}=\boldsymbol{\delta z}+\mathcal{L}_{\xi^\mu}\left(\boldsymbol{\delta z}\right)+\frac{1}{2}\mathcal{L}^2_{\xi^\mu}\left(\boldsymbol{\delta z}\right).
\eea
Expanding $(N\mathcal{C})$ to second order, we first arrive at
\bea
    \mathcal{L}_{\xi^\mu}\left(\boldsymbol{\delta z}\right)=\xi^0\left\{\boldsymbol{\delta z},\left(N+\delta N\right)\mathcal{C}^{(1)}+N\mathcal{C}^{(2)}\right\},
\eea
where $N$ now stands for the background lapse and $\delta N$ for its fluctuations. Given that $\mathcal{C}^{(1)}=\boldsymbol{C}^{(1)}_a\boldsymbol{\delta z}_a$ and $\mathcal{C}^{(2)}=(1/2)\boldsymbol{C}^{(2)}_{ab}\boldsymbol{\delta z}_a\boldsymbol{\delta z}_b$, it boils down to
\bea
    \mathcal{L}_{\xi^\mu}\left(\boldsymbol{\delta z}_a\right)=\xi^0\left(N+\delta N\right)\left[\boldsymbol{\Omega}\right]_{ab} \boldsymbol{C}^{(1)}_b+\xi^0N\left[\boldsymbol{\Omega C}^{(2)}\right]_{ab}\boldsymbol{\delta z}_b.
\eea
Injecting this in $\mathcal{L}^2_{\xi^\mu}\left(\boldsymbol{\delta z}\right)$ and restricting to second order, we obtain
\bea
    \mathcal{L}^2_{\xi^\mu}\left(\boldsymbol{\delta z}_a\right)=\xi^0\mathcal{L}_{\xi^\mu}(\delta N)\left[\boldsymbol{\Omega}\right]_{ab} \boldsymbol{C}^{(1)}_b+\left(\xi^0N\right)^2\left\{\left[\boldsymbol{\Omega C}^{(2)}\right]_{ab}\boldsymbol{\delta z}_b,\boldsymbol{C}^{(1)}_c\boldsymbol{\delta z}_c\right\},
\eea
where the first term comes from the Lie derivative of the Lagrange multiplier, and the second from nested Poisson brackets reading $\left\{\left\{\boldsymbol{\delta z}_a,N\mathcal{C}^{(2)}\right\},\mathcal{C}^{(1)}\right\}$ (other nested Poisson brackets vanish or are at higher orders). The Lie derivative of the fluctuations in the lapse function is obtained as follows (see \cite{Artigas:2023kyo} for details). One plugs $\boldsymbol{\delta z}+\mathcal{L}_{\xi^\mu}(\boldsymbol{\delta z})$ into the Hamilton equations at linear order. Because gauge transformations are canonical, it results in new Hamilton equations which are identical to the first ones with gauge-transformed Lagrange multipliers. By identification, it leads to $\mathcal{L}_{\xi^\mu}(\delta N)=\partial_\tau(N\xi^0)$ which is in perfect agreement with a direct calculations of the Lie derivative given in Eq. (\ref{eq:GT lapse 1}). Computing the remaining Poisson brackets then yields
\bea
    \mathcal{L}^2_{\xi^\mu}\left(\boldsymbol{\delta z}_a\right)=\xi^0\partial_\tau(N\xi_{(1)}^0)\left[\boldsymbol{\Omega}\right]_{ab} \boldsymbol{C}^{(1)}_b+\left(\xi^0N\right)^2\left[\boldsymbol{\Omega C}^{(2)}\boldsymbol{\Omega}\right]_{ab}\boldsymbol{C}^{(1)}_b,
\eea
Combining everything, one then finds 
\bea
\widetilde{\boldsymbol{\delta z}}&=&\,\boldsymbol{\delta z}+\left[\xi^0N+\xi^0\delta N+\frac{1}{2}\xi^0\partial_\tau(N\xi^0)\right]\boldsymbol{\Omega C}^{(1)}+\frac{1}{2}\left(\xi^0N\right)^2\boldsymbol{\Omega C}^{(2)}\boldsymbol{\Omega C}^{(1)} \nonumber \\
&\,&\,+\left(\xi^0N\right)\boldsymbol{\Omega C}^{(2)}\boldsymbol{\delta z},
\eea
where we use matrix-matrix and matrix-vector multiplication to lighten the expression. Finally, we expand $\xi^0$, $\boldsymbol{\delta z}$ and $\delta N$ up to second order, \ie $\xi^0=\xi^0_{(1)}+\xi^0_{(2)}$ and similarly for the phase-space  and the fluctuations of the lapse. Injecting in the above expression and identifying the first and second order, we arrive at
\bea    
    \widetilde{\boldsymbol{\delta z}}^{(1)}&=&\,\boldsymbol{\delta z}^{(1)}+\xi^0_{(1)}N\boldsymbol{\Omega C}^{(1)}, \\
    \widetilde{\boldsymbol{\delta z}}^{(2)}&=&\,\boldsymbol{\delta z}^{(2)}+\left[\xi^0_{(2)}N+\xi^0_{(1)}\delta N^{(1)}+\frac{1}{2}\xi^0_{(1)}\partial_\tau\left(N\xi^0_{(1)}\right)\right]\boldsymbol{\Omega C}^{(1)}+\frac{1}{2}\left(\xi^0_{(1)}N\right)^2\boldsymbol{\Omega C}^{(2)}\boldsymbol{\Omega C}^{(1)} \nonumber \\
&\,&\,+\left(\xi^0_{(1)}N\right)\boldsymbol{\Omega C}^{(2)}\boldsymbol{\delta z}^{(1)}.
\eea
It matches exactly the expressions derived from a direct calculation of the Lie derivative of the ADM variables, Eqs. (\ref{eq:GT1vect}) \&\, (\ref{eq:GT2bisvect}).

Let us now briefly comment on the expression that would have been obatined from \cite{Dom_nech_2018}. Here, they implement the Lie derivative of phase-space variables using the following Poisson bracket 
\bea
    \mathcal{L}_{\xi^\mu}\left(F\right)=\left\{F,\xi^0\mathcal{C}+\xi^i\mathcal{D}_i\right\}.
\eea
Hence, there is no contribution from the Lagrange multipliers. It results that gauge transformations are unmodified at the linear order but reads at second order as follows
\bea
\widetilde{\boldsymbol{\delta z}}^{(2)}&=&\,\boldsymbol{\delta z}^{(2)}+\xi^0_{(2)}N\boldsymbol{\Omega C}^{(1)}+\left(\xi^0_{(1)}N\right)^2\boldsymbol{\Omega C}^{(2)}\boldsymbol{\Omega C}^{(1)}+\left(\xi^0_{(1)}N\right)\boldsymbol{\Omega C}^{(2)}\boldsymbol{\delta z}^{(1)}.
\eea
As one could have expected, there is no contribution from fluctuations in the lapse function nor from its Lie derivative.

\subsection*{Third order}
We now extend the above to the third order. In this case, $e^{\mathcal{L}_{\xi^\mu}}\boldsymbol{\delta z}$ should be expanded to the third order which gives
\bea
    \boldsymbol{\delta z}\to\widetilde{\boldsymbol{\delta z}}=\boldsymbol{\delta z}+\mathcal{L}_{\xi^\mu}\left(\boldsymbol{\delta z}\right)+\frac{1}{2}\mathcal{L}^2_{\xi^\mu}\left(\boldsymbol{\delta z}\right)+\frac{1}{6}\mathcal{L}^3_{\xi^\mu}\left(\boldsymbol{\delta z}\right). \label{eq:GT3rdgen}
\eea
Similarly, the scalar constraint should be expanded up to the third order too, \ie
\bea    
N\mathcal{C}\to\left(N+\delta N\right)\boldsymbol{C}^{(1)}_a\boldsymbol{\delta z}_a+\frac{1}{2}\left(N+\delta N\right)\boldsymbol{C}^{(2)}_{ab}\boldsymbol{\delta z}_a\boldsymbol{\delta z}_b+\frac{1}{6}N\boldsymbol{C}^{(3)}_{abc}\boldsymbol{\delta z}_a\boldsymbol{\delta z}_b\boldsymbol{\delta z}_c,
\eea
where $\boldsymbol{C}^{(3)}_{abc}\equiv\partial_a\partial_b\partial_c\mathcal{C}^{(0)}$.\footnote{Where we remind the reader of footnote \ref{fn:third order constraint multifield} in which we explain that this is valid for single field models or multifield models with a flat field space metric.} This last quantity is symmetric under permutations of any two of its indices. Extending the above calculations to the third order yields
\bea    
\mathcal{L}_{\xi^{\mu}}\left(\boldsymbol{\delta z}_a\right)&=&\,\xi^0\left(N+\delta N\right)\left[\boldsymbol{\Omega C}^{(1)}\right]_a+\xi^0\left(N+\delta N\right)\left[\boldsymbol{\Omega C}^{(2)}\right]_{ab}\boldsymbol{\delta z}_b \nonumber \\ 
&\,&\,+\frac{\xi^0}{2}N\boldsymbol{\Omega}_{ab}\boldsymbol{C}^{(3)}_{bcd}\boldsymbol{\delta z}_c\boldsymbol{\delta z}_d, \\
\mathcal{L}^2_{\xi^{\mu}}\left(\boldsymbol{\delta z}_a\right)&=&\,\xi^0\partial_\tau\left(N\xi^0\right)\left[\boldsymbol{\Omega C}^{(1)}\right]_a+\left(\xi^0\right)^2\left(N^2+2N\delta N\right)\left[\boldsymbol{\Omega C}^{(2)}\right]_{ab}\left[\boldsymbol{\Omega C}^{(1)}\right]_a \nonumber \\
&\,&\,+\xi^0\partial_\tau\left(N\xi^0\right)\left[\boldsymbol{\Omega C}^{(2)}\right]_{ab}\boldsymbol{\delta z}_b +\left(\xi^0N\right)^2\left[\boldsymbol{\Omega C}^{(2)}\boldsymbol{\Omega C}^{(2)}\right]_{ab}\boldsymbol{\delta z}_b \nonumber \\
&\,&\,+\left(\xi^0N\right)^2\boldsymbol{\Omega}_{ab}\boldsymbol{C}^{(3)}_{bcd}\left[\boldsymbol{\Omega C}^{(1)}\right]_c\boldsymbol{\delta z}_d, \\
\mathcal{L}^3_{\xi^{\mu}}\left(\boldsymbol{\delta z}_a\right)&=&\,3N\left(\xi^0\right)^2\partial_\tau\left(N\xi^0\right)\left[\boldsymbol{\Omega C}^{(2)}\right]_{ab}\left[\boldsymbol{\Omega C}^{(1)}\right]_b+\left(\xi^0N\right)^3\left[\boldsymbol{\Omega C}^{(2)}\boldsymbol{\Omega C}^{(2)}\right]_{ab}\left[\boldsymbol{\Omega C}^{(1)}\right]_b \nonumber \\
&\,&\,+\left(\xi^0N\right)^3\boldsymbol{\Omega}_{ab}\boldsymbol{C}^{(3)}_{bcd}\left[\boldsymbol{\Omega C}^{(1)}\right]_c\left[\boldsymbol{\Omega C}^{(1)}\right]_d.
\eea
The above equations are injected in Eq. (\ref{eq:GT3rdgen}) and we further expand the gauge parameter, the phase-space variables and the fluctuations in the lapse function up to the third order. It boils down to
\bea
    \widetilde{\boldsymbol{\delta z}}^{(3)}_a&=&\,\boldsymbol{\delta z}^{(3)}_a+\left\{\xi^0_{(3)}N+\xi^0_{(2)}\left[\delta N^{(1)}+\frac{1}{2}\partial_\tau\left(N\xi^0_{(1)}\right)\right]+\xi^0_{(1)}\left[\delta N^{(2)}+\frac{1}{2}\partial_\tau\left(N\xi^0_{(2)}\right)\right]\right\}\left[\boldsymbol{\Omega C}^{(1)}\right]_a \nonumber \\
    &\,&\,+N\xi^0_{(1)}\left\{N\xi^0_{(2)}+\xi^0_{(1)}\left[\delta N^{(1)}+\frac{1}{2}\partial_\tau\left(N\xi^0_{(1)}\right)\right]\right\}\left[\boldsymbol{\Omega C}^{(2)}\right]_{ab}\left[\boldsymbol{\Omega C}^{(1)}\right]_b \nonumber \\
    &\,&\,+\frac{1}{6}\left(\xi^0_{(1)}N\right)^3\left\{\left[\boldsymbol{\Omega C}^{(2)}\boldsymbol{\Omega C}^{(2)}\right]_{ab}\left[\boldsymbol{\Omega C}^{(1)}\right]_b+\boldsymbol{\Omega}_{ab}\boldsymbol{C}^{(3)}_{bcd}\left[\boldsymbol{\Omega C}^{(1)}\right]_c\left[\boldsymbol{\Omega C}^{(1)}\right]_d\right\} \nonumber \\
    &\,&\,+\left\{\xi^0_{(2)}N+\xi^0_{(1)}\left[\delta N^{(1)}+\frac{1}{2}\partial_\tau\left(N\xi^0_{(1)}\right)\right]\right\}\left[\boldsymbol{\Omega C}^{(2)}\right]_{ab}\boldsymbol{\delta z}^{(1)}_b \nonumber \\
    &\,&\,+\frac{1}{2}\left(\xi^0_{(1)}N\right)^2\left\{\left[\boldsymbol{\Omega C}^{(2)}\boldsymbol{\Omega C}^{(2)}\right]_{ab}\boldsymbol{\delta z}^{(1)}_b+\boldsymbol{\Omega}_{ab}\boldsymbol{C}^{(3)}_{bcd}\left[\boldsymbol{\Omega C}^{(1)}\right]_c\boldsymbol{\delta z}^{(1)}_d\right\} \nonumber \\
    &\,&\,+\xi^0_{(1)}N\left[\boldsymbol{\Omega C}^{(2)}\right]_{ab}\boldsymbol{\delta z}^{(2)}_b+\frac{1}{2}\xi^0_{(1)}N\boldsymbol{\Omega}_{ab}\boldsymbol{C}^{(3)}_{bcd}\boldsymbol{\delta z}^{(1)}_c\boldsymbol{\delta z}^{(1)}_d,
\eea
that we organize in terms of powers of the gauge-parameters, \ie the three first lines are cubic in $\xi^0$, the forth and fifth are quadratic in $\xi^0$ and linear in $\boldsymbol{\delta z}$, and the last line is linear in $\xi^0$ and quadratic in $\boldsymbol{\delta z}$.

\bibliographystyle{JHEP}
\bibliography{ref}

\end{document}